%% file: sir-hawkes.tex
\PassOptionsToPackage{svgnames}{xcolor}

\documentclass[sigconf]{acmart}

\usepackage{booktabs} 
\usepackage{etoc} 

\usepackage{amsmath,bbm,dsfont}
\usepackage{amsfonts}
\usepackage{amssymb}

\usepackage{verbatim}
\usepackage{graphicx}
\usepackage{subfig}
\usepackage{epstopdf}
\DeclareGraphicsExtensions{.pdf,.eps,.png,.jpg,.tif,.tiff,.ps}
\graphicspath{{img//}}

\usepackage{soul}
\usepackage[colorinlistoftodos]{todonotes}
\usepackage[svgnames]{xcolor}

\newcommand{\overbar}[1]{\mkern 1.5mu\overline{\mkern-1.5mu#1\mkern-1.5mu}\mkern 1.5mu}

\usepackage{xspace}
\newcommand{\Active}{{\sc ActiveRT}\xspace}
\newcommand{\News}{{\sc News}\xspace}
\newcommand{\Seismic}{{\sc Seismic}\xspace}

\definecolor{navy}{rgb}{0.1, 0.1, 0.8}
\definecolor{gray}{rgb}{0.4, 0.4, 0.4}
\definecolor{myblue}{rgb}{.8, .8, 1}
\definecolor{olive}{rgb}{0.1, 0.5, 0.1}

\newcommand{\eat}[1]{}
\newcommand{\rev}[1]{{#1}}

\newcommand{\verify}[1]{{}}
\newcommand{\NOTE}[2]{ }
\newcommand{\TODO}[2]{}
\newcommand{\nb}[1]{}

\newcommand{\fp}[2]{\frac{\partial #1}{\partial #2}}

\newcommand{\secmoveup}{\vspace{-.0mm}} 
\newcommand{\eqmoveup}{\vspace{0.0mm}} 
\newcommand{\captionmoveup}{\eqmoveup\vspace{-0.0mm}}   

\newcommand{\squishlist}{
 \begin{list}{$\bullet$}
  { \setlength{\itemsep}{0pt}
     \setlength{\parsep}{3pt}
     \setlength{\topsep}{3pt}
     \setlength{\partopsep}{0pt}
     \setlength{\leftmargin}{1.5em}
     \setlength{\labelwidth}{1em}
     \setlength{\labelsep}{0.5em} } }

\newcommand{\squishlisttwo}{
 \begin{list}{$\bullet$}
  { \setlength{\itemsep}{0pt}
    \setlength{\parsep}{0pt}
    \setlength{\topsep}{0pt}
    \setlength{\partopsep}{0pt}
    \setlength{\leftmargin}{1.5em}
    \setlength{\labelwidth}{1.5em}
    \setlength{\labelsep}{0.5em} } }

\newcommand{\squishend}{
  \end{list}  }

\copyrightyear{2018}
\acmYear{2018} 
\setcopyright{iw3c2w3}
\acmConference[WWW 2018]{The 2018 Web Conference}{April 23--27, 2018}{Lyon,
France}
\acmBooktitle{WWW 2018: The 2018 Web Conference, April 23--27, 2018, Lyon, France}
\acmPrice{}
\acmDOI{10.1145/3178876.3186108}
\acmISBN{978-1-4503-5639-8/18/04}

\begin{document}
\etocdepthtag.toc{mtchapter} 

\title{SIR-Hawkes: Linking Epidemic Models and Hawkes Processes for Information Diffusion in a Finite Population}
\title{SIR-Hawkes: Linking Epidemic Models and Hawkes Processes for Finite Population Diffusion Modeling}
\title{SIR-Hawkes: Linking Epidemic Models and Hawkes Processes to Model Diffusions in Finite Populations}


\author{Marian-Andrei Rizoiu}
\orcid{orcid.org/0000-0003-0381-669X}
\affiliation{%
  \institution{ANU \& Data61 CSIRO}
  \city{Canberra}
  \country{Australia}
}

\author{Swapnil Mishra}
\affiliation{%
  \institution{ANU \& Data61 CSIRO}
  \city{Canberra}
  \country{Australia}
}

\author{Quyu Kong}
\affiliation{%
  \institution{ANU \& Data61 CSIRO}
  \city{Canberra}
  \country{Australia}
}

\author{Mark Carman} 
\affiliation{%
 \institution{Monash University}
 \city{Melbourne} 
 \country{Australia}
}

\author{Lexing Xie}
\affiliation{%
  \institution{ANU \& Data61 CSIRO}
  \city{Canberra}
  \country{Australia}
}

\renewcommand{\shortauthors}{Rizoiu et al.}
\renewcommand{\shorttitle}{SIR-Hawkes}

\begin{abstract}
	\input{0-abstract}
\end{abstract}

%
%
%

\maketitle

\input{1-introduction}

\input{2-prerequisites}

\input{3-linking-models}

\input{4-final-size-distribution}

\input{5-estimating-N}

\input{6-results}

\input{7-related-work}

\input{8-conclusion}

\vspace{0.2cm}
{\noindent \small
\textbf{Acknowledgments.}
This material is based on research sponsored by the Air Force Research Laboratory, under agreement number FA2386-15-1-4018. We thank the National Computational Infrastructure (NCI) for providing computational resources, supported by the Australian Government.
We thank Aditya Krishna Menon for insightful discussions.
}

\bibliographystyle{ACM-Reference-Format}
\bibliography{biblio} 

\input{appendix}


\end{document}

%% file: 0-abstract.tex
%
Among the statistical tools for online information diffusion modeling, both epidemic models and Hawkes point processes are popular choices.
The former originate from epidemiology, and consider information as a viral contagion which spreads into a population of online users.
The latter have roots in geophysics and finance, view individual actions as discrete events in continuous time, and modulate the rate of events according to the self-exciting nature of event sequences.
Here, we establish a novel connection between these two frameworks.
Namely, the rate of events in an extended Hawkes model is identical to the rate of new infections in the Susceptible-Infected-Recovered (SIR) model after marginalizing out recovery events -- which are unobserved in a Hawkes process.
This result paves the way to apply tools developed for SIR to Hawkes, and vice versa.
It also leads to HawkesN, a generalization of the Hawkes model which accounts for a finite population size.
Finally, we derive the distribution of cascade sizes for HawkesN, inspired by methods in stochastic SIR.
Such distributions provide nuanced explanations to the general unpredictability of popularity:
the distribution for diffusion cascade sizes tends to have two modes, one corresponding to large cascade sizes and another one around zero.

%% file: 1-introduction.tex

\section{Introduction}


The research community has long been aware of the importance of the word-of-mouth phenomenon in information dissemination and in shaping user behavior in online and offline environments.
In this paper, we study how information spreads online by modeling its underlying mechanism, i.e. how it passes from individual to individual.
The aim is to link individual actions to collective-level measures, such as popularity or fame.

This work addresses three open questions concerning two classes of approaches mainly used for modeling online diffusions: \emph{epidemic models} and \emph{Hawkes point processes}.
The first open question regards the relationship between \rev{these} two models.
Epidemic models emerged from the field of epidemiology, and consider information as a viral contagion which spreads within a population of online users; 
Hawkes models have been mainly used in finance and geophysics, and view individual broadcasts of information as events in a stochastic point process.
\textbf{Despite having the origins in different disciplines, these two models describe the stochastic series of discrete events; is there an inherent connection between them?}
The second question is about designing more expressive diffusion models.
Hawkes processes are the de facto modeling choice for social media processes, mainly because they can be easily customized to account for social factors such as the influence of users~\cite{Zhao2015,Gomez-Rodriguez2016}, the length of ``social memory''~\cite{Mishra2016,Shen2014} and the inherent content quality~\cite{Martin2016}.
\textbf{Can we employ notions from epidemic models to design a Hawkes process more adept at describing online diffusions?}
\rev{The third question concerns predicting the final size of the cascade, which intuitively reflects the popularity of the underlying message.
Previous work~\cite{Shen2014,Zhao2015,Rizoiu2017,Mishra2016} predict a single value for the expected future popularity, however it is well known that popularity is hard to predict. 
There are many random factors lead to high variance in prediction~\cite{Watts2011}.
\textbf{Can we compute the size distribution, to explain the high variance and hence the unpredictability?}
}

In this work, we address all three questions above, by drawing for the first time the connection between epidemic models and point processes, validating it both theoretically and also empirically on three large publicly available datasets of retweet cascades.

We answer the first question by studying the previously unexplored link between the Susceptible-Infected-Recovered (SIR) epidemic model~\cite{Kermack1927} and the Hawkes \rev{processes}.
\rev{The key to the link is in the modeling of the word-of-mouth process: 
we regard each new each new broadcast from one user to another as an event in Hawkes, and analogous to a new infection in SIR.}
Starting from this observation, we show that the rate of events in \rev{an extended Hawkes} model is identical to the rate of new infections in the SIR model, \rev{after} taking the expectation over recovery events -- which are unobserved in the Hawkes process.
This is significant, as it indicates that tools developed for one approach can be applied to the other.

To answer the second question, we propose HawkesN, an extension of the Hawkes model with a finite population.
The Hawkes process~\cite{Hawkes1971} has no upper limit for the number of events that may occur.
This is hardly a realistic assumption for social media processes such as information diffusion, which relies on a finite underlying population of humans, each broadcasting a message a finite number of times.
We introduce a parameter $N$, denoting the \rev{finite total size of the population}, and we modulate the event rate by the available population.
\rev{We study the estimation of $N$ from data and we construct a lower bound statistic to detect when parameter $N$ does not have a valid solution.}
We show that the HawkesN model explains better longer event sequences.

To address the third question, we construct a probability distribution over future cascade size
by applying a Markov chain technique developed for SIR to a diffusion cascade which has been partially observed and fitted using HawkesN.
Based on our observations on a large sample of real diffusion cascades, we also provide a nuanced explanation for the main-stream belief that popularity is unpredictable.
The distribution shows two peaks:
the larger peak corresponds to the cascade extinguishing quickly after its beginning;
the smaller peak corresponds to a large cascade size.
At the beginning of the cascade it is impossible to distinguish between the two cases, however the posterior probability distribution after observing a prefix of the cascade can be updated to account for the observed events.

\textbf{The main contributions of this work include:}
\begin{itemize}
	\item We show a previously unexplored connection between two different classes of approaches -- epidemic models and Hawkes point processes -- by showing that the rate of events in HawkesN is identical to the expected rate of new infections in SIR after marginalizing out recovery events.

	\item We introduce HawkesN -- a novel class of Hawkes processes in which event intensity is modulated by the remaining population size -- and we show it generalizes better to unseen data than the state-of-the-art modeling. 
	
	\item We study the estimation of population size from observed data and we construct a lower bound statistic to detect when parameter $N$ does not have a valid solution.
	
	\item We use a Markov chain tool from epidemic model theory to predict the distribution of the final size of a cascade.
	We provide a nuanced explanation for the main-stream belief that popularity is unpredictable.
	
	\item We construct \Active -- a new Twitter cascades benchmark dataset, publicly available (together with the HawkesN simulation and fitting R code) at: \texttt{\url{https://github.com/computationalmedia/sir-hawkes}}
\end{itemize}

%% file: 2-prerequisites.tex

\secmoveup 
\section{Prerequisites}

\rev{In this section, we briefly review a few key concepts of the Poisson and Hawkes~\cite{Hawkes1971} point processes (in Sec.~\ref{subsec:hawkes-model}),
and of the SIR epidemic model and its bivariate process formulation (in Sec.~\ref{subsec:sir-model}).}

\secmoveup 
\subsection{Poisson and Hawkes processes}
\label{subsec:hawkes-model}

\textbf{The Poisson processes.}
A point process is a random process whose realizations consists of event times $t_1,t_2, \ldots$ 
\cite{Daley2008}, where $t_j$ denotes the time of occurrence of the $j$-th event. 
\rev{In a \emph{homogeneous} Poisson processes, the inter-arrival times $\tau_j = t_j - t_{j-1}$ are random variables \textit{i.i.d.} exponentially distributed with parameter $\lambda$ -- 
also called the \emph{event rate} of the Poisson process.
In \emph{non-homogeneous} Poisson processes, the event rate is a deterministic time-continuous function $\lambda(t)$, which} defines the probability of an event occurring in the infinitesimal interval around time $t$.
Formally:
\begin{align} \label{eq:non-homog-poisson}
	\mathds{P}(N_{t+h}=n+m\,|\,N_t=n)&=\lambda(t) h+o(h) & \text{when } m=1 \nonumber\\
	\mathds{P}(N_{t+h}=n+m\,|\,N_t=n)&=o(h) &\text{when } m>1 \nonumber \\
	\mathds{P}(N_{t+h}=n+m\,|\,N_t=n)&=1-\lambda(t) h + o(h) &\text{when } m=0 
	\eqmoveup
\end{align}
where $o(h)$ is a function so that $\lim_{h \downarrow 0}\frac{o(h)}{h} = 0$;
\rev{$N_t$ is the counting process associated with the point process, i.e. a random variable which counts the number of events up to (and including) time $t$.}

\textbf{The Hawkes process}~\cite{Hawkes1971} is a self-exciting point
process, in which 
\rev{each previous event occurred at the time $t_j < t$ generates new events at the rate $\phi(t - t_j)$ -- also called the \emph{kernel} of the Hawkes process.} 
The event rate \rev{of} a Hawkes process is \rev{a stochastic function dependent on previous event times,} defined as:
\begin{equation} \label{eq:phi-hawkes}
	\lambda(t) = \mu + \sum_{t_j < t} \phi(t-t_j)
	\eqmoveup
\end{equation}
which models the following process~\cite{Laub2015}:
\rev{a new event either enters the system at the background rate $\mu$;
or it is generated by a previous event at the rate of the corresponding kernel function.}

\subsection{The SIR Model}
\label{subsec:sir-model}

%

The Susceptible-Infected-Recovered (SIR) model defines three classes of individuals \rev{(also known as compartments)}: 
those \emph{susceptible} to infection, 
those currently \emph{infected} (and therefore infectious) and 
those \emph{recovered} from the infection and no longer infective.
\rev{SIR models the following process:
when a susceptible individual meets an infectious individual, the former becomes infected at the rate $\beta$;
infected individuals recover from the infection at a constant rate $\gamma$.}

\rev{\textbf{Deterministic SIR.}
In the deterministic SIR, the individuals and their assignment to each of the three compartments are not observed.
The temporal dynamics of the sizes of each of the compartments are} governed by the following ordinary differential equations~\cite{Allen2008}: 
\begin{align}
	\frac{dS(t)}{dt} &= -\beta \frac{S(t)}{N} I(t) \label{eq:sir-ds} \\
	\frac{dI(t)}{dt} &= \beta \frac{S(t)}{N} I(t) - \gamma I(t) \label{eq:sir-di} \\
	\frac{dR(t)}{dt} &= \gamma I(t). \label{eq:sir-dr}
	\eqmoveup
\end{align}
\rev{$S(t)$, $I(t)$ and $R(t)$ are deterministic functions, denoting the sizes at time $t$ of the susceptible, infected and recovered populations, respectively;
$N = S(t)+I(t)+R(t)$ is the total population size.}



There are a number of assumptions made by the SIR model.
Firstly, it assumes that the population is homogeneous 
and individuals meet any other individual uniformly at random.
Secondly, it assumes that all rates are constant: 
\rev{the infection rate $\beta$} and the recovery rate $\gamma$.
Thirdly, it assumes that the population has no births and no deaths -- \rev{i.e. $N$ is constant throughout the unfolding of the epidemic.
The last} assumption holds when the speed of the epidemic outpaces considerably the speed of change in the population -- e.g., an average retweet diffusion only lasts minutes, compared to years of expected activity of a user on Twitter.

%

\textbf{Stochastic SIR.}
\rev{Several stochastic formulations of the SIR model have been proposed~\cite{Allen2008}, which model the}
behavior of independent and identically distributed agents.
The actions of the agents are described by the same set of holistic rules defined in Eq.~\eqref{eq:sir-ds}-\eqref{eq:sir-dr} \rev{and the same assumptions detailed above}~\cite{Bobashev2007}.
\rev{One such stochastic formulation of SIR is the}
bivariate point process \rev{representation}~\cite{Yan2008}, in which two types of events occur: \emph{infection events} and \emph{recovery events}.
The $j$-th infected individual in the SIR process gets infected at time $t^I_j$ and recovers at time $t^R_j$.
\rev{Therefore,} to each infection event corresponds a recovery event.
\rev{$S_t$, $I_t$ and $R_t$ are discrete random variable taking integer values.
They are the stochastic counterparts of $S(t)$, $I(t)$ and $R(t)$ respectively.
Eq.~\eqref{eq:sir-ds}-\eqref{eq:sir-dr} can be written as stochastic differential equations, but for ease of following we will keep referring to Eq.~\eqref{eq:sir-ds}-\eqref{eq:sir-dr} in the rest of this paper.
We define the \emph{time to recovery} (i.e. the time the individual $j$ is infectious)} as: $\tau_j = t^R_j - t^I_j$.
\rev{From Eq.~\eqref{eq:sir-dr} results that} the times to recovery are distributed exponentially with parameter $\gamma$, and infections last on average $\frac{1}{\gamma}$ units of time.

Let $C_t$ be the counting process of the \emph{infection process} and $R_t$ the counting process of the \emph{recovery process}.
\rev{Note that $C_t = N - S_t$ is total number of occurred infections (regardless if they are still infectious) and it is distinct from $I_t$ (number of infectious at time $t$).}
%
%
\rev{Let} $\mathcal{H}_t$ \rev{be} the history of the \rev{bivariate} epidemic process up to time $t$, \rev{i.e. $\mathcal{H}_t = \{ t^I_1, t^I_2, \ldots, t^R_1, t^R_2, \ldots \}$.}
It can be shown that the rate of \emph{new infections} $\lambda^I(t)$ and the rate of \emph{new recoveries} $\lambda^R(t)$ are:
\begin{equation} \label{eq:sir-bivariate-rates}
	\lambda^I(t) = \beta \frac{S_t}{N} I_t ; \;\; \lambda^R(t) = \gamma I_t.
	\eqmoveup
\end{equation} 
\rev{We sketch the proof for the previous statement.
\citet{Yan2008} derives the probability of a new infection at time $t$ given $\mathcal{H}_t$ as:}
\begin{align*}
	\mathds{P}(C_{t+\delta t} - C_t = 1 | \mathcal{H}_t) &= \beta \frac{S_t}{N}I_t \delta t + o(\delta t) \nonumber \\
	\mathds{P}(C_{t+\delta t} - C_t > 1 | \mathcal{H}_t) &= 0 \nonumber \\
	\mathds{P}(C_{t+\delta t} - C_t = 0 | \mathcal{H}_t) &= 1 - \beta \frac{S_t}{N}I_t\delta t + o(\delta t),    \label{eq:intesity-Ct}
	\eqmoveup
\end{align*} 
Given Eq.~\eqref{eq:non-homog-poisson}, the new infections process is a temporal point process of intensity $\beta \frac{S_t}{N}I_t$.
\rev{$\lambda^R(t)$ is derived similarly.}
%

Fig.~\ref{fig:sir-two-processes} illustrates an SIR \rev{realization} as a bivariate point process: 
five infection events occur at times $t^I_1,.., t^I_5$ (shown in red);
five recovery events occur at $t^R_1, ..t^R_5$ (shown in blue).
The middle panel of Fig.~\ref{fig:sir-two-processes} shows the size of the infectious population \rev{$I_t$} over time.
Each new infection event increments \rev{$I_t$}, and each new recovery decreases \rev{$I_t$} by one.
The bottom panel of Fig.~\ref{fig:sir-two-processes} shows the \rev{corresponding} new infection and new recovery rates.
Initially, $\lambda^I(t)$ is significantly  higher than $\lambda^R(t)$. 
As the number of susceptible individuals \rev{$S_t$} gets depleted, the term \rev{$\frac{S_t}{N}$ in Eq.~\eqref{eq:sir-bivariate-rates}} inhibits $\lambda^I(t)$ which becomes zero after the fifth infection ($S_t = 0, t \geq t^I_5$).
The new recovery rate also becomes zero after the last infected individual recovers ($I_t = 0, t \geq t^R_5$).

\rev{\textbf{The connection between deterministic and stochastic SIR}
is that} the mean behavior of the stochastic process asymptotically approaches that of the deterministic process~\citep{Allen2008,Yan2008}.
\rev{The connection between the deterministic and the stochastic population sizes is $S(t) = \mathds{E}_{\mathcal{H}_t}\left[ S_t \right]$, $I(t) = \mathds{E}_{\mathcal{H}_t}\left[ I_t \right]$ and $R(t) = \mathds{E}_{\mathcal{H}_t}\left[ R_t \right]$~\cite{Allen2008}.}
Our own results simulating the two variants are presented in an online supplement~\cite{supplemental}.

As will be elaborated in \rev{Sec.~\ref{subsec:linking-two-models}}, \rev{the bivariate point process} SIR \rev{formulation provides} the link to the Hawkes point processes.

\input{fig2-sir-as-two-processes}

%% file: fig2-sir-as-two-processes.tex

\begin{figure}[tbp]
	\centering
	\includegraphics[width=0.47\textwidth]{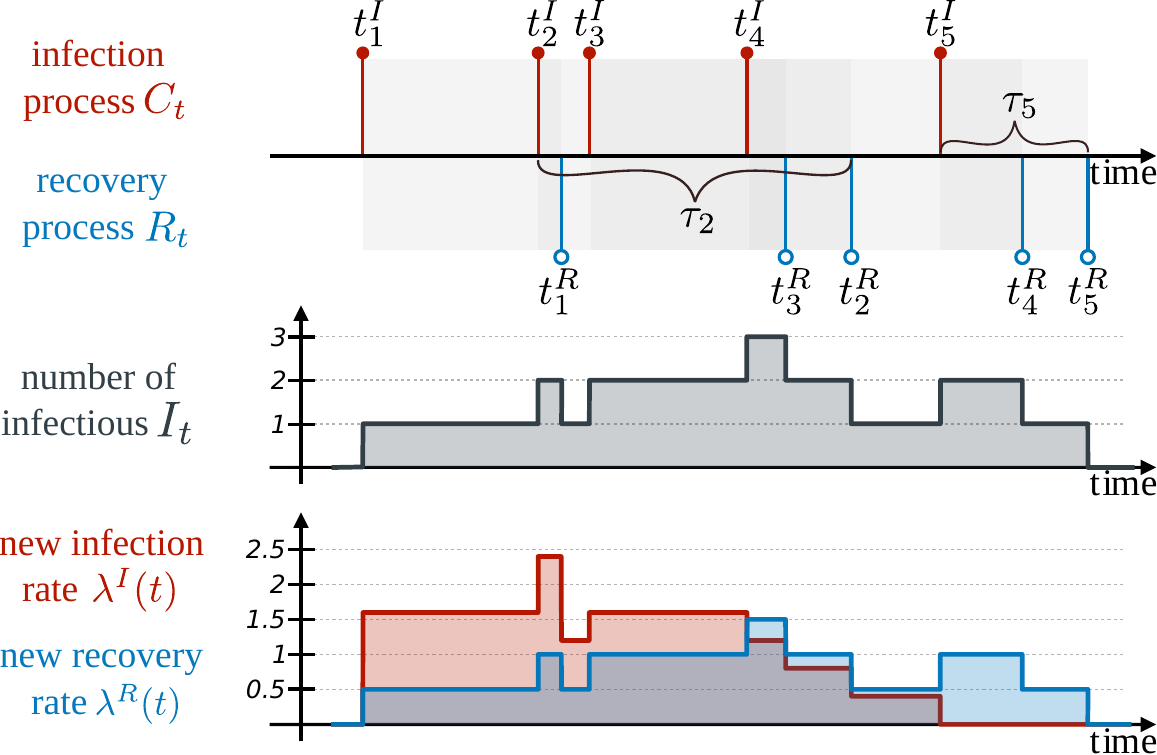}
	
	\caption{
		\rev{An illustration of SIR as a bivariate point process: the infection process and the recovery process.}
		\emph{(top panel)} \rev{The $j^{th}$ individual} gets infected at time $t^I_j$ and recovers at $t^R_j$. 
		\rev{The \emph{time to recovery} $\tau_j = t^R_j - t^I_j$ is the period the individual stays infectious.}
		\emph{(middle panel)} The size of the infectious population $I(t)$ over time.
		\emph{(lower panel)} The infection rate $\lambda^I(t)$ and the recovery rate $\lambda^R(t)$ for the SIR parameters: $N = 5$, $\beta = 2$, $\gamma = 0.5$.
	}
	\label{fig:sir-two-processes}
	\captionmoveup
\end{figure}

%% file: 3-linking-models.tex

\section{Linking epidemic models and Hawkes processes}

\rev{We first propose \emph{HawkesN}, a generalization of the Hawkes process with finite population (in Sec.~\ref{subsec:hawkesN}) and we show the connection between HawkesN and the SIR epidemic model (in Sec.~\ref{subsec:linking-two-models}).}

\subsection{HawkesN: a process in finite population}
\label{subsec:hawkesN}

\rev{We generalize the Hawkes model~\cite{Hawkes1971} to account for finite population sizes.
Intuitively cascades not only follow self-exciting word of mouth diffusions, but they are also limited by the size of the relevant community. }
The effect of \rev{introducing the finite population size $N$} is that \rev{the event rate at time $t$} is modulated by the available population.
To the best of our knowledge, no prior work on modeling social processes using Hawkes models had accounted for a finite underlying population. 

The event rate function in HawkesN is defined as:

\begin{equation} \label{eq:hawkesN}
	\lambda^H(t) = \left( 1  -  \frac{N_t}{N} \right) \left[ \mu + \sum_{t_j < t} \phi(t - t_j) \right], 
	\eqmoveup	
\end{equation}
where $\phi(t - t_j)$ can be the same kernel function used with Hawkes, 
and $N_t$ is the counting process associated with the point process.
\rev{Both $\lambda^H(t)$ and $N_t$ are right-continuous functions.}
%
The term $1 - \frac{N_t}{N}$ scales the event rate at time $t$ with the proportion of the events which can still occur after time $t$.
When $t = 0$, we have $\lambda^H(t) = \mu$.
When $N_t = N$, we have $\lambda^H(t) = 0$, i.e., there will be no more new events when the pool of users who can act is exhausted.
\rev{When $N \rightarrow \infty$, Eq.~\eqref{eq:hawkesN} simplifies to Eq.~\eqref{eq:phi-hawkes}.
In other words, the Hawkes process is a special case of HawkesN with infinite population.}

Fig.~\ref{fig:HawkesN} illustrates the HawkesN process for an information diffusion in a population of five users ($N = 5$).
Each user takes an action at most once, represented as event time $t_j, j=1..5$. 
The corresponding counting process $N_t$ is shown in the middle plot.
Events $t_2 .. t_5$ are considered to have been triggered by event $t_1$.
The bottom panel compares \rev{the \emph{offspring rates} -- the rate of events generated by the first event at $t_1$ -- for Hawkes (denoted as $\phi_1(t)$) and HawkesN (denoted as $\phi^H_1(t)$).}
In HawkesN, the population modulates the event rate by decreasing it after each new event and the event rate becomes zero after $t_5$.
\rev{The Hawkes process does not take into account the population size, i.e. it will have $\phi(t)>0$ in Eq.~\ref{eq:phi-hawkes} even after $t_5$.}

\rev{We use the exponential kernel function for HawkesN:}
\begin{equation} \label{eq:kernel-function-hawkesN}
	\phi(\tau) = \kappa \theta e^{-\theta\tau}
	\eqmoveup
\end{equation}
$\kappa$ is a \rev{scaling factor} and $\theta$ is the parameter of the exponential function which models the decay of \emph{social memory}.
The exponential kernel is a common choice in literature~\cite{Mishra2016,Zarezade2017,Zhao2015,Shen2014,Bao2015,Ding2015,Gao2015}.
\rev{Other kernels have been used with Hawkes}, including power-law functions~\cite{Helmstetter2002,Crane2008,Mishra2016,Kobayashi2016} and Rayleigh functions~\cite{Wallinga2004}.
Using HawkesN with non-exponential kernel functions is left for future work.
%

\input{fig1-hawkesN}

\subsection{Linking HawkesN and SIR}
\label{subsec:linking-two-models}

We now present our main result, Theorem~\ref{theorem:expected-equivalence}, which links stochastic SIR and the HawkesN process.

\textbf{Intuition.}
When modeling information diffusion, both SIR and HawkesN model the same phenomenon: users come into contact with the diffused content, which they further broadcast to other users.
Each new broadcast is modeled as a new event in HawkesN, and as a new infection in SIR.
The key to linking HawkesN and SIR models is the \rev{conceptual similarity} between an event in HawkesN and a new infection in SIR. 
In HawkesN, past events generate new events at the rate $\phi(t)$, which is exponentially time-decaying in Eq.~\eqref{eq:kernel-function-hawkesN}.
In SIR, an infectious individual $j$ infects susceptible individuals at a rate of $\frac{\beta S_t}{N}$ during \rev{the time it is infectious} $\tau_j$, which is exponentially distributed with parameter $\gamma$ (discussed in Sec.~\ref{subsec:sir-model}).

\begin{theorem} \label{theorem:expected-equivalence}
	Suppose the new infections in a stochastic SIR process of parameters $\{\beta, \gamma, N\}$
	follow a point process of intensity $\lambda^I(t)$.
	Suppose also the events in a HawkesN process with parameters $\{\mu, \kappa, \theta, N\}$ have the intensity $\lambda^H(t)$ (Eq~\ref{eq:hawkesN}).
	\rev{Let $\mathcal{T} = \{\tau_1, \tau_2, \ldots\}$ be the set of the times to recovery of the infected individuals in SIR.}
	The expectation of $\lambda^I(t)$ over $\mathcal{T}$ is equal $\lambda^H(t)$:
	\begin{equation*} 
		\mathds{E}_\mathcal{T}[ \lambda^I(t)] = \lambda^H(t),
		\eqmoveup
	\end{equation*}
	when $\mu = 0$, $\beta = \kappa \theta$, $\gamma = \theta$.
\end{theorem}
\noindent Note that both $\mathds{E}_\mathcal{T}[ \lambda^I(t)]$ and $\lambda^H(t)$ are random functions, as they depend on the random infection times $t^I_j$ (for SIR) and the random event times $t_j$ (for HawkesN).
The expectation only removes the randomness from the recovery times $t^R_j$ in SIR.

The rest of this section proves this theorem.


\textbf{The expected new infection rate.}
\rev{We express $S_t$ and $I_t$ in Eq.~\eqref{eq:sir-bivariate-rates} using indicator functions of the infection event times and the times to recovery:}
\begin{align}
	S_t &= N - C_t = N - \sum_{j \geq 1} \mathds{1}(t^I_j < t) \nonumber \\
	I_t &= C_t - R_t = \sum_{j \geq 1} \mathds{1}(t^I_j < t, t^R_j > t) = \sum_{t^I_j < t} \mathds{1}(t^I_j + \tau_j > t) . \label{eq:I-S-stochastic}
	\eqmoveup
\end{align}
We examine a point process consisting only of the infection events $\{t^I_j\}$.
The event rate in this process is obtained by marginalizing out times of recovery: 
\begin{align}
	\mathds{E}_\mathcal{T} \left[ \lambda^I(t) \right] 
	&\overset{\eqref{eq:sir-bivariate-rates},\eqref{eq:I-S-stochastic}}{=} \mathds{E}_\mathcal{T} \left[ \beta \frac{S_t}{N} \sum_{t^I_j < t} \mathds{1}(t^I_j + \tau_j > t) \right] \nonumber \\
	&= \sum_{t^I_j < t} \mathds{E}_\mathcal{T} \left[ \beta \frac{S_t}{N} \mathds{1}(t^I_j + \tau_j > t) \right] \nonumber \\
	&= \sum_{t^I_j < t}  \int_{0}^{\infty}\beta \frac{S_t}{N} \mathds{1}(t^I_j + \zeta > t) r(\zeta) d\zeta \nonumber \\
	&= \sum_{t^I_j < t}  \beta \frac{S_t}{N} \int_{t-t^I_j}^{\infty} r(\zeta) d\zeta , \label{eq:exp-lambda-i}
	\eqmoveup
\end{align}
where $r(\zeta)$ is the exponential probability distribution function for the time to recovery (cf. Sec.~\ref{subsec:sir-model}).
Knowing $S_t = N - C_t$, we obtain:
\begin{equation} \label{eq:SIR-expectation}
	\mathds{E}_\mathcal{T} \left[ \lambda^I(t) \right]  = \left(1 - \frac{C_t}{N} \right) \sum_{t^I_j < t} \beta e^{-\gamma (t - t^I_j)} .
	\eqmoveup
\end{equation}

We can see that Eq.~\eqref{eq:hawkesN} (with the exponential kernel in Eq.~\eqref{eq:kernel-function-hawkesN}) and Eq.~\eqref{eq:SIR-expectation} are identical when $N_t = C_t$ and under the parameter equivalence in Theorem~\ref{theorem:expected-equivalence}.
That is to say, the new infection point process in an SIR model and the HawkesN point-process with no background event rate are described by the same conditional intensity.
This completes the proof of Theorem~\ref{theorem:expected-equivalence}.
We also demonstrate the equivalence empirically, through simulation and subsequent parameter fitting, in the online supplement~\cite{supplemental}.

%% file: fig1-hawkesN.tex

\begin{figure}[tbp]
	\centering
	\includegraphics[width=0.47\textwidth]{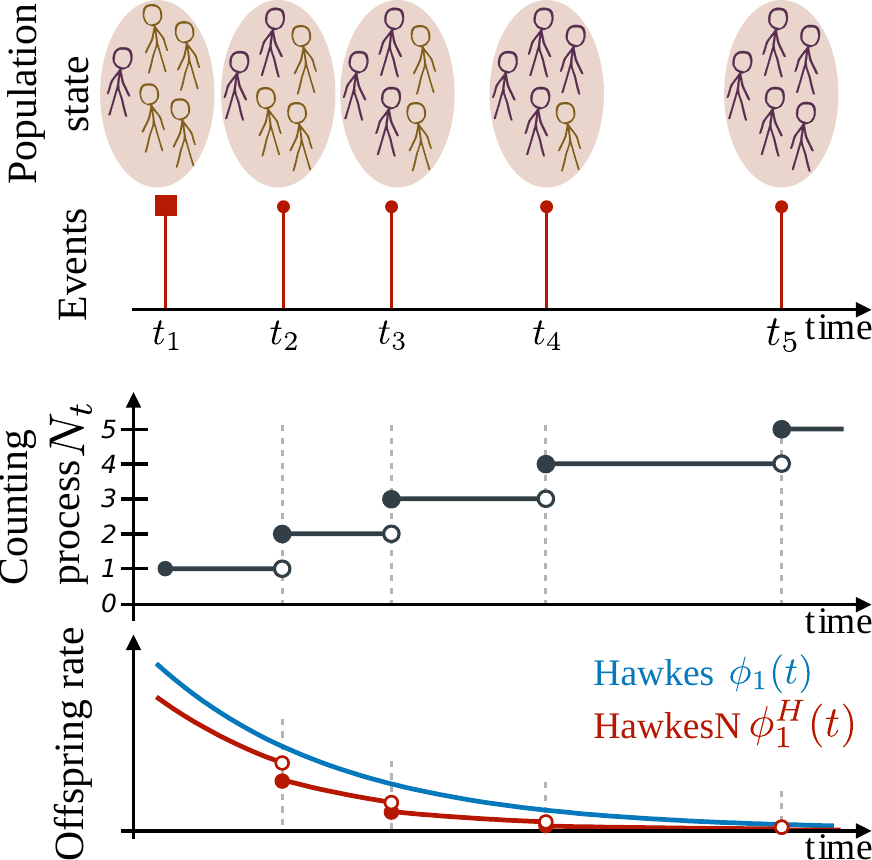}
	
	\caption{
		\rev{An example diffusion illustrating the finite population effects in self-exciting process.}
		\emph{(top panel)} 
		\rev{An event refers to the $j^{th}$ user taking an action at time $t_j$ (e.g. posting a tweet).}
		%
		The state of the user population is shown at each time $t_j$:
		\textcolor{Purple}{purple users} have performed the past observed \rev{actions}; 
		\textcolor{Chocolate}{orange users} are yet to perform any action.
		\emph{(middle panel)} The counting process $N_t$ increases by one with each event;
		\emph{(lower panel)} 
		The \rev{\emph{offspring rate} $\phi_1(t)$ -- the} rate of events generated by this first event at time $t_1$, modeled by Hawkes and by HawkesN \rev{(denoted by $\phi_1^H(t)$)}.
	}
	\label{fig:HawkesN}
	\captionmoveup
\end{figure}

%% file: 4-final-size-distribution.tex
\input{fig3-markov-states-probs}

\section{Diffusion Size Distribution}
\label{sec:diffusion-size-distribution}

We compute the probability distribution of the final size of an information diffusion cascade which has been partially observed and fitted using HawkesN, using a Markov chain technique developed for SIR.
In Sec.~\ref{subsec:size-in-SIR} we review known results on the final size distribution of an SIR epidemic.
In Sec.~\ref{subsec:size-in-HawkesN} we employ the equivalence shown in Theorem~\ref{theorem:expected-equivalence} to compute the final size distribution of a cascade modeled with HawkesN.

\subsection{Epidemic size distributions in SIR}
\label{subsec:size-in-SIR}

The final size of an infection is defined as the total number of individuals that have been infected (and recovered) during the epidemic. 
Estimating the final size while the epidemic is in its early stages is a well-studied problem in epidemiology.
In this section, we review a solution to this problem for the stochastic SIR model.

\textbf{SIR as a Markov chain.}
\rev{The stochastic SIR introduced in Sec.~\ref{subsec:sir-model} can be formulated as a}
bivariate continuous time-homogeneous Markov chain~\cite{Allen2008}.
Each state\rev{ $\sigma \in \Sigma$} is uniquely defined by the ordered pair \rev{of random variables} $\{S_t, I_t\}$, denoting the sizes of the susceptible and \rev{the} infected populations. 
$\Sigma$ is the finite space of all possible states, visually represented in Fig.~\ref{fig:markov-chain} as a triangle on the two-dimensional surface with the number of susceptible on the x-axis and the number of infected on the y-axis.
From a given state $\{S_t = s, I_t = i\}$ \rev{(denoted from here on as $\{s, i\}$)}, there are only two other \rev{states in which the system can transition depending on the type of event that occurs:}
a new infection arrives, and the system transitions $\{s, i\} \rightarrow \{s-1, i+1\}$ with the probability $p(\{s-1,i+1\} | \{s,i\})$;
or a new recovery arrives, and the system transitions $\{s, i\} \rightarrow \{s, i-1\}$ with the probability $p(\{s,i-1\} | \{s,i\})$
(shown in the figure by the red and blue arrows, respectively).

\rev{\textbf{Transition matrix and probabilities.}}
The epidemic ends when $I_t = 0$, i.e. no more \rev{infectious} individuals exist to propagate the epidemic.
Consequently, the states $\{s, 0\}$ are \emph{absorbing states} -- once the system arrives in one of these states, it does not transition to any other state.
\rev{From a non-absorbing state $\{s, i\}$, new infections are observed at the rate} $\frac{\beta}{N} s i$ and new recoveries at the rate of $\gamma i$ (see Eq.~\eqref{eq:sir-bivariate-rates}).
\rev{We obtain the transition probabilities:}
\begin{align}
	\eqmoveup
	p(\{s-1,i+1\} | \{s,i\}) &= \frac{\frac{\beta}{N} s i}{\frac{\beta}{N} s i + \gamma i} = \frac{\beta s}{\beta s + N \gamma } , \text{ for } i > 0 \nonumber \\
	p(\{s,i-1\} | \{s,i\}) &= \frac{\gamma i}{\frac{\beta}{N} s i + \gamma i} = \frac{N \gamma}{\beta s + N \gamma} , \text{ for } i > 0. \label{eq:transition-probs}
	\eqmoveup
\end{align}
\rev{Suppose that the states in $\Sigma$ are ordered from 1 to $|\Sigma|$.
We define the \emph{transition matrix} $T = [t_{kl}]$ of size $|\Sigma| \times |\Sigma|$, where $t_{kl} = p (\sigma_k = \{ s_k, i_ k\} | \sigma_l = \{s_l, i_l\})$ is the probability of transitioning from state $\sigma_l$ to state $\sigma_k$.
From Eq.~\eqref{eq:transition-probs} we obtain:
\begin{equation} \label{eq:transition-matrix}
t_{kl} = 
\left\{
\begin{array}{ll}
	\frac{\beta s_l}{\beta s_l + N \gamma } &,s_k = s_l - 1, i_k = i_l + 1, i_l > 0 \\
	\frac{ N \gamma}{\beta s_l + N \gamma } &,s_k = s_l, i_k = i_l - 1, i_l > 0 \\
	1 &, s_k = s_l, i_k = i_l = 0 \\
	0 & , \text{otherwise}
\end{array}
\right. .
\end{equation} 
Note that the sum of each column $j$ in the transition matrix $M$ is equal to 1, as it contains the probabilities of transitioning from $\sigma_l$ to another state.}

\rev{\textbf{Probability state vector and size distribution.}
Let $\pi$ be the probability state vector $\pi \in \mathds{R}^{|\Sigma| \times 1}$, with the $l^{th}$ position of $\pi$ giving the probability that the system is currently found in state $\sigma_l$.
Let $\pi_0 = [0, .., 0, 1, 0, .., 0]$ be the initial probability vector, with the value of 1 corresponding to $\tilde{\sigma}$ the \emph{initial state}, and zero everywhere else.
Starting from $\pi_0$, we compute the probability state vector after one transition as $\pi^{(1)} = M \times \pi_0$.
$\pi^{(2)} = M^2 \times \pi_0$ gives the probabilities after two transitions, $\pi^{(3)} = M^3 \times \pi_0$ after three etc.
Given that there are at most $N-1$ infection events and $N$ recovery events in an SIR realization, the system is guaranteed to converge after $2N-1$ steps~\cite{Allen2008}.
At convergence, all states except the absorbing states have a probability of zero in $\pi^{(2N-1)}$.
We denote as $P(s)$ the value in $\pi^{(2N-1)}$ for the state $\{s, 0\}, s = 0, 1, \ldots, N-I_0$.
}

From an initial state $\tilde{\sigma}$ the system can finish in any of the absorbing states $\{s, 0\}$ with the probability $P(s)$.
The distribution of final size \rev{of the diffusion} is that of the random variable $N' = N - s$.


\subsection{Cascade size distribution in HawkesN}
\label{subsec:size-in-HawkesN}

\rev{Given Theorem~~\ref{theorem:expected-equivalence}, computing the probability distribution over the final size of the cascade after observing an arbitrary number of events, conceptually amounts to changing the initial state $\tilde{\sigma} = \{\tilde{s}, \tilde{i}\}$ and using the method described in Sec.~\ref{subsec:size-in-SIR}.
Suppose we observed $l$ events in HawkesN, we have $\tilde{s} = N - l$.
The recovery events are not observed and the exact size of the infected population $\tilde{i}$ is not known.
We compute its expectation over times to recovery $\mathcal{T}$ as:
\begin{equation} \label{eq:expectation-i-tilde}
	\eqmoveup
	\mathds{E}_\mathcal{T}[\tilde{i}] = \mathds{E}_\mathcal{T} \left[ \sum_{j = 1}^l \mathds{1}(t^I_j + \tau_j > t_l) \right] \overset{Eq.\eqref{eq:exp-lambda-i},\eqref{eq:SIR-expectation}}{=} \sum_{j = 1}^l e^{- \gamma (t_l-t^I_j)}
	\eqmoveup
\end{equation}
and we run the method in Sec.~\ref{subsec:size-in-SIR} starting from the initial state $\tilde{\sigma} = \{ N - l, \mathds{E}_\mathcal{T}[\tilde{i}]\}$.
$\mathds{E}_\mathcal{T}[\tilde{i}]$ in Eq.~\eqref{eq:expectation-i-tilde} is a real number, that we round to the closest integer.

In our discussion in Sec.~\ref{subsec:cascade-size}, we study two probability distributions: 
the \emph{apriori} distribution is computed starting from the initial state $\tilde{\sigma} = \{ N - 1, 1\}$ and it is dependent on model parameters only.
The \emph{aposteriori} distribution is the size distribution after observing $l$ events, it is computed starting from $\tilde{\sigma} = \{ N - l, \mathds{E}_\mathcal{T}[\tilde{i}]\}$ and it is dependent on model parameters and the observed event times $t_1, t_2, \ldots, t_l$ (cf. Eq.~\eqref{eq:expectation-i-tilde}).}

%% file: fig3-markov-states-probs.tex
\begin{figure}[tbp]
	\centering
	\includegraphics[width=0.31\textwidth]{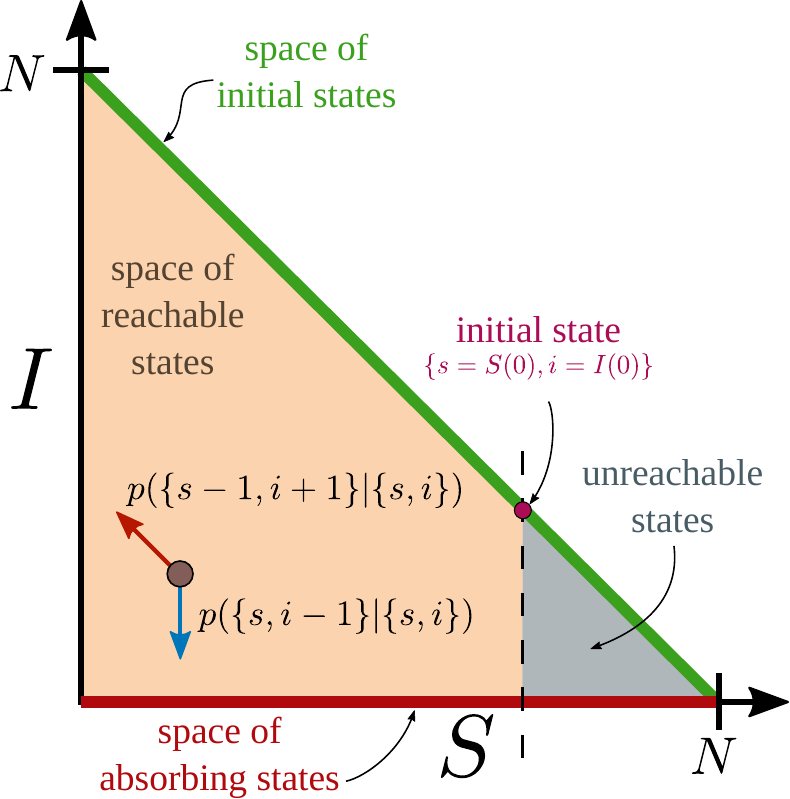}
	
	\caption{
		Visualization of the space of states of the SIR bivariate Markov Chain $\{s, i\}$.
		$s, i \in \mathds{N}$ and $s + i \leq N $, in other words the space of valid states sits under the green line $s+i = N$. 		
		The initial state in an SIR epidemic is always on the \textcolor{ForestGreen}{green line} ($S(0) + I(0) = N$). 
		Given an initial state $\{s = S(0), i = I(0)\}$ (shown by the \textcolor{DarkMagenta}{magenta circle}), the \textcolor{DarkOrange}{orange area} shows the space of reachable states and the \textcolor{gray}{gray area} depicts the \emph{unreachable states}.
		From a state $\{s,i\}$ the system can to $\{s-1,i+1\}$ 
		(new infection); and to $\{s,i-1\}$ 
		(new recovery).
		The absorbing states ($\{s, 0\}$) is shown with a \textcolor{red}{red line}.
	}
	\label{fig:markov-chain}
	\captionmoveup
\end{figure}

%% file: 5-estimating-N.tex
\secmoveup
\section{Fitting HawkesN to data}
\label{sec:fitting-params-hawkesN}

In epidemiology, the size of population $N$ is usually considered a fixed known parameter -- e.g. the number of people in a community.
For online diffusions, it could be possible to estimate $N$ from past diffusions (as discussed in Sec.~\ref{sec:discussion}).
\rev{However,} in this section we analyze the case when $N$ is not known beforehand and needs to be estimated from \rev{observed} data.

\secmoveup
\subsection{The likelihood of HawkesN}
\label{subsec:fitting-hawkesn}

Let $\{t_1, t_2, \ldots, t_n\}$ be  a set of \rev{event times} assumed to have been generated from a HawkesN process described in Sec.~\ref{subsec:hawkesN}.
When modeling diffusion cascades, it is typically assumed that every event apart from \rev{$t_1$} is a reaction to the first event, i.e.,
the background intensity is zero $\mu = 0, \forall t > 0$~\cite{Mishra2016}.
\rev{We estimate the remaining HawkesN parameters $\{ \kappa, \theta, N \}$ by maximizing the log-likelihood function of the point process}
(see the online supplement~\cite{supplemental} or \citet[Ch.~7.2]{Daley2008}):
\begin{equation} \label{eq:hawkes-ll}
	\mathcal{L}(\kappa, \theta, N) = \sum_{j=1}^n \log\lambda^H\left(t_j\right) - \int_0^{t_n} \lambda^H(\tau)\mathrm{d}\tau.
\end{equation}
We further detail the integral term:
\begin{align}
	\int_0^{t_n} \lambda^H(\tau)\mathrm{d}\tau =& \int_0^{t_n} \left( 1  -  \frac{N_{t}}{N} \right) \sum_{t_j < t} \phi(t - t_j) dt \nonumber \\
	=& \sum_{j = 0}^{n-1} \int_{t_j}^{t_n} \left( 1  -  \frac{N_{t}}{N} \right) \phi(t - t_j) dt \nonumber \\
	=& \sum_{j = 0}^{n-1} \sum_{l=j}^{n-1} \frac{N - l}{N} \int_{t_l}^{t_{l+1}} \phi(t - t_j) dt \nonumber \\
	^{cf.~\eqref{eq:kernel-function-hawkesN}} =& \kappa \sum_{j=0}^{n-1 } \sum_{l=j}^{n-1} \frac{N - l}{N} \left[e^{-\theta(t_l - t_j)} - e^{-\theta(t_{l+1} - t_j)} \right]. \label{eq:ll}
	\eqmoveup
\end{align}
Eq.~\eqref{eq:hawkes-ll} is a non-linear objective 
\rev{and there} are a few natural constraints for each of the model parameters, namely: $\theta>0$, $\kappa>0$ and $N \geq n$.
We use the mathematical modeling language AMPL~\cite{fourer1987ampl}, which \rev{provides an interface to different tools for continuous optimization, including automatic gradient computation and solvers.}
We choose as solver Ipopt~\cite{Wachter2006}, \rev{a common choice in literature for large problems with} non-linear objectives. 
More details can be found in the online supplement~\cite{supplemental}.

\secmoveup
\subsection{Estimating population size $N$}
\label{subsec:identifiability}

\rev{Here we examine the case when the population size $N$ is the only unknown. 
The purpose is to identify how difficult it is to retrieve the value of $N$ from data.
Having a value of $N$ for which the derivative is zero is a necessary condition for a local maximum in the log-likelihood function in Eq.~\eqref{eq:hawkes-ll}.} 
%
%
We write the derivative of the log-likelihood with respect to $N$:
\begin{align}
	\frac{\partial \mathcal{L}}{\partial N} =& \sum^n_1 \frac{(\mu(t) + \sum_{t_j < t} \phi(t-t_j)) \frac{j - 1}{N^2}}{(\mu(t) + \sum_{t_j < t} \phi(t-t_j))\frac{N-j + 1}{N}} \nonumber \\
	&- \kappa \sum^{n-1}_{j=0} \sum^{n-1}_{l=j} \frac{l}{N^2} \left[ e^{-\theta(t_l-t_j)} - e^{-\theta (t_{l+1} - t_j)} \right] \nonumber \\
    =& \sum^n_1 \frac{j - 1}{N(N - j+1)} -  \frac{1}{N^2}\sum^{n-1}_{j=0} \sum^{n-1}_{l=j} l\kappa \left[ e^{-\theta(t_l-t_j)} - e^{-\theta (t_{l+1} - t_j)} \right] . \label{eq:dLL}
\end{align}
Knowing that $N - j + 1 \leq N, \forall j = 1..n$, we construct a lower bound for $\frac{\partial \mathcal{L}}{\partial N}$:
\begin{equation} \label{eq:test-bound}
	\frac{\partial \mathcal{L}}{\partial N} \ge \frac{1}{N^2}\underbrace{ \left(\frac{(n-1)n}{2} - \sum^{n-1}_{j=0} \sum^{n-1}_{l=j} l\kappa\left[e^{-\theta(t_l-t_j)} - e^{-\theta (t_{l+1} - t_j)}\right]\right) }_{\mathcal{S}(\kappa, \theta, \{t_1, t_2, \ldots, t_n\})}
\end{equation}
\rev{We define the right hand side of Eq.~\eqref{eq:test-bound} $\mathcal{S}(\kappa, \theta, \{t_1, t_2, \ldots, t_n\})$ as a statistic of the observed} event times $t_j$ and \rev{of fixed} model parameters $\kappa$ and $\theta$.
\rev{The statistic does not depend on $N$.}
\rev{Given parameters and a} set of event times $t_j$, when the statistic $\mathcal{S}(\kappa, \theta, \{t_1, t_2, \ldots, t_n\}) > 0$ it is guaranteed that the log-likelihood function $\mathcal{L}$ keeps monotonically increasing with $N \geq n$, and \rev{no valid solution exists for $N$.}

\begin{table}[tbp]
	\centering
	\caption{Percentage of non-valid $N$ solutions and accuracy of detection using statistic $\mathcal{S}$.}
	\begin{tabular}{rrrrrrr}
  		\toprule
 		Percentage observed & $5\%$ & $10\%$ & $20\%$ & $40\%$ & $80\%$  \\ 
  		\midrule
  		Non-valid $N$ (SIR) & $4\%$ & $0\%$ & $0\%$ & $0\%$ & $0\%$ \\ 
		Non-valid $N$ (HawkesN) & $56\%$ & $42\%$ & $37\%$ & $18\%$ & $0\%$\\ 
		\midrule
		valid $N$ roots & $44$ & $58$ & $63$ & $82$ & $100$\\
		found using $\mathcal{S} < 0$ & $40$ & $57$ & $63$ & $82$ & $100$ \\ 
   \bottomrule
\end{tabular}
	\label{tab:un-identifiable-N}
	\captionmoveup
\end{table}

\begin{figure}[tbp]
	\centering
	\includegraphics[width=0.48\textwidth]{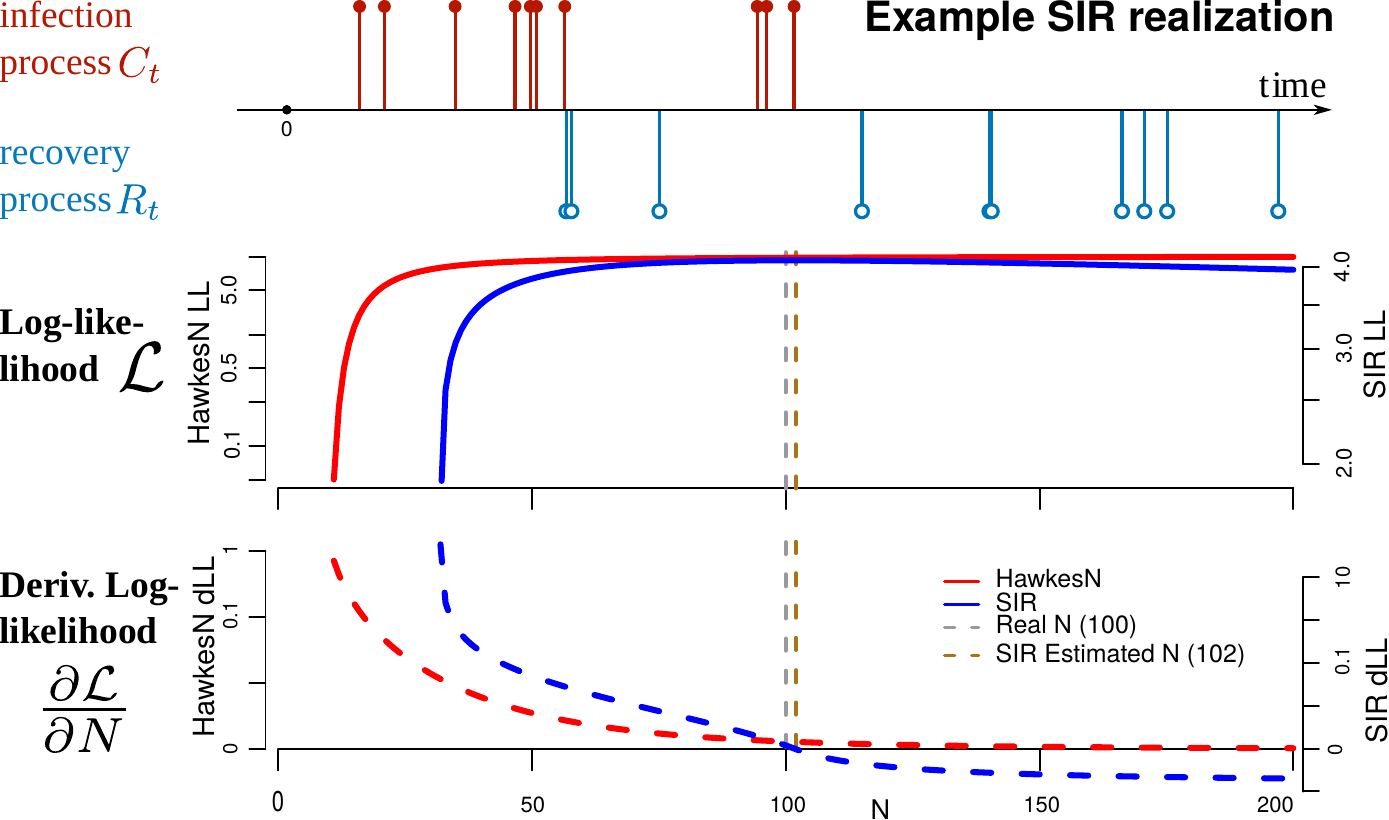}
	
	\caption{
		\emph{(top panel)} Example of an simulated SIR process realization, containing 10 infection events and 10 recovery events.
		\emph{(middle panel)} The log-likelihood as a function of $N$ for HawkesN (over the infection events, in red) and for SIR (in blue).
		\emph{(bottom panel)} The derivative of the log-likehood w.r.t. $N$.
		For HawkesN, the derivative is always positive and the log-likelihood monotonically increases, and \rev{there is no valid $N$ solution}.
		For SIR, the log-likelihood has a maximum at 102, close to the simulated value $N = 100$.
		\rev{Statistic value for this realization $\mathcal{S} = 13.77$}.
	}
	\label{fig:un-identifiable-N}
	\captionmoveup
\end{figure}

\begin{figure}[tbp]
	\centering
	\includegraphics[width=0.45\textwidth]{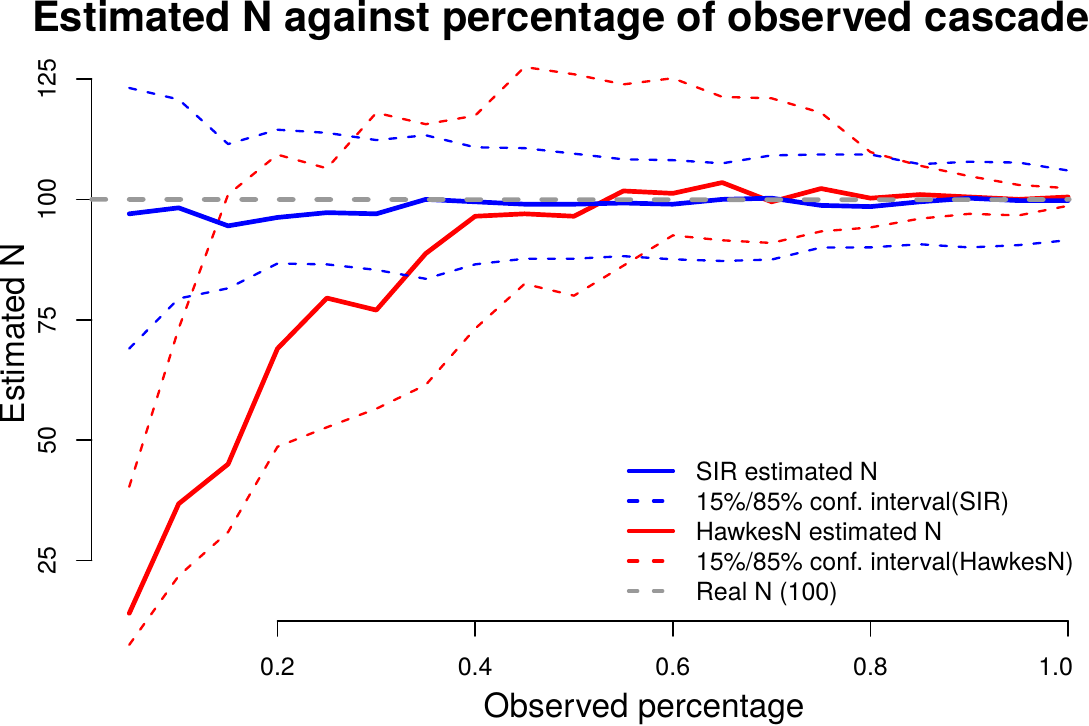}
		
	\caption{
		Median and 15\%/85\% confidence intervals for estimating $N$ using the maximum log-likelihood using HawkesN (\textcolor{red}{red color}) and SIR (\textcolor{blue}{blue color}).
	}
	\label{subfig:fit-N-HawkesN-SIR}
	\captionmoveup
\end{figure}

\textbf{Difficulty of estimating $N$.}
\rev{We illustrate the difficulty of estimating $N$ by simulation, and we show that this is dependent on the number of observed events in the cascade.}
Starting \rev{from the set of parameters $\mu = 0, \kappa = 5, \theta = 0.2, N = 100$}, we simulate \rev{100} realizations using stochastic SIR.
Assuming fixed all parameters except $N$, we study \rev{the validity and the quality of estimating} $N$ over increasingly longer prefixes of each cascade containing a percentage of all events in the range $[5\%, 100\%]$.
SIR observes both infection and recovery events in each prefix, while HawkesN observes infection events only.

\rev{We implement a numerical procedure for finding $N$:
we divide the range $[0, 200]$ into 1000 intervals and we numerically search each interval for a root for Eq.~\ref{eq:dLL} using \texttt{uniroot} in R.
This is a slow procedure which provides a ground truth against which we compare the statistic $\mathcal{S}$.
}
Table~\ref{tab:un-identifiable-N} shows for how many cascades \rev{there is no valid solution for $N$} for SIR (first row) and HawkesN (second row).
Five observed percentages are shown $5\%, 10\%, 20\%, 40\%$ and $80\%$.
For HawkesN, $56\%$ of cascades \rev{do not have a valid} $N$ after observing $5\%$ of the events.
\rev{We observe that, when more than $50\%$ of the cascade is observed, a solution for $N$ exists.}
For SIR however, only $10\%$ of the cascade have a \rev{non-valid} $N$ at the beginning of the cascades, and \rev{all cascades have valid solutions to $N$ once more than $10\%$ is observed}.
\rev{This indicates that it is more difficult to estimate $N$ in HawkesN than in SIR.
The bottom two rows of Table~\ref{tab:un-identifiable-N} show how many cascades have valid solutions for $N$ and for how many of these the statistic $\mathcal{S}$ is negative.
$\mathcal{S} < 0$ mis-identifies only 4 valid solutions (out of 44) after observing $5\%$ of the cascade and it identifies all valid solutions for percentages greater than $10\%$.
Note that $\mathcal{S}$ is a lower bound for the log-likelihood and it is guaranteed to find all non-valid solutions.}

Fig.~\ref{fig:un-identifiable-N} shows an example \rev{of a cascade with 10 infection events and 10 recovery events.
No valid solution exists for $N$ when using HawkesN} -- the log-likelihood function is monotonically increasing.
\rev{When the recovery events are observed in SIR, $N$ has a feasible solution close to the ground truth.}
%
This indicates that the timing of the recovery events (not observed in HawkesN) embeds information about the size of the population.
\rev{Fig.~\ref{subfig:fit-N-HawkesN-SIR} confirms this conclusion, showing that SIR estimates correctly $N$ even at the beginning of cascades, whereas HawkesN requires observing around $50\%$ of the the cascade to estimate $N$ correctly.}
For the full details of the simulation and additional analytic analysis, please consult the \citet{supplemental}.

%% file: 6-results.tex

\input{fig5-holdout-log-likelihood}

\input{fig7-size-distribution}

\secmoveup
\section{Experiments and results}
In this section, we investigate the performances of HawkesN \rev{on three Twitter diffusion datasets (described in Sec.~\ref{subsec:datasets}).}
We evaluate the generalization performance of HawkesN (in Sec.~\ref{subsec:explain-holdout}) and
we profile cascade size distributions and we provide a new explanation for the perceived popularity unpredictability (Sec.~\ref{subsec:cascade-size}).

\secmoveup
\subsection{Datasets}
\label{subsec:datasets}

\begin{table}[tbp]
	\centering
	\setlength{\tabcolsep}{4pt}
	\caption{Datasets profiling: number of cascades and number of tweets tweets; min, mean and median cascade size.}
	\begin{tabular}{rrrrrr}
	  \toprule
	 & \#cascades & \#tweets & Min. & Mean & Median \\ 
	  \midrule
		\Active~\cite{Rizoiu2017} & 41,411 & 8,142,892 & 20 & 197 & 41 \\ 
	  	\Seismic~\cite{Zhao2015} & 166,076 & 34,784,488 & 50 & 209 & 111 \\ 
	  	\News~\cite{Mishra2016} & 20,093 & 3,252,549 & 50 & 162 & 90 \\ 
	  \bottomrule
	\end{tabular}
	\label{tab:dataset-profiling}
	\captionmoveup
\end{table}

We use three datasets of retweet diffusion cascades in Twitter, used in previous work.
For each tweet in each cascade, we have information about the time offset of the retweet and the number of followers of the user posting the retweet.
The \Active dataset was collected by \citet{Rizoiu2017} during 6 months in 2014. 
It contains more than 41k retweet cascades related to more than 13k Youtube videos, each cascade containing at least 20 tweets.
The \Seismic dataset was collected by \citet{Zhao2015}. It contains a sample of all tweets during a month (i.e. using the firehose Twitter API restricted access), further filtered so that the length of each cascade is greater than 50.
The \News dataset was collected by \citet{Mishra2016} over a period of four moths in 2015.
They selected tweets containing links to news articles, by tracking the official twitter handles of popular news outlets, such NewYork Times, or CNN.
Each cascade contains at least 50 tweets.
Table~\ref{tab:dataset-profiling} summarizes these datasets.

\secmoveup
\subsection{Generalization to unobserved data}
\label{subsec:explain-holdout}

\rev{All three datasets described in the previous section also contain user information for each tweet.
The tweets are pairs $\{m_j, t_j\}, t=1,\ldots,n$, where $m_j$ is the number of followers of the user having emitted tweet $j$ at time $t_j$.
In this section, we choose to use the modified exponential kernel function proposed by~\citet{Mishra2016}, which also accounts for the number of followers for a user:
$\phi(\tau) = \kappa m^\eta \theta e^{-\theta\tau}$.
More details about the marked HawkesN and its equivalence with SIR are found in the~\citet{supplemental}.
}

We empirically validate HawkesN by studying how it generalizes to unseen data.
We compare HawkesN with the Hawkes model for information diffusion, proposed by \citet{Mishra2016}.
We adopt the setup in~\cite{Zhao2015,Shen2014,Bao2015,Ding2015,Gao2015,Rizoiu2017}: 
\rev{the first few events in a diffusion are observed and used to fit the models.
Hawkes is fitted as described in~\cite{Mishra2016}, and HawkesN is fitted as described in Sec.~\ref{subsec:fitting-hawkesn}.
The population size $N$ is also fitted from data.}
We measure the holdout likelihood, i.e. the likelihood of the events in the unobserved period.
The \rev{lower the negative} holdout likelihood, the better the model generalizes to unseen data.
We report the \emph{per event} holdout negative likelihood, to render the results comparable across holdout sets containing different numbers of events.
Given \rev{the analysis in Sec.~\ref{subsec:identifiability}},
we chose to observe a given proportion of each cascade, to render the results comparable across cascades of different length.

Fig.~\ref{subfig:hawkesn-increasing-perc} shows the generalization performances of HawkesN, when varying the percentage of observed events from 10\% to 95\%.
Consistent with \rev{the conclusions in Sec.~\ref{subsec:identifiability}}, we observe a high variance of performance when observing less than 40\% of each cascade.
The basic Hawkes model shows less variance at lower percentages (shown in the online supplement~\cite{supplemental}).
Plots (b) to (d) in Fig.~\ref{fig:holdout-ll} show the generalization performance of Hawkes and HawkesN, on the three datasets, for the observed percentages of 40\% and 80\%.
Visibly, \rev{HawkesN has a consistently lower median value for the negative log-likelihood than Hawkes for higher observed percentages.}
The mean negative log-likelihood values are comparable for HawkesN and Hawkes on \News and \Seismic.
On \Active the mean of HawkesN is higher -- likely due to Youtube videos behaving differently, with some old ones (e.g. Music) still being shared. 

For higher observed percentages, the mean negative log-likelihood improves for HawkesN and it degrades for Hawkes.
This indicates that the the modulation factor ($1 - \frac{Nt}{N}$ in Eq.~\ref{eq:hawkesN}) helps improve likelihood, and HawkesN fits longer event sequences better.

\secmoveup
\subsection{Explaining popularity unpredictability}
\label{subsec:cascade-size}

In this section, we study the probability distribution of \rev{population} size for real-life cascades.
Both left and right plots in Fig.~\ref{fig:size-distribution} show the same cascade \rev{from the \News dataset}, with the HawkesN parameters fit on 27 and 47 events respectively (here $N$ is a meta-parameter fixed at $N = 80)$.
The \emph{apriori} probability size distribution -- the distribution after observing only the first event -- shows two maxima: one around very small values of cascade size, and one around $\sim 65$.
This provides the following explanation for the general perceived unpredictability of online popularity.
\rev{For cascades showing a bi-modal \emph{apriori} size distribution, }
there are two likely outcomes: either it dies out early or it reaches a large size compared to the maximum population $N$.
At time $t = 0$ is it impossible to differentiate between the two outcomes.
The situation is different after observing a number of events.
The \emph{aposteriori} probability distribution (shown in Fig.~\ref{fig:size-distribution} with a blue line) reflects the information gained from the observed events and it shows a single maximum towards the higher size values.
The more events we observe, the higher the likelihood of the \rev{true value of} cascade size.
\rev{We also observe that the size distribution gets narrower as we observe more events, i.e. there is less uncertainty in cascade size prediction.}
This provides another explanation to why autoregressive popularity prediction approaches~\cite{Chang2014,Pinto2013,Szabo2010} achieve higher results.
\rev{Online popularity has been previously claimed to be unpredictable}
~\cite{Martin2016}, however as far as we know this \rev{is the first explanation for it, based on analytical results on size distributions}.

%% file: fig5-holdout-log-likelihood.tex

\begin{figure*}[tbp]
	\centering
	\newcommand\myheight{0.16}
	\subfloat[] {
		\includegraphics[page=1,height=\myheight\textheight]{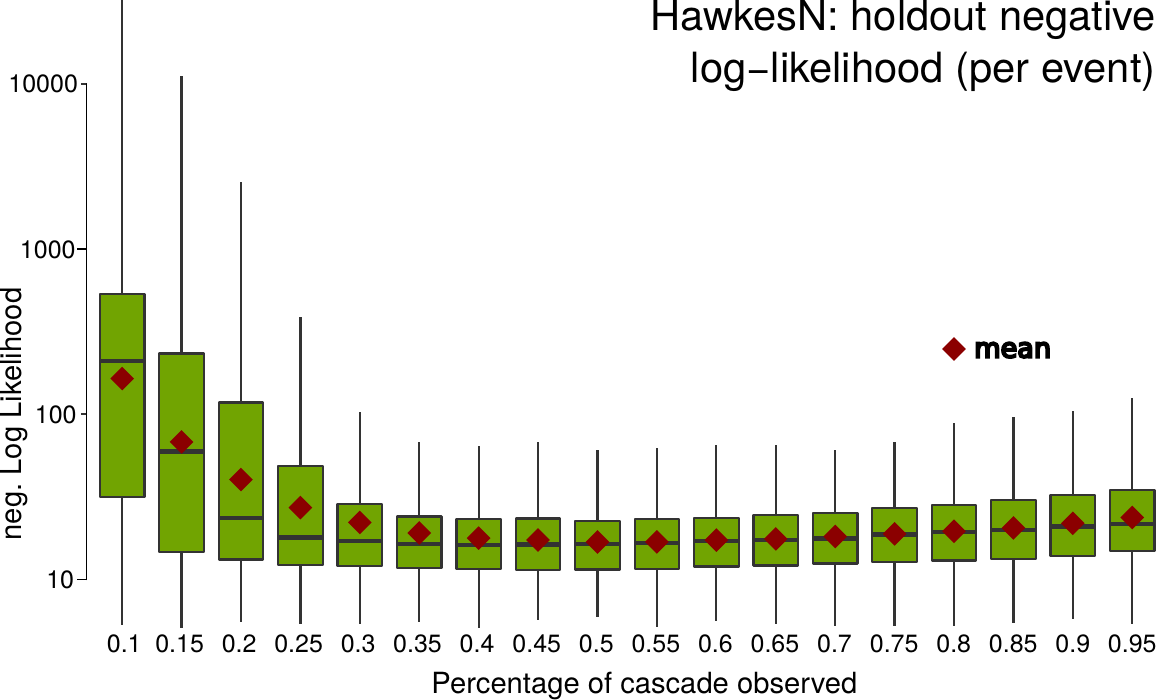}
		\label{subfig:hawkesn-increasing-perc}
	}
	\subfloat[] {
		\includegraphics[page=2,height=\myheight\textheight]{fig5}
	}
	\subfloat[] {
	\includegraphics[page=3,height=\myheight\textheight]{fig5}
	}
	\subfloat[] {
		\includegraphics[page=4,height=\myheight\textheight]{fig5}
	}
	\caption{ 
		Performances of HawkesN explaining unobserved data, using holdout negative log likelihood.
		The performance over all cascades in a dataset are summarized using boxplots, lower is better \emph{(a)}.
        The percentage of observed events in each cascade used to train HawkesN is varied between 10\% and 95\%. We use 1000 cascades randomly sampled from \News.
        \emph{(b), (c) and (d)} The performances on all cascades of \Active, \Seismic and \News, for Hawkes and HawkesN, when observing 40\% and 80\% of each cascade.
	}
	\label{fig:holdout-ll}
	\captionmoveup
\end{figure*}

%% file: fig7-size-distribution.tex

\begin{figure}[htbp]
	\centering
	\includegraphics[width=0.48\textwidth]{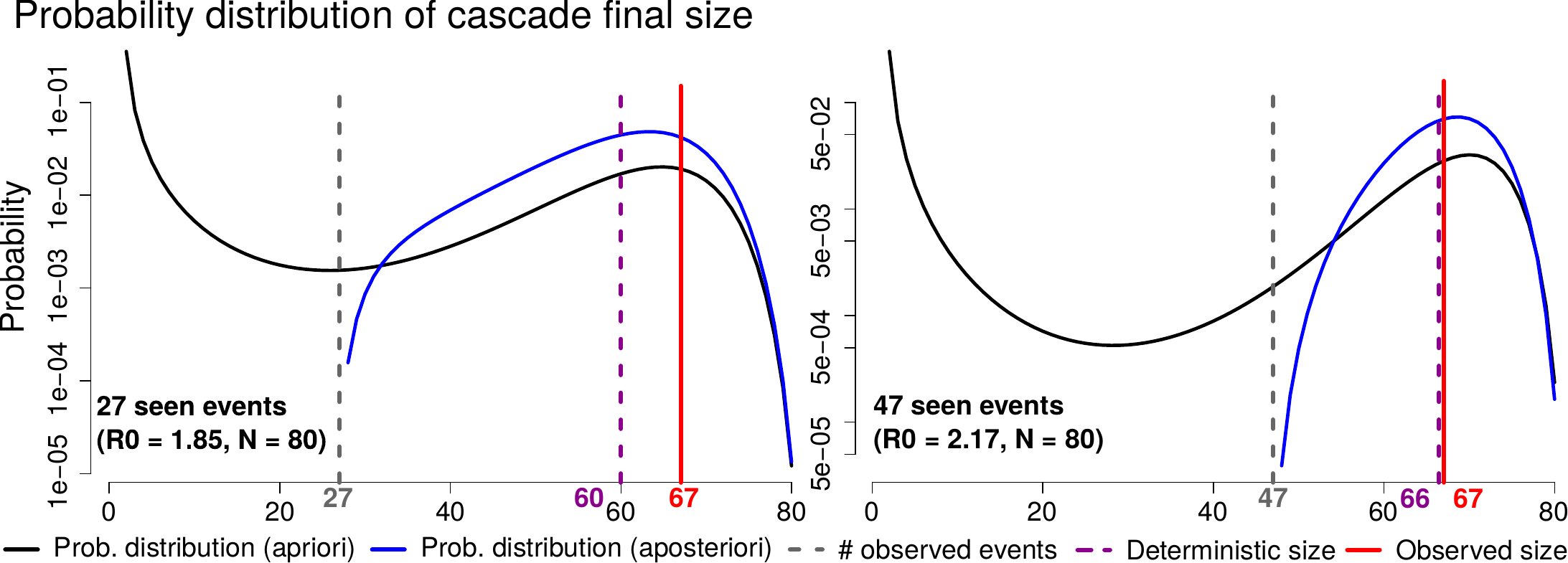}
	\caption{ 
		Final size probability distribution for a \News cascade.
		HawkesN was fitted on the first 27 events \emph{(left)} and 47 events \emph{(right)}.
		The black line shows the \emph{apriori} distribution, after seeing a single event.
		the \textcolor{blue}{blue line} shows the \emph{apostoriori} distribution, after seeing the observed events.
		Shown with vertical lines: the number of observed events in \textcolor{gray}{gray dashed}, the deterministic prediction in \textcolor{magenta}{magenta dashed} and the actual observed cascade size in \textcolor{red}{red}.
	}
	\label{fig:size-distribution}
	\captionmoveup
\end{figure}

%% file: 7-related-work.tex

\secmoveup
\section{Related Work}

We structure \rev{some of the} related work \rev{in the field of social media analysis} into two broad categories, based on the used framework: 
point process approaches and epidemic models.

\textbf{Point process approaches.}
\rev{Point-process based generative models are a popular choice for} popularity modeling~\cite{Crane2008,Ding2015,Yu2017} and prediction~\cite{Shen2014,Zhao2015,Rizoiu2017,Mei2017}.
In their seminal work, \citet{Crane2008} \rev{linkes} popularity bursts and decays \rev{to the effects of a Hawkes self-exciting process}.
More sophisticated models have been proposed to model and simulate popularity in microblogs~\cite{Yu2017} and videos~\cite{Ding2015}.
These approaches successfully account for the social phenomena which modulate online diffusion: the ``rich-get-richer'' phenomenon and social contagion.
Certain models can output an estimate for the total size of a retweet cascade.
\citet{Shen2014} employ reinforced Poisson processes, modeling three phenomena: fitness of an item, a temporal relaxation function and a reinforcement mechanism;
while SEISMIC~\cite{Zhao2015} employs a double stochastic process, one accounting for infectiousness and the other one for the arrival time of events.
Our work differs from the above in two aspects.
First it proposes a generalization of the Hawkes model, which operates in a finite population -- which is a more realistic assumption.
Second, it outputs a size probability distribution and it explains the perceived unpredictability of online popularity~\cite{Martin2016}.

A recently emerging body of work employs Stochastic Differential Equations to formulate Hawkes point processes.
RedQueen~\cite{Zarezade2017} and Cheshire~\cite{Zarezade2017b} are two algorithms aimed at optimizing social influence, which they formulate as a stochastic optimal control problem.
\citet{Wang2016} use stochastic control and reinforcement learning to address the user activity guiding problem and feedback in social systems.
Later, \citet{Wang2017} use the stochastic differential equation model to link the microscopic event data and macroscopic inference, and to approximate its probability distribution.
The similarity between the above and our work is at the level of tools, by using stochastic calculus to link the event-level to the event rate and compute expected quantities.
However, none of the above links point processes to epidemic models.
The advantage of our solution is that it enables to leverage the mature tools in epidemic models to the field of information diffusion.

\textbf{Epidemic model approaches.}
Despite being developed for the field of epidemiology, epidemic models have been applied to information diffusion problems through the analogy of information spread as a disease.
Classic epidemic models were early applied in the knowledge and scientific theory diffusion study~\cite{Goffman1971} and latter employed in many areas, such as economic and finance time series analysis~\cite{Shtatland2008}.
\citet{Pastor-Satorras2001} applied SIS (Susceptible-Infected-Susceptible) epidemic model to simulate computer virus transmission over the Internet.
A series of studies analyzed the spread of rumors in complex networks based on an epidemic model~\cite{Zanette2002,Moreno2004,Trpevski2010,Gruhl2004}.
More recently, \citet{Woo2016} modeled topic diffusion in web forums using an SIR model; \citet{Martin2016} fitted an epidemic model to retweet cascades and used the computed basic reproduction number to theorize the unpredictability of online popularity.
\rev{ 
\citet{Bauckhage2015} use a stochastic SIR model to characterize attention dynamics of viral videos.
\citet{Goel2015} apply large-scale agent based SIR simulation on a random network to study the virality on Twitter diffusions.
\citet{feng2015competing} propose a fractional SIR model in which the infection probability of a node is proportional to its fraction of infected neighbors and apply it on Sina Weibo data.
}
However, these work do not leverage tools specific to epidemic models (e.g. the probability distribution of size), nor do they link to point process models as our work does.


%% file: 8-conclusion.tex

\secmoveup
\section{Conclusion}
\label{sec:discussion}

In this work, we present a previously unexplored connection between Hawkes point processes and SIR epidemic models.
First, we establish a novel connection between these two frameworks by linking the rate of events in an extended Hawkes model to the rate of new infections in the Susceptible-Infected-Recovered (SIR) model after marginalizing out recovery events -- which are unobserved in a Hawkes process.
This paves the way to applying tools developed for one approach to the other approach.
It also leads to HawkesN, an extension of the Hawkes process with a finite number of events.
Finally, we present a novel method to compute the probability distribution of the final size of a cascade after observing its initial unfolding using HawkesN, which is based on a Markov chain tools developed for SIR.
We use the probability of cascade size on a large sample of real cascades to provide a nuanced explanation for the general unpredictability of popularity.


\textbf{Assumptions, limitations and future work.}
This work assumes a fixed population (users don't enter, nor do they exit). 
A link could be drawn between evolving populations in SIR and $\mu(t) \neq 0$ in HawkesN.
The current work assumes that the maximum population size $N$ is estimated for each cascade, while observing i.
Future work could use other observed similar cascades to infer the size of a ``thematic neighborhood'' before a cascade starts unfolding.
Finally, allowing for user-specific behavior in the SIR model or kernel functions other than the exponential function requires more advanced SIR formulations, such as an agent-based formulation.

%% file: appendix.tex
%
\newpage
\appendix
\etocdepthtag.toc{mtappendix}
\etocsettagdepth{mtchapter}{none}
\etocsettagdepth{mtappendix}{subsection}
\etoctocstyle{1}{Contents (Appendix)}
\tableofcontents‎‎

\section{Inter-event time probabilities in non-homogeneous Poisson processes}
\label{ap-subsec:probs-NHPP}

In this section, we revisit the Non-Homogeneous Poisson Process (NHPP) and we compute the formula for the probabilities of observing inter-arrival times.
We also show that NHPP is a non-Markovian process and we derive a simple proof for the formula for the log-likelihood of a NHPP, which is widely used in CS literature, but an accessible proof of which is currently missing.

\subsection{Inter-arrival times probabilities}

Here we compute the probability of observing $t_i$ -- the arrival of an event.
We denote by $\tau_i$ the inter-arrival time between event $i-1$ and event $i$.
It follows that $\tau_i = t_i - t_{i-1}$ and $t_i = \sum_1^j \tau_j$.
We study in parallel the Homogeneous Poisson Process (HPP) and NHPP.
For ease of understanding, we further consider the two cases when $i = 1$ and $i > 1$.

\textbf{The arrival of the first event $t_1$}.
In a HPP of intensity $\lambda$, the probability of having no events in the time interval $[0, t)$ is:
\begin{equation} \label{eq:CCDF-waiting-time}
	\mathds{P}[t_1 \ge t] = e^{-\lambda t} \enspace .
\end{equation}
This can be interpreted as the probability of waiting at least $t$ units of time until the first event.
Consequently, Eq.~\eqref{eq:CCDF-waiting-time} is the CCDF (Complementary Cumulative Distribution Function) of the waiting time until the first event.
The PDF is $ PDF = \fp{}{t} (1 - CCDF) = - \fp{}{t} CCDF$.
Consequently \emph{the waiting time to the first event in a HPP is distributed exponentially}, with parameter $\lambda$:
\begin{equation}
	\mathds{P}[t_1 = t] = - e^{-\lambda t} \fp{-\lambda t}{t} = \lambda e^{-\lambda t} \enspace.
\end{equation}

For a NHPP with the event rate $\lambda(t)$, we first define the function $\Lambda(t) = \int_0^t \lambda(\tau) d\tau$.
The inverse relation between $\lambda(t)$ and $\Lambda(t)$ is $\lambda(t) = \fp{}{t} \Lambda(t)$.
We have:
\begin{equation} \label{eq:NHPP-wait-0-t}
	\mathds{P}[t_1 \ge t] = e^{-\Lambda(t)} \enspace ,
\end{equation}
and we compute
\begin{equation} \label{eq:NHPP-waiting-t1}
	\mathds{P}[t_1 = t] = \fp{}{t} e^{-\Lambda(t)} = - e^{-\Lambda(t)} \fp{}{t} \Lambda(t) = \lambda(t) e^{-\Lambda(t)} \enspace.
\end{equation}
Note that \emph{the waiting time to the first event is not exponentially distributed in the case of NHPP}.
An intuitive interpretation of Eq.~\eqref{eq:NHPP-waiting-t1} is that the probability of observing an event at time $t$ is the product of 
the probability of observing an event in the infinitesimal time interval  $[t, t + \partial t]$ -- equal to the event rate $\lambda(t)$ --
and the probability having observed no event in $[0, t]$ -- as defined in Eq.\eqref{eq:NHPP-wait-0-t}.
%
%

\textbf{The arrival of $t_2, t_3, \ldots, t_n$}.
For a HPP of rate $\lambda$, the probability of not observing an event in the interval $[t, t + s]$ -- after having observed a first event at time $t_1 = t$ -- is:
\begin{equation*}
	\mathds{P}[t_2 - t_1 \ge s | t_1 = t] = e^{- \lambda (t+s - t)} = e^{- \lambda s} \enspace .
\end{equation*}
does not depend of $t$. By denoting $\tau_2 = t_2 - t_1$ and $\tau_1 = t_1$, we obtain
\begin{equation} \label{eq:HPP-inter-arrival-times-prob-distribution}
	\mathds{P}[\tau_2 = s | \tau_1 = t] = \lambda e^{- \lambda s} \Longrightarrow \mathds{P}[\tau_i = s ] = \lambda e^{- \lambda s} \enspace.
\end{equation}
Inter-arrival times in a HPP are exponentially distributed with parameters $\lambda$, and the probability of observing a $\tau_i$ does not depend on the previous inter-arrival times $\tau_1, \tau_2, \ldots, \tau _{i-1}$.
This property is called \emph{memorylessness} -- and it is equivalent to the Markovian property~\cite{Allen2008} -- as the next state of the process depends only on the current state and not on the past.

For the NHPP of rate $\lambda(t)$, we have
\begin{align}
	\mathds{P}[t_2 - t_1 \ge s | t_1 = t] &= e^{ \Lambda (t) - \Lambda(t+s)} \nonumber \\
	\Rightarrow \;  \mathds{P}[t_2 - t_1 = s | t_1 = t] &= \fp{}{s} \mathds{P}[t_2 - t_1 \ge s | t_1 = t] \nonumber \\
	 &= \lambda(t+s) e^{ \Lambda (t) - \Lambda(t+s)} \nonumber
\end{align}
$\Lambda (t) - \Lambda(t+s)$ can be interpreted as the minus area under the curve of $\lambda(t)$. 
We can further show that
\begin{equation} \label{eq:NHPP-inter-event-times-prob}
	\mathds{P}[\tau_{i+1} = s | \mathcal{H}_i] = \lambda(t_i+s) e^{ \Lambda (t_i) - \Lambda(t_i+s)} 
\end{equation}
where $\mathcal{H}_i = \{ t_1, t_2, \ldots, t_i\}$ is the history of the process up to event $t_i$. 
Note that when $\lambda(t) = \lambda$ -- i.e. a HPP -- we have $\Lambda(t) = \lambda t$ and Eq.~\ref{eq:HPP-inter-arrival-times-prob-distribution} and~\ref{eq:NHPP-inter-event-times-prob} are identical.
We can express Eq.~\eqref{eq:NHPP-inter-event-times-prob} in terms of event times (rather than inter-event times):
\begin{equation} \label{eq:NHPP-event-times-prob}
	\mathds{P}[t_{i+1} | \mathcal{H}_i] = \lambda(t_{i+1}) e^{ \Lambda (t_i) - \Lambda(t_{i+1})} 
\end{equation}

\subsection{Two follow-up conclusions}

We study the Markovian property of NHPP and we derive its likelihood function.

\textbf{NHPP is not Markovian}.
One direct consequence of Eq~\eqref{eq:NHPP-inter-event-times-prob} is that inter-arrival times in a NHPP are not exponentially distributed.
We further study if the process is memoryless -- i.e. if it has the Markovian property.
For this, we compute the join probability of having an event in the interval $[0, t]$ and a second event in $[t, s]$.
\begin{align}
	\mathds{P}[t_1 = t, t_2 = t + s] &= \mathds{P}[t_1 = t] \mathds{P}[t_2 = t + s | t_1 = t] \nonumber \\ 
	&= \lambda(t) \lambda(t + s) e^{- \Lambda(t + s)} \label{eq:non-markov}
\end{align}
which shows that $t_2$ is not independent of $t_1$.
The implication is that the next state of a NHPP -- i.e. $t_{i+1}$ -- is dependent on all previous states -- $t_j, j \in [1 \dots i]$.
\emph{This shows that NHPP is not Markovian.}
Note that this is a general results, for non-specific functions $\lambda(t)$
Specific functions $\lambda(t)$ can be constructed so that the NHPP becomes Markovian.

As a sanity check, we write Eq.~\eqref{eq:non-markov} for a HPP.
We obtain
\begin{align}
	\mathds{P}[t_1 = t, t_2 = t + s] &= \lambda ^2 e^{-\lambda (t+s)} \nonumber \\
	&= \lambda e^{-\lambda t} \lambda e^{- \lambda s} = \mathds{P}[t_1 = t] \mathds{P}[t_2 = t + s]
\end{align}
therefore the inter-arrival times $\tau_1$ and $\tau_2$ are independent and exponentially distributed -- as expected.

\textbf{The likelihood function for NHPP}.
Given $\mathcal{H}_i$, which includes the parameter of the process $\theta$ and the history of the process up to event $t_i$, the probability of an event at time $t_{i+1}$ is defined (according to Eq.~\eqref{eq:NHPP-event-times-prob} as the probability of observing an event at time $t_{i+1}$ -- $\Lambda(t_{i+1})$ -- and the probability of not having observed any event in the interval $[t_i, t_{i+1}]$.

We construct the likelihood function as
\begin{align}
	Likelihood(\theta) &= \mathds{P}[t_1, t_2, \dots, t_n | \theta] \nonumber \\
	&= \mathds{P}[t_1 | \theta] \mathds{P}[t_2 | t_1, \theta] \mathds{P}[t_3 | t_2, t_1, \theta] \ldots \mathds{P}[t_n | t_{n-1}, \ldots t_1, \theta] \nonumber \\
	&= \prod_{i = 1}^n \mathds{P}[t_i | \mathcal{H}_{i-1}] = e^{-\Lambda(t_1)+\Lambda(t_1)-\Lambda(t_2) + \ldots - \Lambda(t_n)} \prod_{i = 1}^n \lambda(t_i) \nonumber \\
	&= \prod_{i = 1}^n \lambda(t_i) e^{-\Lambda(t_n)} \nonumber
\end{align}
Finally, we derive the expression of the log-likelihood widely used in literature:
\begin{align} 
	log(Likelihood(\theta)) &= \sum_{i = 1}^n log \left( \lambda(t_i) \right) - \Lambda(t_n) \nonumber \\
	&= \sum_{i = 1}^n log \left( \lambda(t_i) \right) - \int_0^{t_n} \lambda(\tau)d\tau . \label{eq:log-likelihood}
\end{align}

\section{Fitting HawkesN with AMPL -- implementation}
We fit the parameters of the HawkesN model to observed data by maximizing the log-likelihood function Eq.~\eqref{eq:log-likelihood}. 
We use AMPL, an industry standard for modeling optimization problems and with a transparent interfaces to powerful solvers.
We start with an introduction of AMPL (Sec.~\ref{subsec:ampl-intro}), we describe our optimization setup and the employed solvers (Sec.~\ref{subsec:optimization-setup}) and we finish with the R interface that we constructed for AMPL (Sec.~\ref{subsec:ampl-r-interface}).

\subsection{AMPL introduction}
\label{subsec:ampl-intro}
Since the first commercial release in 1993, AMPL -- which stands for A Mathematical Programming Language -- has provided a convenient interface between mathematic modelers and implemented solvers~\cite{fourer1987ampl}. 
It now also offers a complete tool set including many solvers for modeling different optimization problems.

Our optimization problem used to involve much more than just deducing log-likelihood functions before utilizing APML. 
Special effort had to be expanded to derive some components because of specific requirements from solving algorithms. 
For example, to apply IPOPT solver to our model estimation, we were required to sketch out all parameter derivatives of log-likelihood functions and Jacobian matrix. 
AMPL, however, allows us to solve the problem by only defining the problem and formulating the constraints.

To run AMPL on models, it needs two parts as input including model files and data files. Model files define the problem, while data files specify constants and initial values for variables. AMPL translator will read in those files and translate them into languages that solvers can understand. 
AMPL is particularly notable for its general syntax, including variable definitions and data structures. 

\subsection{Used solvers and optimization setup}
\label{subsec:optimization-setup}
AMPL supports a comprehensive set of solvers including solvers for linear programming, quadratic programming and non-linear programming~\cite{fourer1993ampl}. 
This link\footnote{\url{http://www.ampl.com/solvers.html}} gives a full list of solvers for AMPL. 

\textbf{Solvers Applied in Implementation}.
We used two solvers in our fitting procedure:
\begin{itemize}
    \item \textbf{LGO}: a \emph{global optimizer} for non-linear problems, which is capable of finding approximate solutions when the problems have multiple local optimal solutions (\cite{pinter1997lgo}). This is also one of the default solvers provided by AMPL.
    \item \textbf{IPOPT}: an open-source large-scale \emph{local optimizer} for non-linear programming, which is released in 2006~\cite{Wachter2006}. 
\end{itemize}
Local solvers rely on improving an existing solution, employing complex techniques to avoid getting stuck in local minima.
They require an initial point from which to start exploring the space of solutions.
Global solvers attempt to search for the optimal solution in the entire space of solutions (one solution would be, for example, to divide the solution space into hyper-squares and apply local optimization in each one of them).
Global solvers tend to find solutions which are not too far from the optimal, but they lack the precision of specialized local solvers
In summary: local solvers achieve solutions very close to the optimal, but run the risk of getting stuck in horrible local optima;
global solvers achieve imprecise solutions close to the optimal.

\textbf{Optimization implementation setup}. 
Our optimization setup is constructed to account for the weaknesses of each class of solvers.
A classical solution to the problem of local optima with local solvers is to repeat the function optimization multiple times, from different starting points.
We generate 8 random sets of initial parameters, within the definition range of parameters, and we use the IPOPT solver using each of these as initial point.
We also combine the global and the local solver: we use LGO to search in the space of solutions for an approximate solution, which we feed into IPOPT as initial point for further optimization.
Lastly, we run IPOPT without any initial parameters, leveraging IPOPT's internal strategy for choosing the starting point based on the parameters' range of definition.
After completing these 10 rounds of optimization, we select the solution with the maximum training log likelihood values.
This tends to be the combination of global and local optimizer (LGO + IPOPT).

\subsection{Interfacing AMPL with R}
\label{subsec:ampl-r-interface}
Our entire code base is using the R language, but AMPL has its own modeling language. 
Therefore, we need to interface between R and AMPL. 
Inspired by a blog post\footnote{\url{https://www.rmetrics.org/Rmetrics2AMPL}}, we implemented our own interface between AMPL and R language. The core ideas are described as follow:
\begin{itemize}
	\item \textbf{Generating model files and data files}: 
	one of the major components of this interface is generating temporary model and data files, which model the problem to be solved and the used data into AMPL language and format. 
	As our experiments involve a large amount of cascades, we prefixed all temporal files with process ids so that running AMPL in parallel becomes possible.
	For I/O speed considerations, all files are created in ram-drives, therefore eliminating the penalty of disk access.
    \item \textbf{Interacting with AMPL}: this is also implemented using files in ram-drives. 
    After the model and data files are generated, we call AMPL via the \texttt{system} command in R.
    AMPL saves the optimization results in files, and our interface extracts and returns the results.
    \item \textbf{Exception handling}: solvers occasionally encounter errors during the optimization process, typically numerical errors due to the precision of float numbers. 
\end{itemize}

\section{Marked HawkesN}

\verify{this section will collect all material around marked Hawkes, and wrap it around nicely.}

\subsection{Kernel function and branching factor}

\textbf{Kernel functions for online diffusions.}
The exponential kernel $\phi(\tau) = \theta e^{- \theta \tau}$ is a popular choice when modeling online social media~\cite{Mishra2016,Zarezade2017,Zhao2015,Shen2014,Bao2015,Ding2015,Gao2015}.
Other kernel choices include power-law functions $\phi(\tau) = (\tau+c)^{-(1+\theta)}$, used in geophysics~\cite{Helmstetter2002} and social networks~\cite{Crane2008,Mishra2016,Kobayashi2016} and the Rayleigh functions $e^{-\frac{1}{2} \theta \tau^2}$, used in epidemiology~\cite{Wallinga2004}.
Here, we choose to use the modified exponential kernel function proposed by~\citet{Mishra2016}, which captures the local influence of user in addition to the temporal decay:
\begin{equation} \label{eq:kernel-function-hawkesN-marked}
	\phi(\tau) = \kappa m^\eta \theta e^{-\theta\tau}
	\eqmoveup
\end{equation}
where $m$ is the local user influence, 
$\eta$ introduces a warping effect for the local user influence,
$\kappa$ is a scaler and 
$\theta$ is the parameter of the exponential function.
When modeling diffusion cascades, it is typically assumed that every event apart from the first one is a reaction to the first event, i.e.,
the background intensity is zero $\mu(t) = 0, \forall t > 0$~\cite{Mishra2016}.

\textbf{Branching factor.}
We define the branching factor of HawkesN as the expected number of children events directly spawned \emph{by the first event of the process}.
For large values of $N$ and fast decaying kernel functions $\phi(t)$, we can approximate $\frac{N_t}{N} \approx 0$ and therefore the branching factor for HawkesN is:
\begin{equation} \label{eq:branching-factor-hawkesn-marked}
	n^\ast \approx \int_1^\infty \int_0^\infty p(m)  \kappa m^\eta \theta e^{-\theta \tau} d\tau dm  = \kappa \frac{\alpha - 1}{\alpha - \eta - 1}
	\eqmoveup
\end{equation}
where $p(m)$ is the distribution of local influence that \citet{Mishra2016} studied on a large sample of tweets, and found to be a power-law of exponent $\alpha = 2.016$.
In HawkesN, the branching factor is indicative of the speed at which the cascade unfolds and its final size distribution (as shown in Sec.~\ref{subsec:cascade-size}).
%

\subsection{Log-likelihood function.}
The parameters of HawkesN can be estimated from observed data using a maximum likelihood procedure.
When modeling diffusion cascades, it is typically assumed that every event apart from the first one is a reaction to the first event, i.e.,
the background intensity is zero $\mu(t) = 0, \forall t > 0$~\cite{Mishra2016}.
Therefore, the HawkesN process is completely defined by three parameters $\{ \kappa, \theta, N \}$.
The log-likelihood of observing a set of events $\{(m_j,t_j), j=1,\ldots,n\}$ in a non-homogeneous Poisson process of event rate $\lambda^H(t)$ is (see the online supplement~\cite{supplemental} or \citet[Ch.~7.2]{Daley2008}):
\begin{equation} \label{eq:hawkes-ll-marked}
	\mathcal{L}(\kappa, \beta, c, \theta) = \sum_{j=1}^n \log\left(\lambda^H\left(t_j^-\right)\right) - \int_0^{t_n} \lambda^H(\tau)\mathrm{d}\tau.
\end{equation}
We detail further the integral term:
\begin{align}
	\int_0^{t_n} \lambda^H(\tau)\mathrm{d}&\tau = \int_0^{t_n} \left( 1  -  \frac{N_{t^-}}{N} \right) \sum_{t_j < t} \phi(t - t_j) dt \nonumber \\
	=& \sum_{j = 1}^{n-1} \int_{t_j}^{t_n} \left( 1  -  \frac{N_{t^-}}{N} \right) \phi(t - t_j) dt \nonumber \\
	=& \sum_{j = 1}^{n-1} \sum_{l=j}^{n-1} \frac{N - l}{N} \int_{t_l}^{t_{l+1}} \phi(t - t_j) dt \nonumber \\
	^{cf.~\eqref{eq:kernel-function-hawkesN}} =& \kappa \sum_{j=1}^{n-1 } (m_j)^\eta \sum_{l=j}^{n-1} \frac{N - l}{N} \left[e^{-\theta(t_l - t_j)} - e^{-\theta(t_{l+1} - t_j)} \right]. \label{eq:ll-marked}
	\eqmoveup
\end{align}
Eq.~\eqref{eq:hawkes-ll} is a non-linear objective that we maximize to find the set of parameters.
There are a few natural constraints for each of the model parameters, namely: $\theta>0$, $\kappa>0$, and $0<\eta<\alpha-1$ for the branching factor to be defined.
We use the mathematical modeling language AMPL~\cite{fourer1987ampl}, which offers a complete set of modeling tools, including automatic gradient computation and support for a large number of solvers.
We choose as solver Ipopt~\cite{Wachter2006}, the state of the art optimizer for non-linear objectives. 
More details can be found in the online supplement~\cite{supplemental}.

\subsection{Equivalence to stochastic SIR}

Denote $\tau=\{\tau_1, \tau_2, \ldots\}$  as times to recovery of infected individuals in SIR; 
$m=\{m_1, m_2, \ldots\}$ as user influences in HawkesN; 
$\alpha$ as the power-law exponent of user influence distribution (Eq.~\eqref{eq:branching-factor-hawkesn-marked}).
We now have a marked equivalent of Theorem~\ref{theorem:expected-equivalence}:
\begin{theorem} \label{theorem:expected-equivalence-marked}
	Suppose the new infections in a stochastic SIR process of parameters $\{\beta, \gamma, N\}$
	follow a temporal point process of intensity $\lambda^I(t)$.
	Suppose also the events in a HawkesN process with parameters $\{\kappa, \eta, \theta, N\}$ have the event intensity $\lambda^H(t)$ (Eq~\ref{eq:hawkesN}).
	The expectation of $\lambda^I(t)$ over all times to recovery $\tau$ is equal 
    to the expectation of $\lambda^H(t)$ over individual event strengths $m$. 
	\begin{equation*} 
		\mathds{E}_\tau[ \lambda^I(t)] = \mathds{E}_m [\lambda^H(t)],
		\eqmoveup
	\end{equation*}
	when $\mu(t) = 0$, $\beta = \kappa \theta \frac{\alpha - 1}{\alpha - \eta - 1}$, $\gamma = \theta$.
\end{theorem}

\TODO{MAR}{Say that the two expectations continue to be random functions, as the expectations only partially remove randomness.
Sources of randomness.}

\TODO{MAR}{also say that Theorem~\ref{theorem:expected-equivalence} is a special case of Theorem~\ref{theorem:expected-equivalence-marked} where $\mu = 0$}

\textbf{Expected event rate in HawkesN over user influence.}
In the stochastic SIR model, the actions of each individual are guided by the same set of global rules, i.e. the differences between individuals are not observed.
The HawkesN model with the kernel defined in Eq.~\eqref{eq:kernel-function-hawkesN} accounts for different local user influences, which are averaged out in Theorem~\ref{theorem:expected-equivalence}. 
We obtain:
\begin{align} \label{eq:HawkesN-expectation}
	\mathds{E}_m\left[\lambda^H(t)\right]
		&= \mathds{E}_m\left[ \left( 1 - \frac{N_t}{N} \right) \left( \mu(t) + \sum_{t_j < t} \kappa m^\eta_j \theta e^{-\theta (t - t_j)} \right) \right] \nonumber \\
		&=  \left( 1 - \frac{N_t}{N} \right) \left( \mu(t) + \sum_{t_j < t} \kappa  \theta e^{-\theta (t - t_j)} \int_1^\infty m^\eta p(m) dm \right) \nonumber \\
		&=  \left( 1 - \frac{N_t}{N} \right) \left( \mu(t) + \sum_{t_j < t} K  \theta e^{-\theta (t - t_j)} \right)
		\eqmoveup
\end{align}
Where $p(m)$ is the distribution of local user influence of parameter $\alpha$ (see Sec.~\ref{subsec:hawkesN}).
As a result, $K$ in Eq.~\eqref{eq:HawkesN-expectation} is $K = \kappa \frac{\alpha - 1}{\alpha - \eta - 1}$. 

We can see that Eq.~\eqref{eq:SIR-expectation} and~\eqref{eq:HawkesN-expectation} are identical when $N_t = C_t$ (i.e. we observed the same random process) and under the parameter equivalence in Theorem~\ref{theorem:expected-equivalence}.
That is to say, the new infection point process in an SIR model is equivalent in expectation with a HawkesN point-process with no background event rate.
This completes the proof of Theorem~\ref{theorem:expected-equivalence}.
We also demonstrate the equivalence empirically, through simulation and subsequent parameter fitting, in the online supplement~\cite{supplemental}.

Corollary~\ref{corollary:n-star-r0} holds for the marked HawkesN process:
\begin{corollary}  \label{corollary:n-star-r0-marked}
	The Basic Reproduction Number of an SIR process and the branching factor of its equivalent HawkesN process (according to Theorem~\ref{theorem:expected-equivalence-marked}) are equal.
	\vspace{-0.3cm}
\end{corollary}

\begin{equation*} 
	\textit{Proof:}\qquad n^\ast \overset{Eq.~\eqref{eq:branching-factor-hawkesn-marked}}{=} \kappa \frac{\alpha - 1}{\alpha - \eta - 1} \overset{Th.~\ref{theorem:expected-equivalence-marked}}{=} \frac{\beta}{\gamma} = \mathcal{R}_0 . 
\end{equation*}

\section{Relation between deterministic SIR and stochastic SIR}

\citet{Allen2008} analyzes in details the relation between the deterministic SIR and the stochastic SIR and shows that the mean behavior of the stochastic version converges asymptotically to the deterministic version.
She shows that the mean of the random function $I(t)$ in the stochastic SIR epidemic process is less than the solution $I(t)$ to the deterministic differential equation in~Eq.\eqref{eq:det-sir-size}.
We study the equivalence of the two flavors of SIR through simulation.
We simulate 100 realizations of the stochastic SIR and the deterministic SIR from the same set of parameters.
Fig.~\ref{fig:stochastic-vs-deterministic-SIR} shows the sizes of the population of Susceptible $S(t)$, Infected $I(t)$, Recovered $R(t)$ and the cumulated infected $C_t$.
For the stochastic version, we show the median and the 2.5\% / 97.5\% percentiles.
This result complements the analysis in Sec.~\ref{subsec:sir-model}.

\input{fig3-stochastic-vs-deterministic-sir}
\input{5-simulation-fitting}

\section{Robustness of fit -- additional graphics}
\label{subsec:robustness-of-fit}

\TODO{MAR}{Moved from main text. Make this section unitary!}
One key question regarding the HawkesN process in the context of modeling information diffusion is the number of events in each cascade that need to be observed for an accurate estimation of the parameters.
This is particularly important when the maximum number of events $N$ is not known in advance and needs to be estimated from data.
Starting from a set of parameters, we simulate 100 realizations.
We fit HawkesN on increasing prefixes of each realization.
Fig.~\ref{fig:equivalence-through-fitting} shows the graphics for the branching factor and parameter $N$ for HawkesN (the graphics for the other parameters are shown in the online supplement~\cite{supplemental}).
For calibration, we perform the same exercise for the basic Hawkes Process and we presents the graphic for its branching factor in Fig.~\ref{subfig:reliability-fit-hawkes}.
We chose to show these parameters as they are highly indicative for the unfolding of the rest of the cascade (as shown in Sec.~\ref{sec:diffusion-size-distribution}).
The basic Hawkes requires observing less the 30\% of the length of the cascade to make reliable estimates.
Our proposed HawkesN model is more sensitive to the amount of available information, and requires observing more than 40\% of the cascade before the median $n^\ast$ and $N$ estimates approach the true values.
This is because we estimate the population size $N$ from observed data.
Alternatively, $N$ could be estimated from past diffusions (discussed in Sec.~\ref{sec:discussion}).

\input{fig4-observed-estimated-quantities-synthetic-data}

\begin{figure}[tbp]
	\centering
	\newcommand\myheight{0.2}
	\subfloat[] {
		\includegraphics[page=1,height=\myheight\textheight]{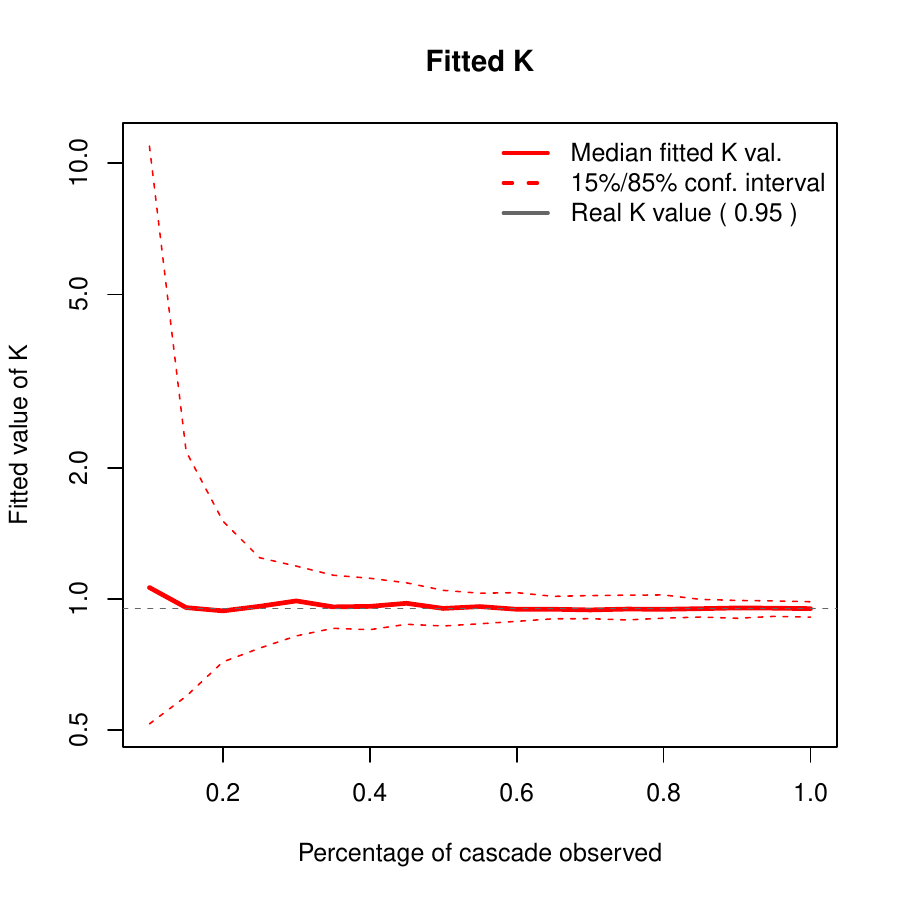}
	}
	\subfloat[] {
		\includegraphics[page=2,height=\myheight\textheight]{robustness-HAWKES}
	}\\
	\subfloat[] {
		\includegraphics[page=3,height=\myheight\textheight]{robustness-HAWKES}
	}
	\subfloat[] {
	\includegraphics[page=1,height=\myheight\textheight]{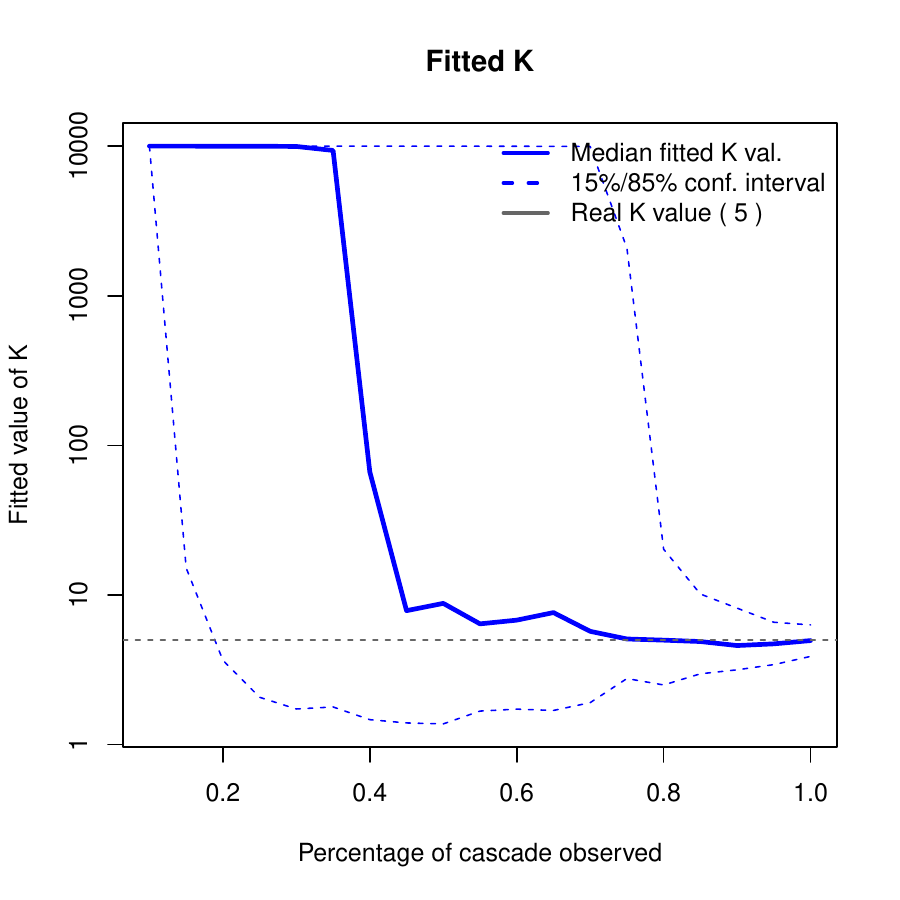}
	}\\
	\subfloat[] {
		\includegraphics[page=2,height=\myheight\textheight]{robustness-HAWKESN}
	}
	\subfloat[] {
		\includegraphics[page=3,height=\myheight\textheight]{robustness-HAWKESN}
	}
	\caption{ 
		 Robustness of estimating parameters $\kappa$, $\beta$ and $\theta$ for Hawkes \emph{(a)-(c)} and HawkesN \emph{(d)-(f)}.
		One set of parameters for each model was simulated 100 times and fitted on increasingly longer prefixes of each simulation.
		One value for parameter is obtained for each fit and the median and the 15\%/85\% percentile values are shown.
	}
	\label{fig:robustness-fit-extra-graphs}
	\captionmoveup
\end{figure}

Fig.~\ref{fig:robustness-fit-extra-graphs} shows the robustness of fit for parameters $\kappa$, $\beta$ and $\theta$ for Hawkes \emph{(a)-(c)} and HawkesN \emph{(d)-(f)}.
This result complements Sec.~\ref{subsec:robustness-of-fit}.

\section{Generalization performance -- Hawkes}
\begin{figure}[tbp]
	\centering
	\newcommand\myheight{0.18}
	\includegraphics[page=1,width=0.45\textwidth]{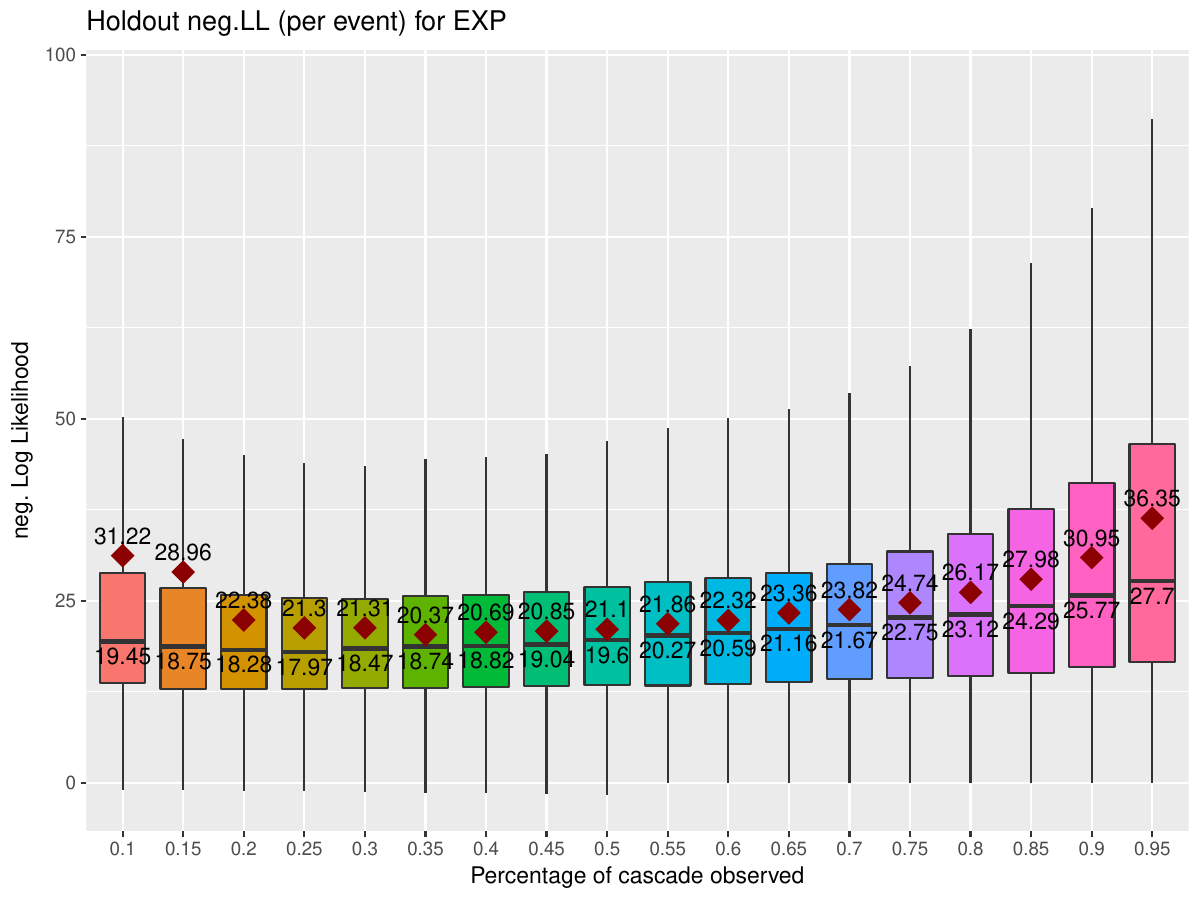}
	\caption{ 
		Performances of Hawkes explaining unobserved data, using holdout negative log likelihood.
		The performance over 1000 randomly sampled cascades in \News are summarized using boxplots, lower is better.
        The percentage of observed events in each cascade used to train Hawkes is varied between 10\% and 95\%.
	}
	\label{fig:news1k-increasing-perc}
	\captionmoveup
\end{figure}

Fig.~\ref{fig:news1k-increasing-perc} shows the generalization performance of Hawkes, for increasing amounts of data.
Each cascade in a random sample of 1000 cascades in \News is observed for increasing periods of time.
This result complements the analysis in Sec.~\ref{subsec:explain-holdout}.

\section{Estimating population size ($N$) in HawkesN}
\label{si-sec:estimating-N}

Total population size ($N$) estimation is an important yet challenging task in our proposed HawkesN model as we assume a fixed population size for each cascade. For this reason, we conducted a bottom-up experiment on understanding the difficulty of population size estimation, including both analytic and empirical study. This experiment is presented in three main steps. In Sec. \ref{hawkesn_step1}, we apply a simplified intensity function and derive its closed-form solution for estimating $N$. Sec. \ref{hawkesn_step2} lists some empirical experiments on the simplified intensity function showing the difficulty of retrieving real $N$ values. Last, we conduct empirical experiments in Sec. \ref{hawkesn_step3} for retrieving $N$ from HawkesN model.

\subsection{Step 1: Analytic Results for A Simplified Intensity Function}
\label{hawkesn_step1}
As introduced before, the intensity function for HawkesN model is shown as Eq.~(\ref{eq:hawkesN}). In order to get an intuition of population size estimation, we only consider a simplified intensity function in this section which is defined as:
\begin{equation}
	\lambda^{H}(t^{-}) = 1 - \frac{N_{t^{-}}}{N}
\end{equation}
where we simply let the kernel function $\phi(t-t_j) = 0$ and immigrant event arrival rate $\mu(t) = 1$.

Eq.~(\ref{eq:hawkes-ll}) defines the likelihood function, from which we can derive the likelihood function for our simply intensity function:
\begin{align} \label{eq:likelihood}
	\mathcal{L}(N) &= \sum^{n}_{j=1}\log(\lambda^{H}(t^{-}_j)) - \int^{t_n}_{0} \lambda^{H}(\tau) d\tau \nonumber \\
    			&= \sum^{n}_{j=1} \left[ \log\left(1 - \frac{j-1}{N}\right) - \int^{t_j}_{t_{j-1}} 1- \frac{j-1}{N} d\tau \right] \nonumber \\
        		&= \sum^{n}_{j=1} \left[ \log\left(1 - \frac{j-1}{N}\right) - \left(1- \frac{j-1}{N}\right)(t_j - t_{j-1}) \right] \nonumber \\
                &= \sum^{n}_{j=1} \left[ \log\left(1 - \frac{j-1}{N}\right) + \frac{j-1}{N}(t_j - t_{j-1}) \right] - \sum^{n}_{j=1} (t_j - t_{j-1}) \nonumber \\
                &= \sum^{n}_{j=1} \left[ \log\left(1 - \frac{j-1}{N}\right) + \frac{j-1}{N}(t_j - t_{j-1}) \right] - t_n
\end{align}
note that we define $t_0 = 0$ and $t_1$ is the event time of the first event.

\subsubsection{Maximum Likelihood Estimates of $N$}

We are interested in computing maximum likelihood estimates (MLEs) of the parameter $N$ given historical event times $\{t_1, ..., t_N\}$. Given Eq.~(\ref{eq:likelihood}), for any optimal solution $N^*$.
We compute the derivative of the Log-Likelihood function:
\begin{align}
	\frac{d\mathcal{L}}{dN} &= \sum^{n}_{j=1} \left[\frac{\frac{j-1}{N^2}}{1 - \frac{j-1}{N}} - \frac{j-1}{N^2}(t_j - t_{j-1}) \right] \nonumber \\
    &= \sum^{n}_{j=1} \frac{j-1}{N^2} \left[\frac{N}{N - j + 1} - (t_j - t_{j-1})\right] = 0 \label{eq:simple_kernel_derivative} \\
    & \Longrightarrow \sum^{n}_{j=1} \frac{j-1}{{N^*}^2 - N^*(i-1)} = \sum^{n}_{j=1} \frac{j-1}{{N^*}^2}(t_j - t_{j-1}) \nonumber \\
    & \Longrightarrow \sum^{n}_{j=1} \frac{N^*(j-1)}{N^* - j+1} = \sum^{n}_{j=1} (j-1)(t_j - t_{j-1})
\end{align}
as this is obscure, we break this down into following simple cases:
\begin{itemize}
	\item \textbf{One event}: apparently when $i=1$, $N$ is unidentifiable.
    \item \textbf{two events}: we get
    \begin{align*}
    	& \frac{N}{N-1} = t_2 - t_1 \\
        & N = \frac{t_1 - t_2}{1 + t_1 - t_2}
    \end{align*}
    In order to keep $N > 0$, $1+t_1-t_2 < 0$. Also we need $N > 2$, thus $1 < t_2 - t_1 < 2$
    \item \textbf{three events}: we get
    \begin{align*}
    	& \frac{N}{N-1} + \frac{2N}{N-2} = t_2 - t_1 + 2(t_3 - t_2) \\
        & \frac{N}{N-1} + \frac{2N}{N-2} = 2t_3 - t_2 - t_1 \\
        & 3N^2 - 4N = (2t_3 - t_2 - t_1)(N^2 - 3N + 2) \\
        & (3 - 2t_3 + t_2 + t_1)N^2 +\\ 
        & \hspace{0.3cm} (6t_3 - 3t_2 - 3t_1 - 4)N -(4t_3 - 2t_2 - 2t_1) = 0 
    \end{align*}
    Thus $N$ has two solutions, $N = \frac{(4-6t_3 + 3t_2 + 3t_1) \pm \sqrt{(2t_3 - t_2 - t_1)^2 + 16}}{6 - 4t_3 + 2t_2 + 2t_1}$, where $2t_3 - t_2 - t_1 \neq 3$.
\end{itemize}

\subsubsection{Identify Likelihood Without Maximum Value}

We note in Eq.~(\ref{eq:simple_kernel_derivative}) that we can separate the variable $N$ and other constants by the following deduction:
\begin{align}
	\frac{d\mathcal{L}}{dN} &= \sum^{n}_{j=1} \frac{j-1}{N^2} \left[\frac{N}{N - j + 1} - (t_j - t_{j-1})\right] \nonumber \\
    						&\overset{j-1 \ge 0}{\ge} \sum^{n}_{j=1} \frac{j-1}{N^2} \left[\frac{N}{N} - (t_j - t_{j-1})\right] \nonumber \\
                            &= \frac{1}{N^2} \sum^{n}_{j=1} (j-1)(1 - t_j + t_{j-1}) \nonumber \\
                            &= \frac{1}{N^2} (\frac{n(n-1)}{2} - \sum^{n-1}_{j=1} (t_n - t_j)) 
\end{align}
We denote 
\begin{equation} \label{eq:test}
	test = \frac{n(n-1)}{2} - \sum^{n-1}_{j=1} (t_n - t_j)
\end{equation}
As $\frac{1}{N^2} > 0$, the constant part, $test$, determines the sign of $\frac{d\mathcal{L}}{dN}$. When $test > 0$, $\frac{d\mathcal{L}}{dN} > 0$ which means the likelihood function is monotonically increase and theoretically there is no valid maximum value within the range. On the other hand, however, if $test < 0$, there might be valid maximum value.

\subsection{Step 2: Experiments on Simplified Intensity Function}
\label{hawkesn_step2}

In this section, we conducted some experiments on simulated cascades using simplified intensity function.

\subsubsection{Empirical Analysis of Valid Roots of $\frac{d\mathcal{L}}{dN} = 0$}

\begin{figure}[tbp]
	\centering
	\newcommand\myheight{0.2}
	\subfloat[] {
		\includegraphics[page=1,height=\myheight\textheight]{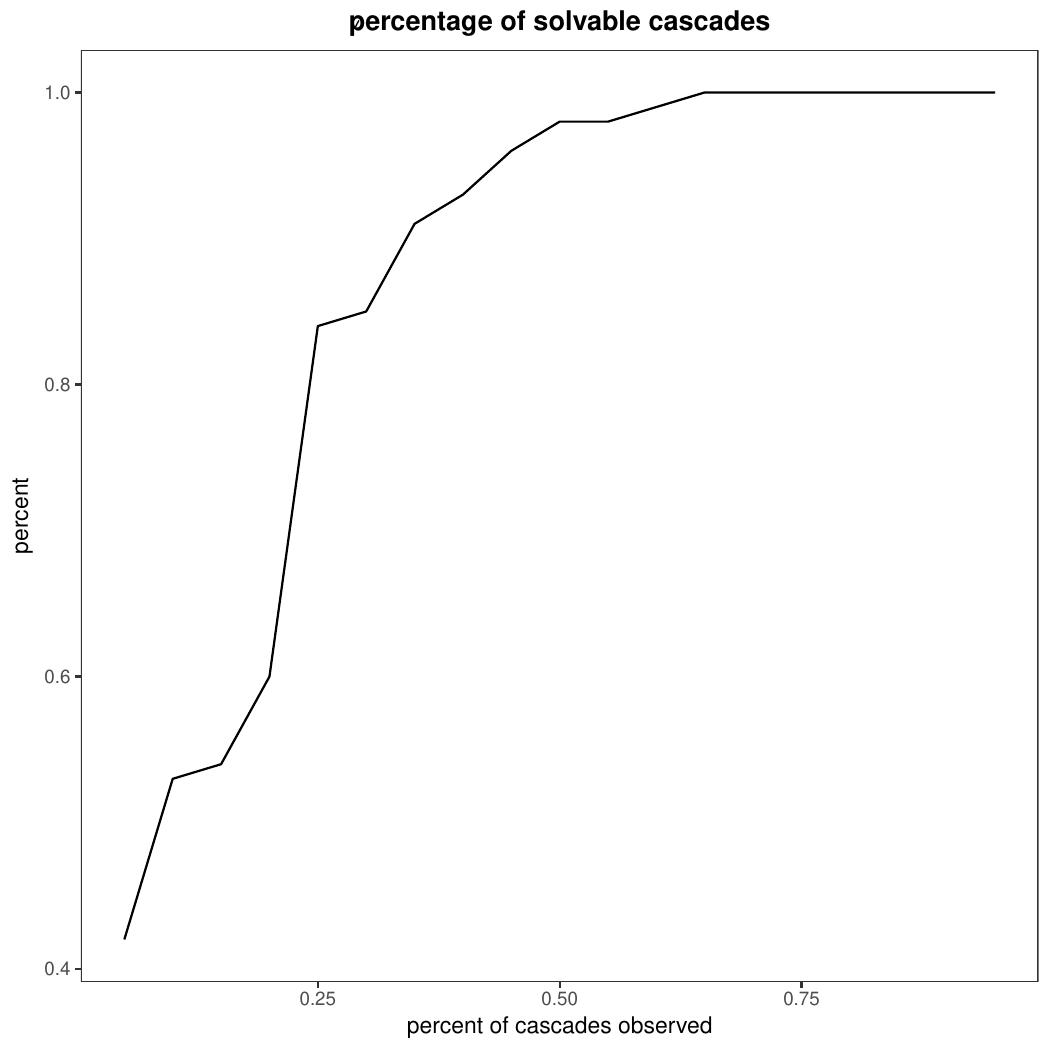}
	}
	\caption{ 
		 Percentages of cascades that have a valid root as percentages of cascades observed increase.
	}
	\label{fig:valid-root-vs-events-observed}
	\captionmoveup
\end{figure}

We define a valid root as an optimal solution $N^*$, such that $\frac{d\mathcal{L}}{dN^*} = 0$ and $N^{*} > n$ where $n$ is the number of events observed. Given a number of events, there might not exist any valid roots for $\frac{d\mathcal{L}}{dN} = 0$. Fig.~(\ref{fig:valid-root-vs-events-observed}) shows the fact that the more events are used for finding roots, the more likely there will be valid roots.

\subsubsection{Empirical Analysis on Number of Valid Roots}

Throughout our experiments for all simulated cascades, there are only two possible cases where there is either one valid root or there is no solution. For this reason, we empirically conclude that there will not exist more than one valid root for $\frac{d\mathcal{L}}{dN} = 0$.

\subsubsection{Correlation between Likelihood Maximum and Root of Likelihood Derivation}

\begin{figure}[tbp]
	\centering
	\newcommand\myheight{0.2}
	\subfloat[] {
		\includegraphics[page=4,height=\myheight\textheight]{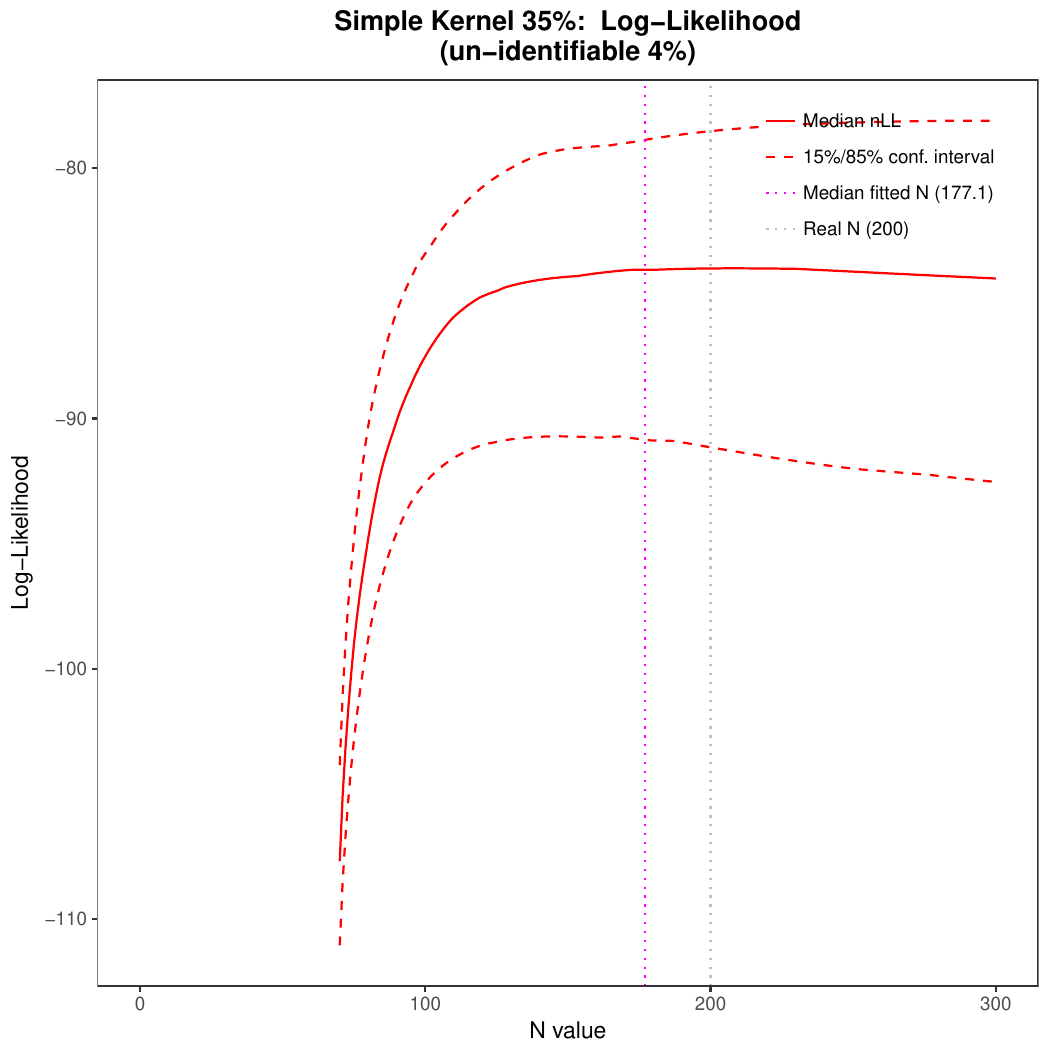}
	}
	\subfloat[] {
		\includegraphics[page=6,height=\myheight\textheight]{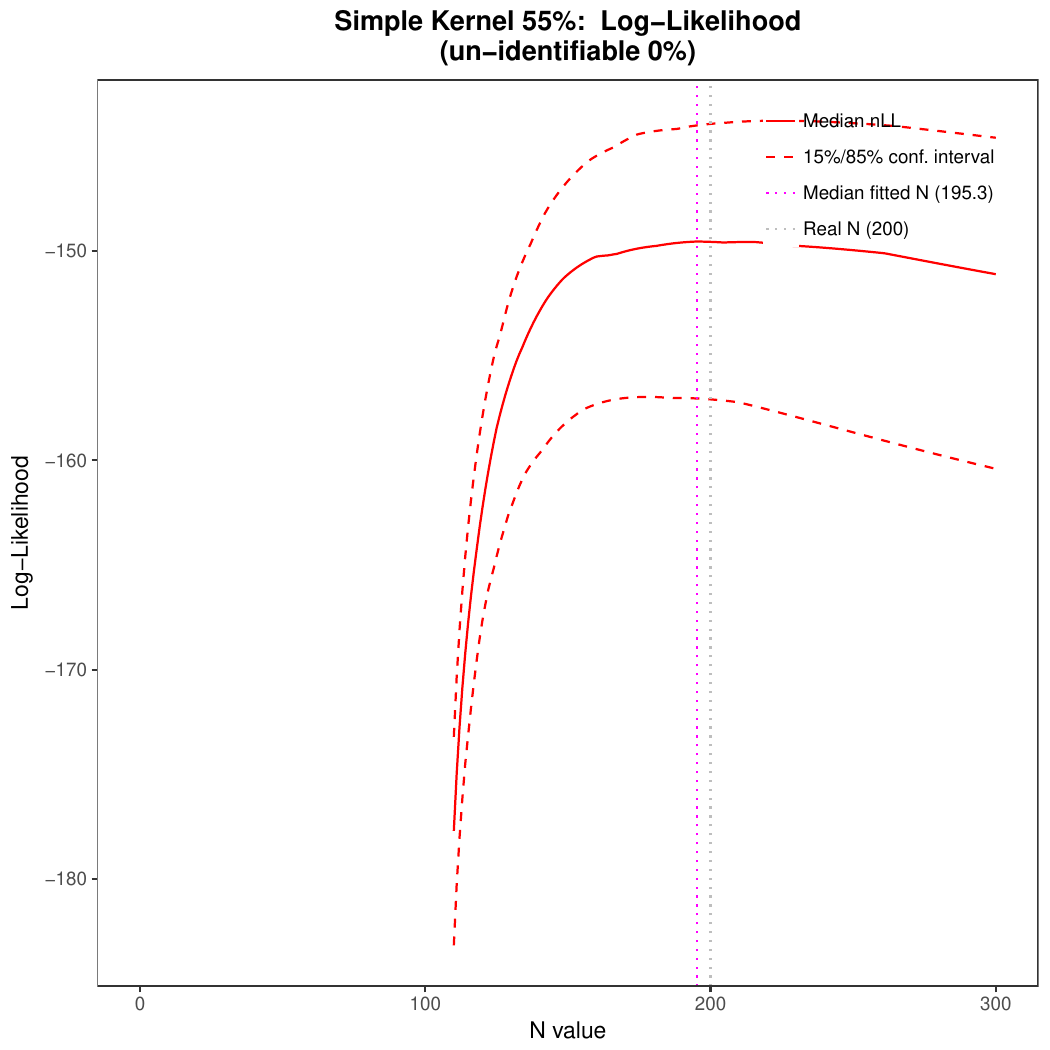}
	}\\
	\subfloat[] {
		\includegraphics[page=8,height=\myheight\textheight]{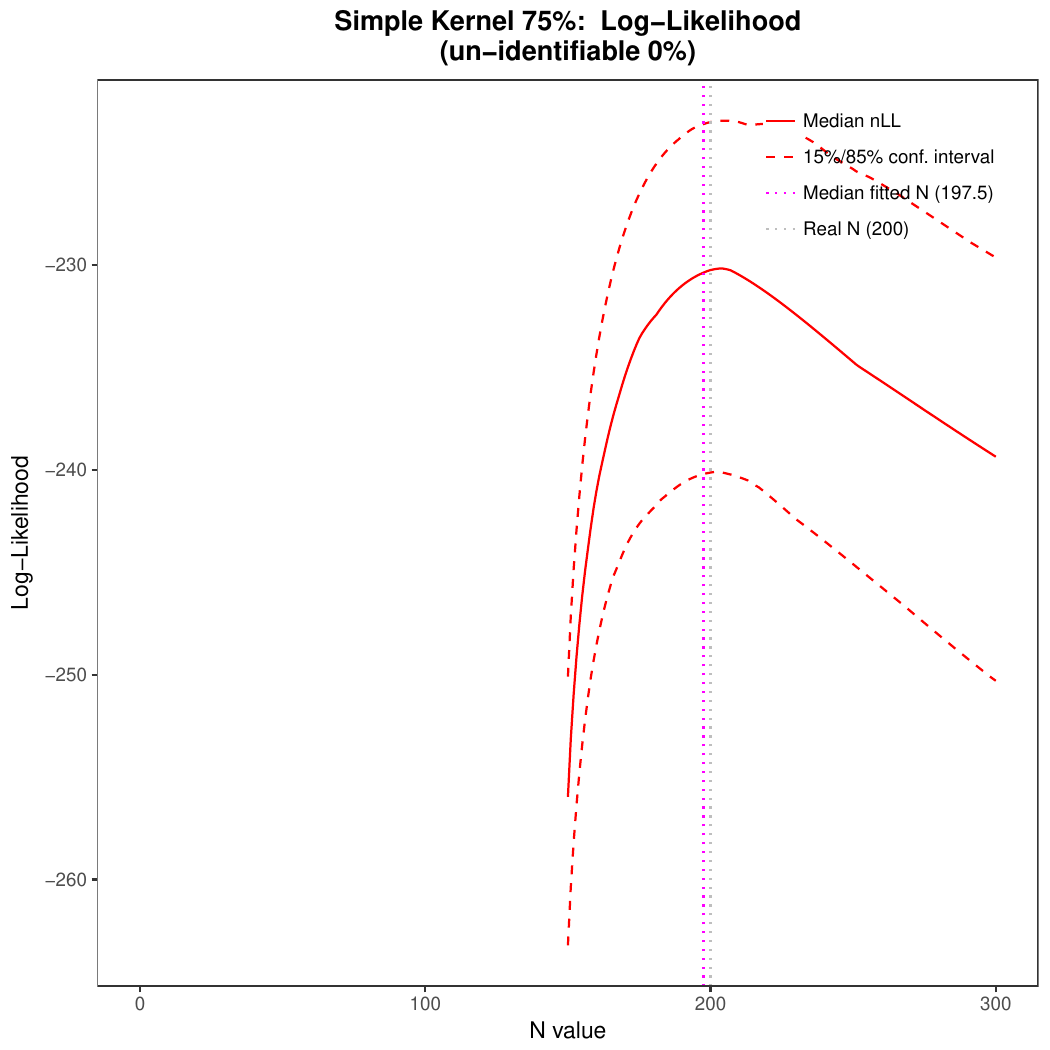}
	}
	\subfloat[] {
	\includegraphics[page=10,height=\myheight\textheight]{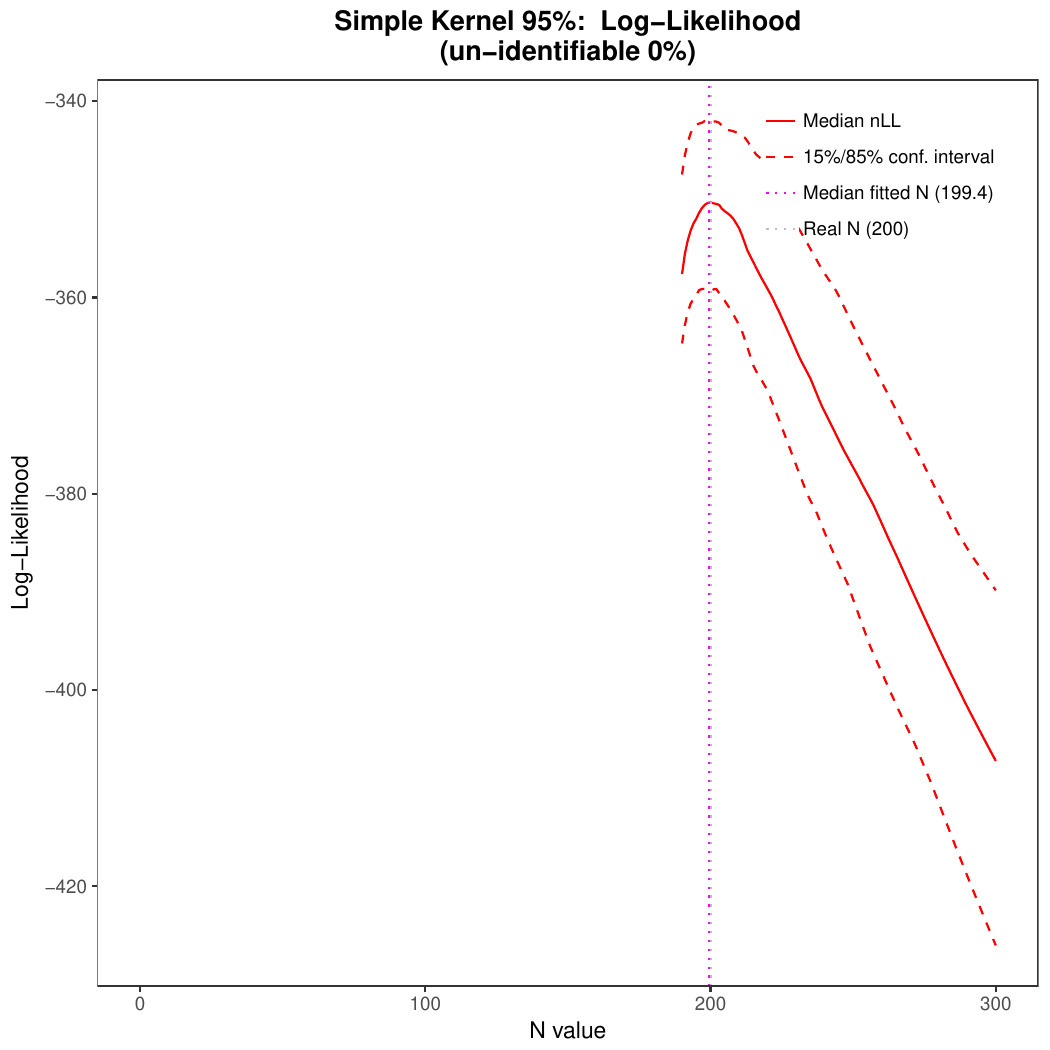}
	}
	\caption{ 
		 Changing of log-likelihood values as values of estimated $N$ change. \emph{(a)-(d)} shows plots for different percentages of cascades are observed.
	}
	\label{fig:estimated-N-to-LL}
	\captionmoveup
\end{figure}

We show the correlation between likelihood values over different $N$ and the valid root we found by the derivative of likelihood. From Fig.~(\ref{fig:estimated-N-to-LL}) we can verify the trend of likelihood values and the correctness of valid roots showing the maximum likelihood values.

\subsubsection{Correctness of Estimated $N$}

\begin{figure}[tbp]
	\centering
	\newcommand\myheight{0.2}
	\subfloat[] {
		\includegraphics[page=4,height=\myheight\textheight]{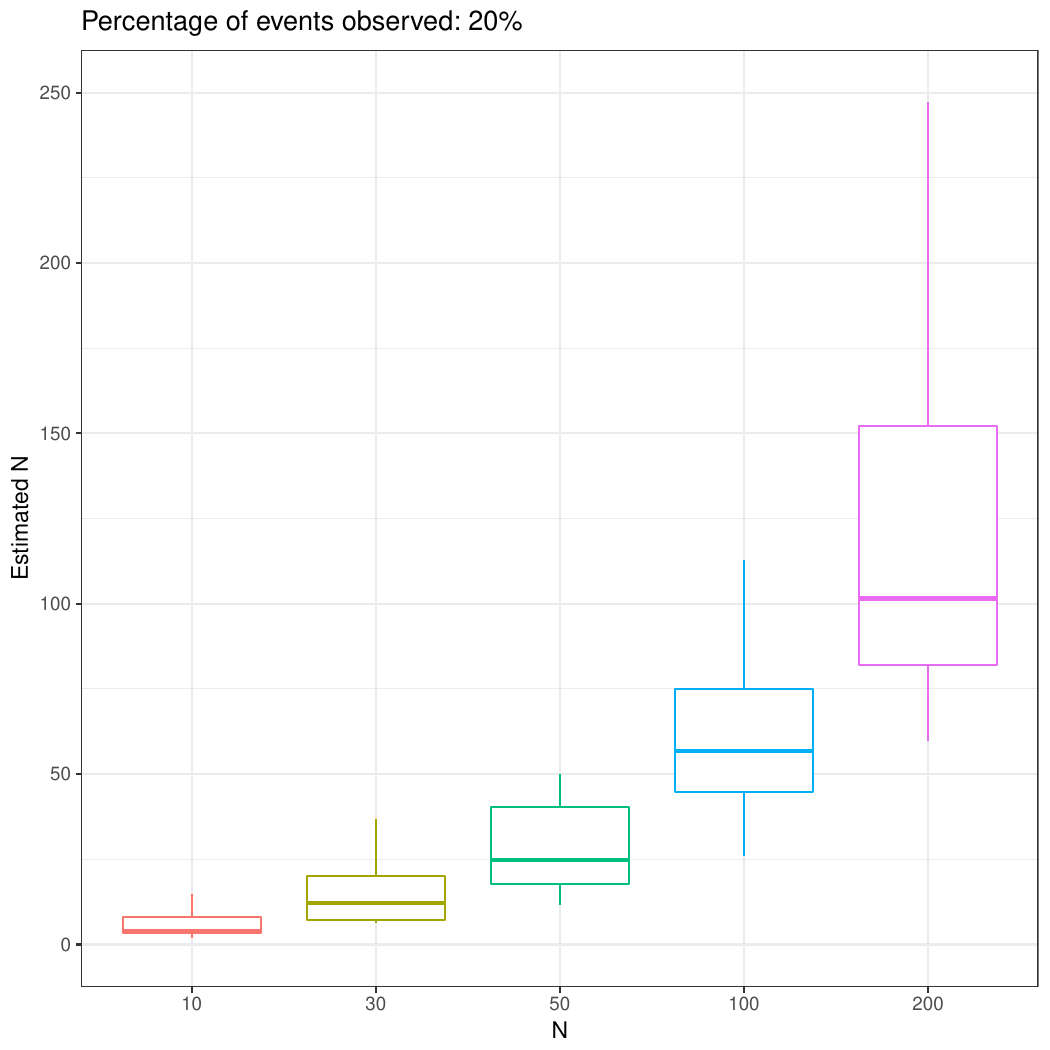}
	}
	\subfloat[] {
		\includegraphics[page=6,height=\myheight\textheight]{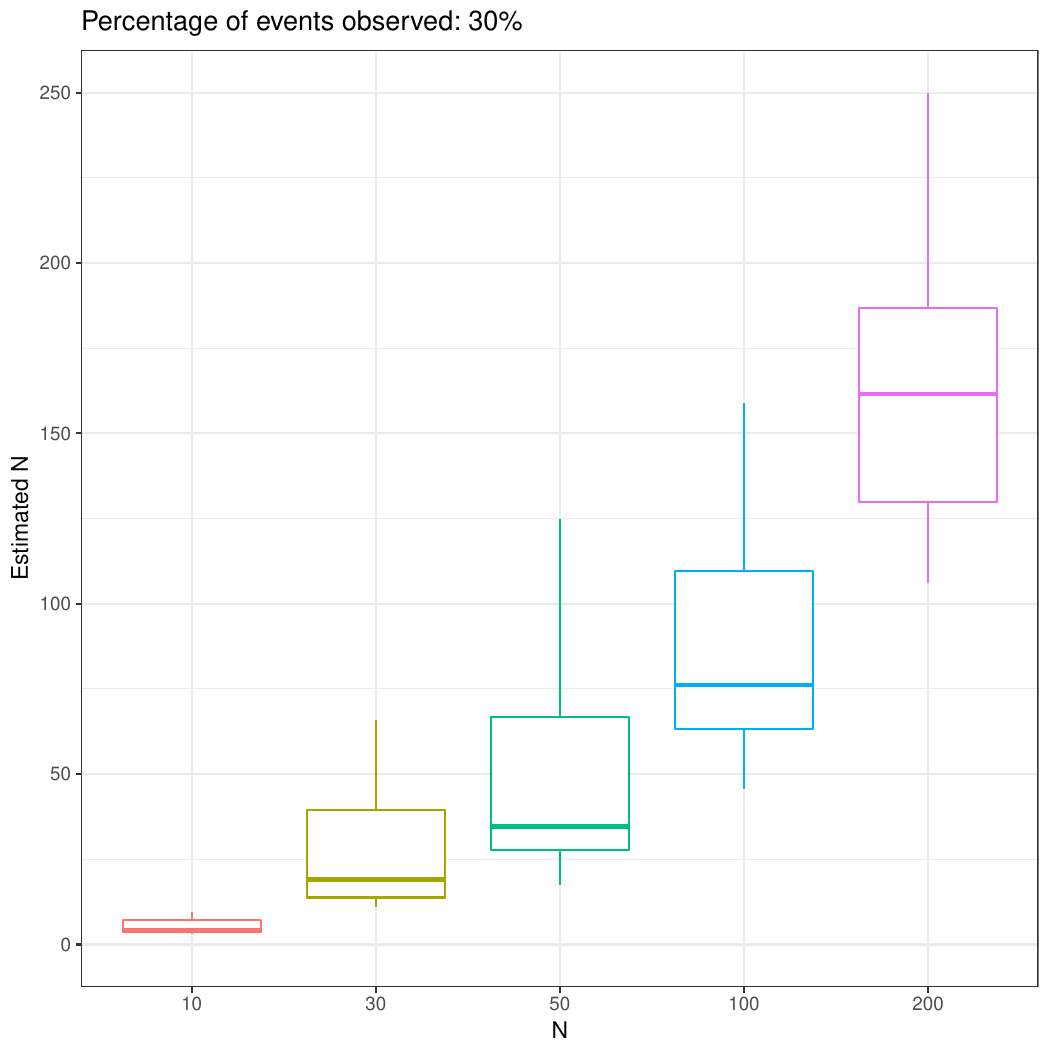}
	}\\
	\subfloat[] {
		\includegraphics[page=8,height=\myheight\textheight]{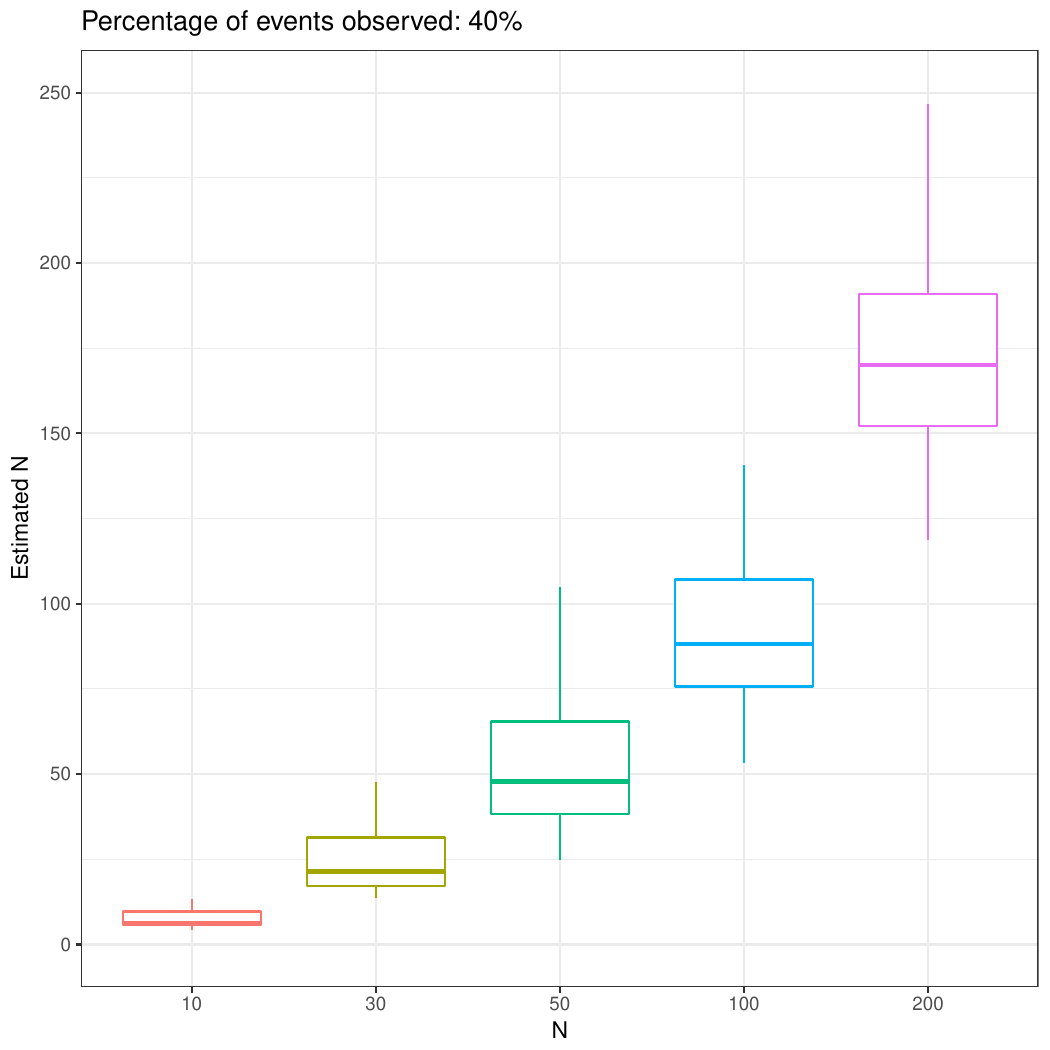}
	}
	\subfloat[] {
	\includegraphics[page=10,height=\myheight\textheight]{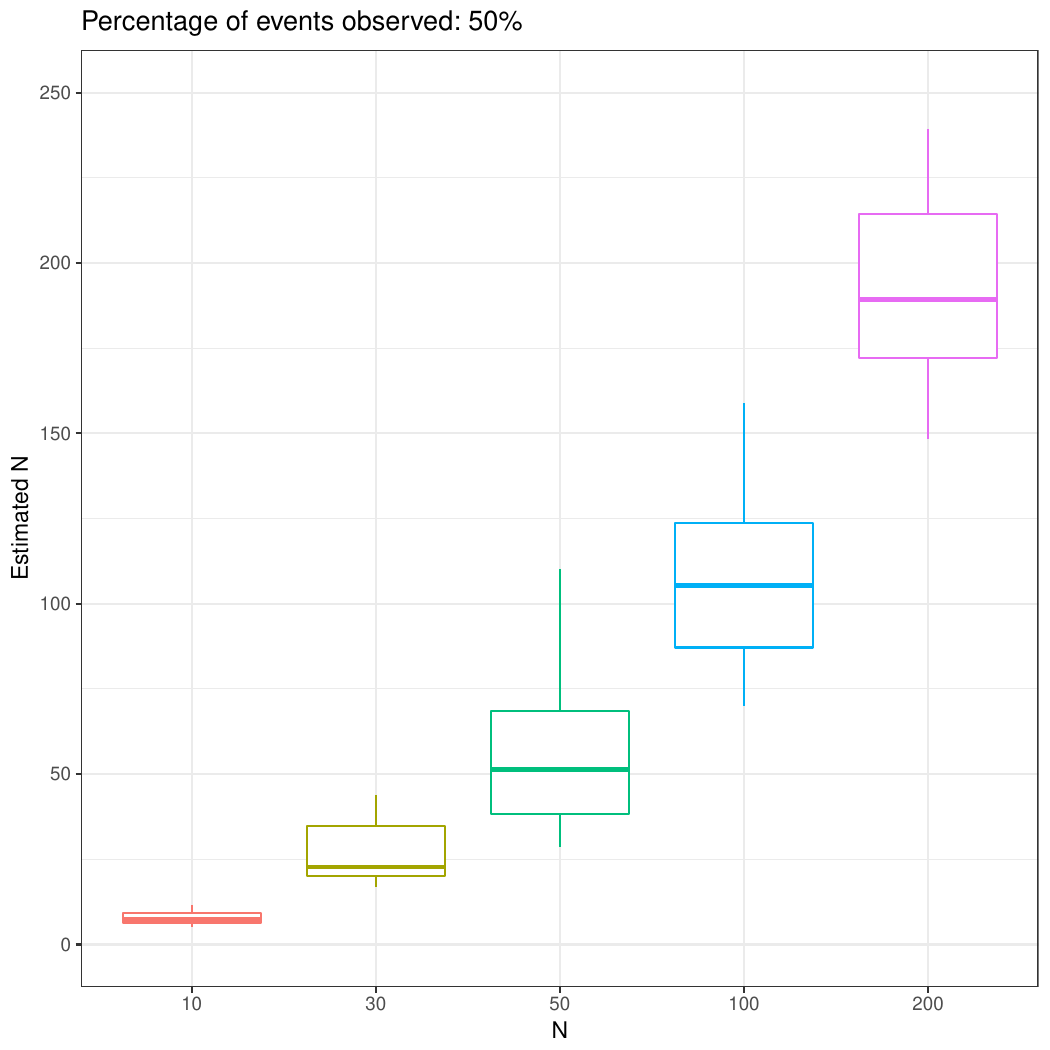}
	}
	\caption{ 
		 Y axis is the estimated $N$ value and X axis is the real $N$ used for simulation. \emph{(a)-(d)} shows plots for different percentages of cascades are observed.
	}
	\label{fig:correctness-of-retrieved-N}
	\captionmoveup
\end{figure}

Fig.~(\ref{fig:correctness-of-retrieved-N}) shows how well does the MLE method retrieve $N$ value when different percentages of cascades are observed. We can find that, to retrieve the real $N$ value, a large part of a cascade is required. 

\subsubsection{Difficulty of Estimating $N$}
\begin{figure}[tbp]
	\centering
	\newcommand\myheight{0.2}
	\subfloat[] {
		\includegraphics[page=2,height=\myheight\textheight]{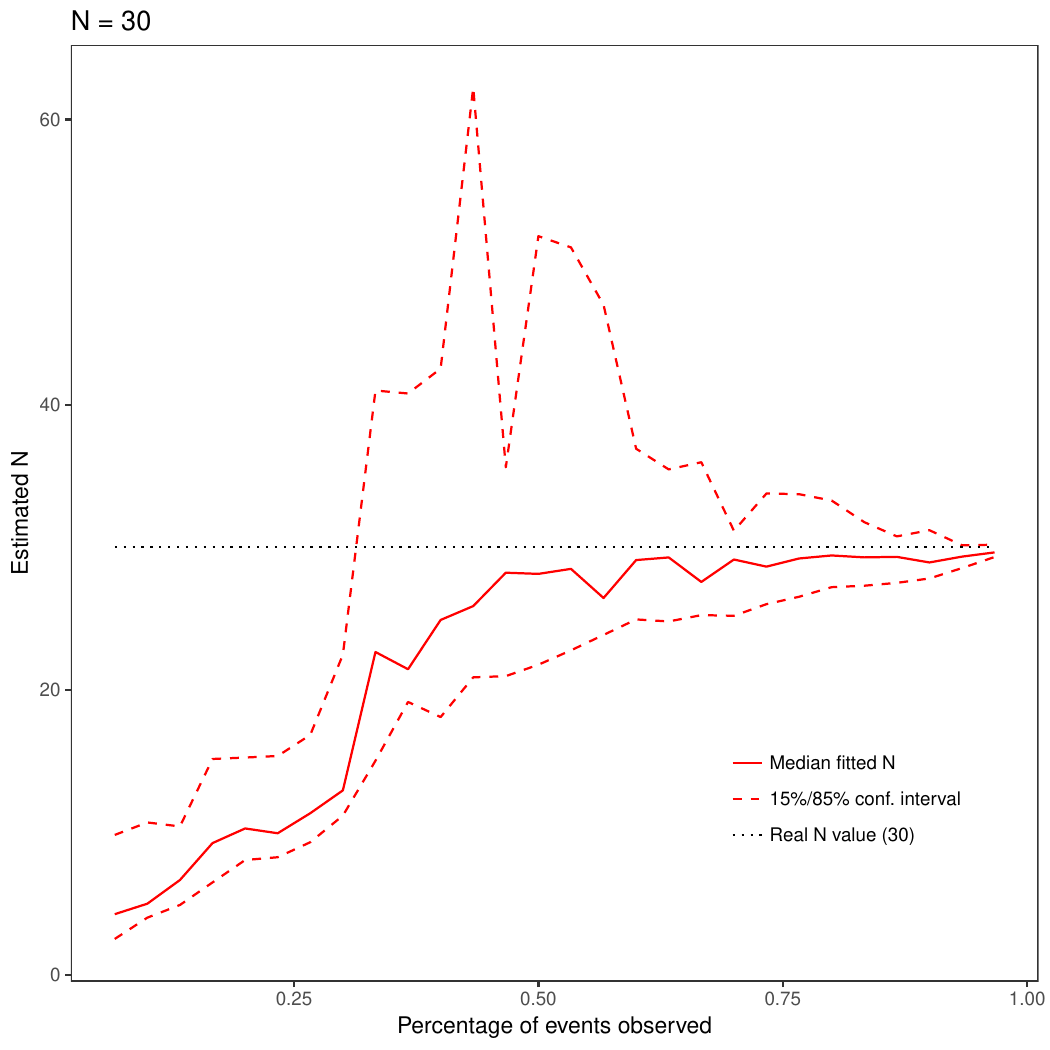}
	}
	\subfloat[] {
		\includegraphics[page=3,height=\myheight\textheight]{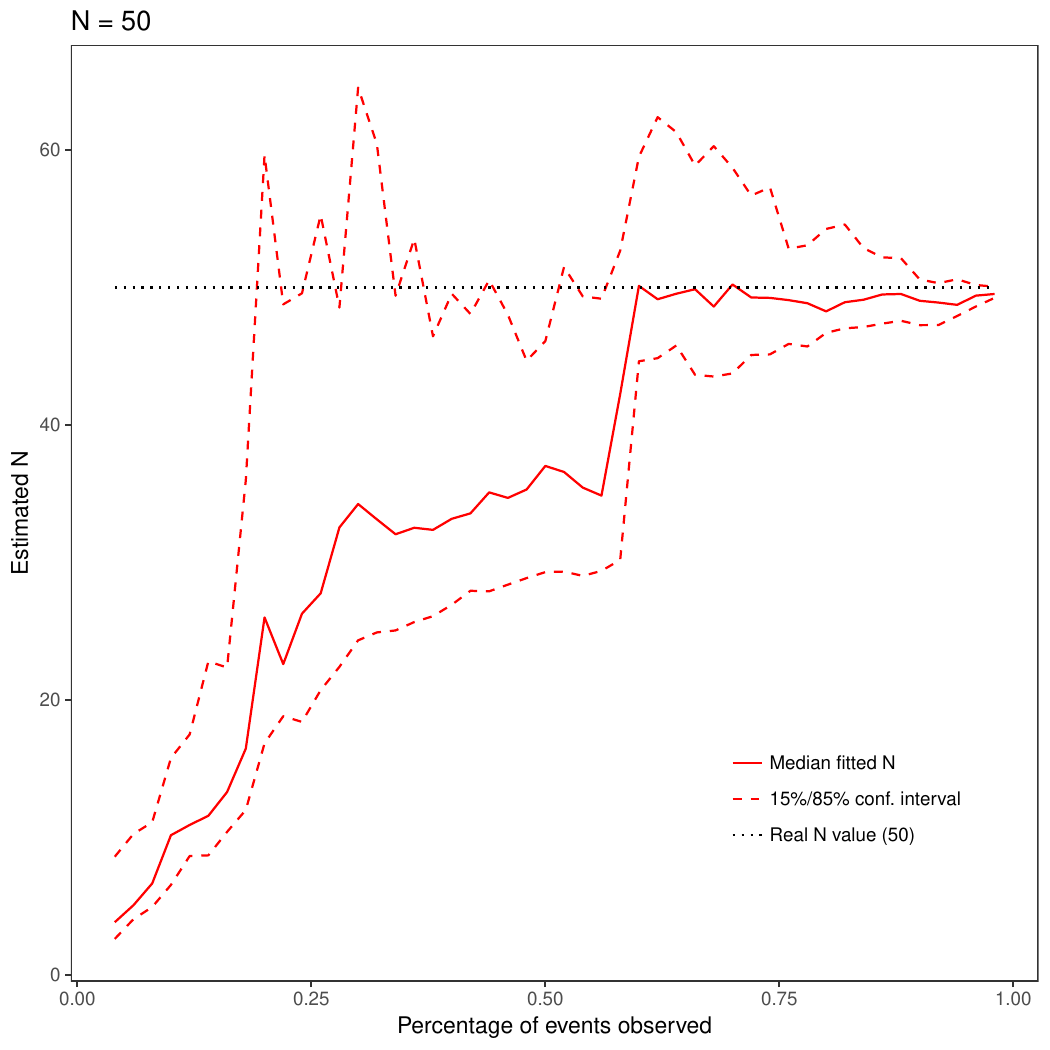}
	}\\
	\subfloat[] {
		\includegraphics[page=4,height=\myheight\textheight]{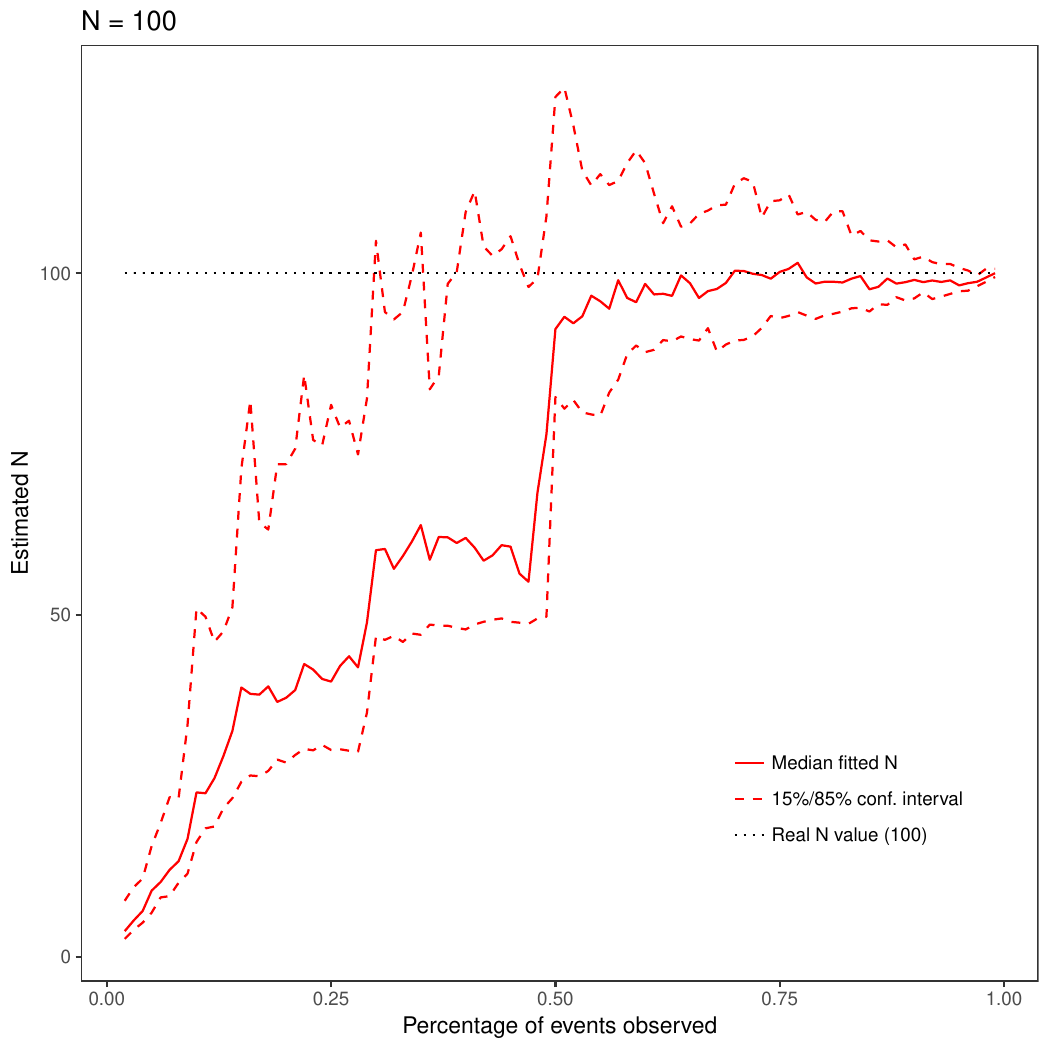}
	}
	\subfloat[] {
	\includegraphics[page=5,height=\myheight\textheight]{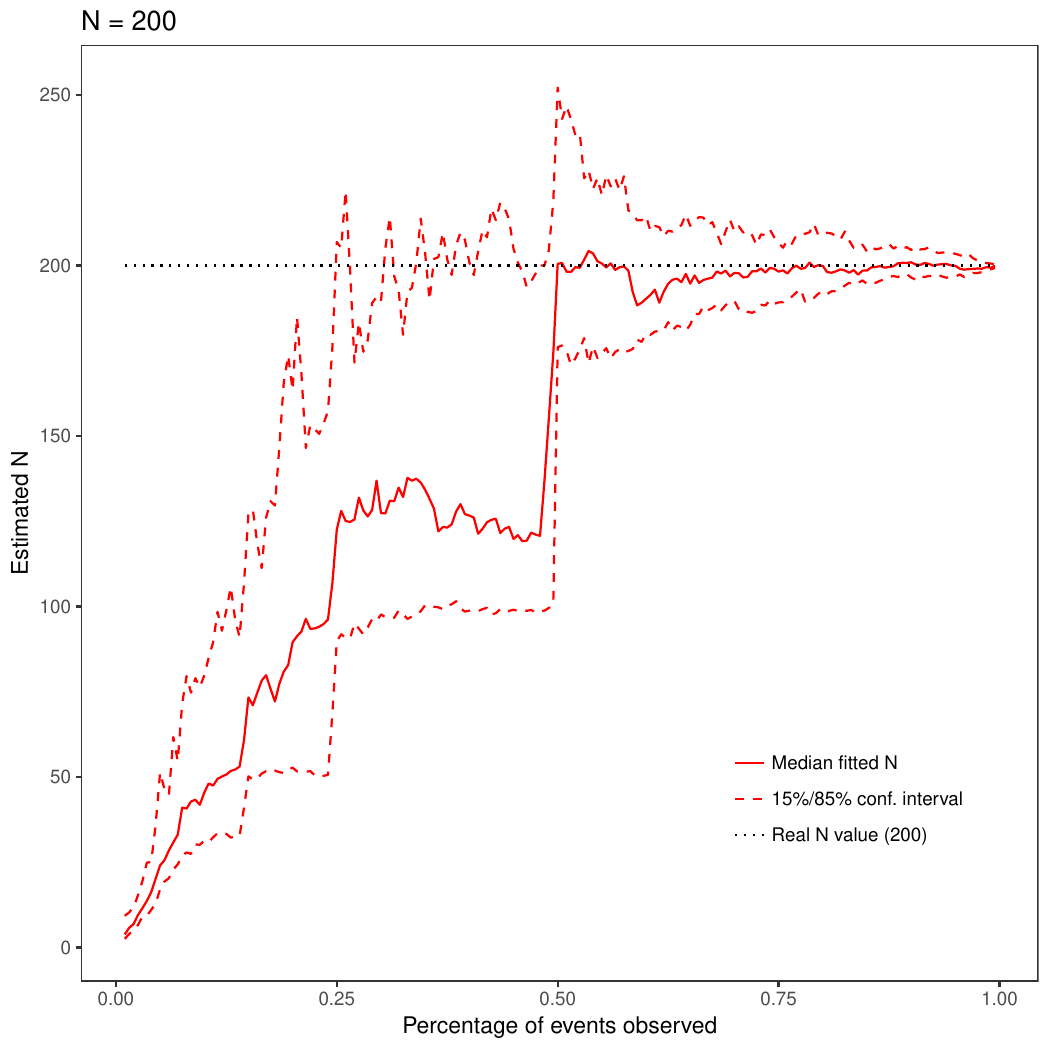}
	}
	\caption{ 
		 X axis is the percentages of a cascade observed and Y axis is the estimated $N$ value. \emph{(a)-(d)} shows plots for different real $N$ values used for simulation.
	}
	\label{fig:hardness-of-retrieved-N}
	\captionmoveup
\end{figure}

Fig.~(\ref{fig:hardness-of-retrieved-N}) shows how hard to retrieve $N$ by estimating early events. From the figure, we found that, to retrieve a correct value, we need more than $50\%$ of event history of a given cascade which means it is quite difficult to estimate $N$.

\subsubsection{Confusion Table of $test$ and Valid Root Existence}

Eq.~(\ref{eq:test}) leads to a way to identify likelihood functions without maximum values in a computational fast way. In order to validate this assumption, we conduct an experiment and generate the confusion table for $test$ and valid root existence. In the experiment, we apply $N=100$ and $200$ simulated cascades. Table~\ref{tab:confusion-table-simple-kernel} shows the informativeness of $test$ indicating the existence of a valid root given different percentages of cascades observed.

\begin{table}[tbp]
	\centering
	\caption{Confusion table of $test$ and Valid Root Existence} 
	\begin{tabular}{r|rrr}
	  \toprule
	   & Valid Root Exists & Valid Root Absents \\ 
 	 \midrule
     	\multicolumn{3}{c}{Percentage of cascades observed: 5\%} \\
     \midrule  
		$test \ge 0$ & 0 &  121 \\ 
		$test < 0$ & 78 &  1 \\ 
     \midrule
     	\multicolumn{3}{c}{Percentage of cascades observed: 20\%} \\
     \midrule  
		$test \ge 0$ & 0 &  71 \\ 
		$test < 0$ & 129 &  0 \\
 	 \midrule
     	\multicolumn{3}{c}{Percentage of cascades observed: 50\%} \\
     \midrule  
		$test \ge 0$ & 0 &  2 \\ 
		$test < 0$ & 198 &  0 \\ 
 	 \midrule
     	\multicolumn{3}{c}{Percentage of cascades observed: 80\%} \\
     \midrule  
		$test \ge 0$ & 0 &  0 \\ 
		$test < 0$ & 200 &  0 \\ 
 	  \bottomrule
	\end{tabular}
	\label{tab:confusion-table-simple-kernel}
	\captionmoveup
\end{table}

\subsection{Step 3: Experiments on HawkesN}
\label{hawkesn_step3}

In this section, we finally take a step further by conducting empirical experiments on our proposed HawkesN model. Throughout all simulations in our experiments, we fix the values of $\kappa, \theta$, after which $N$ becomes the only variable in HawkesN for estimating.

\subsubsection{Correlation between Likelihood Maximum and Root of Likelihood Derivation}

\begin{figure}[tbp]
	\centering
	\newcommand\myheight{0.15}
	\subfloat[] {
		\includegraphics[page=4,height=\myheight\textheight]{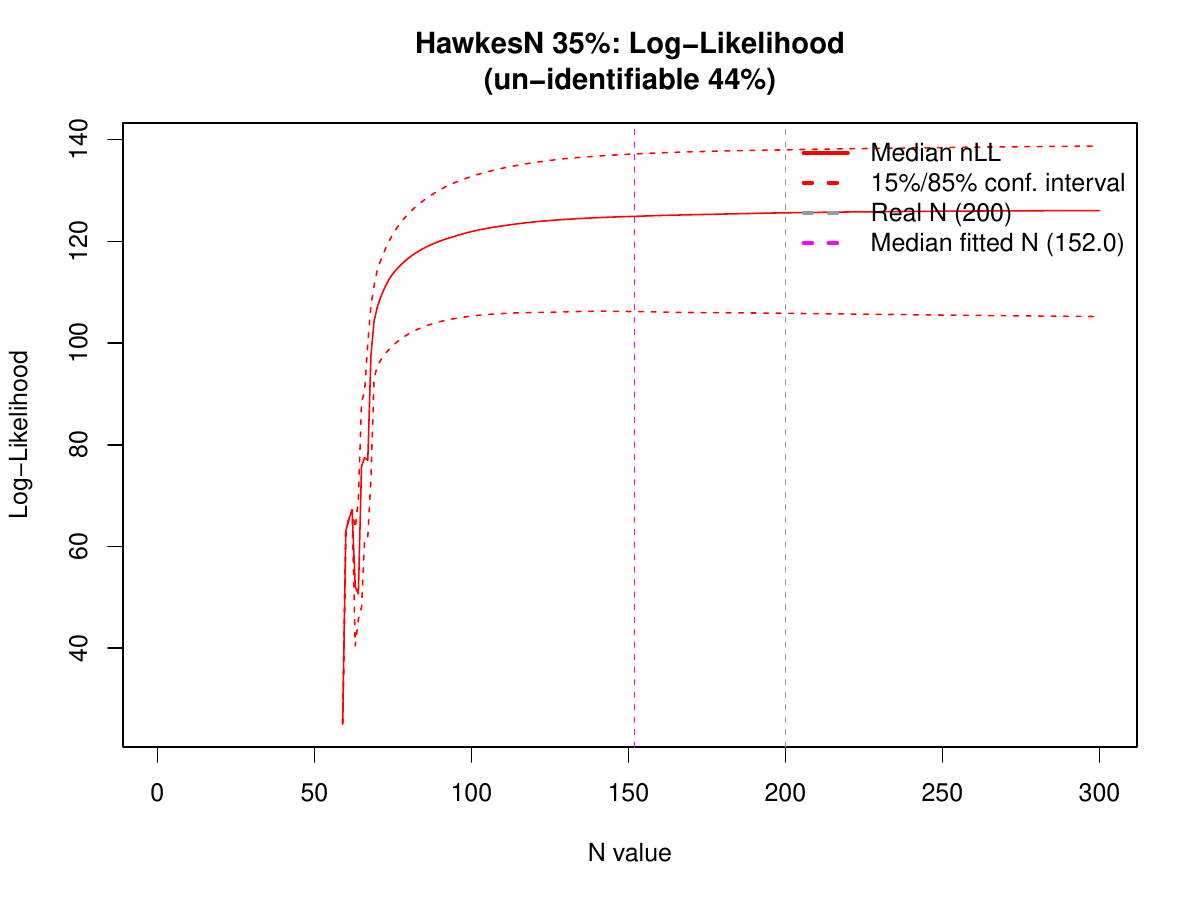}
	}
	\subfloat[] {
		\includegraphics[page=6,height=\myheight\textheight]{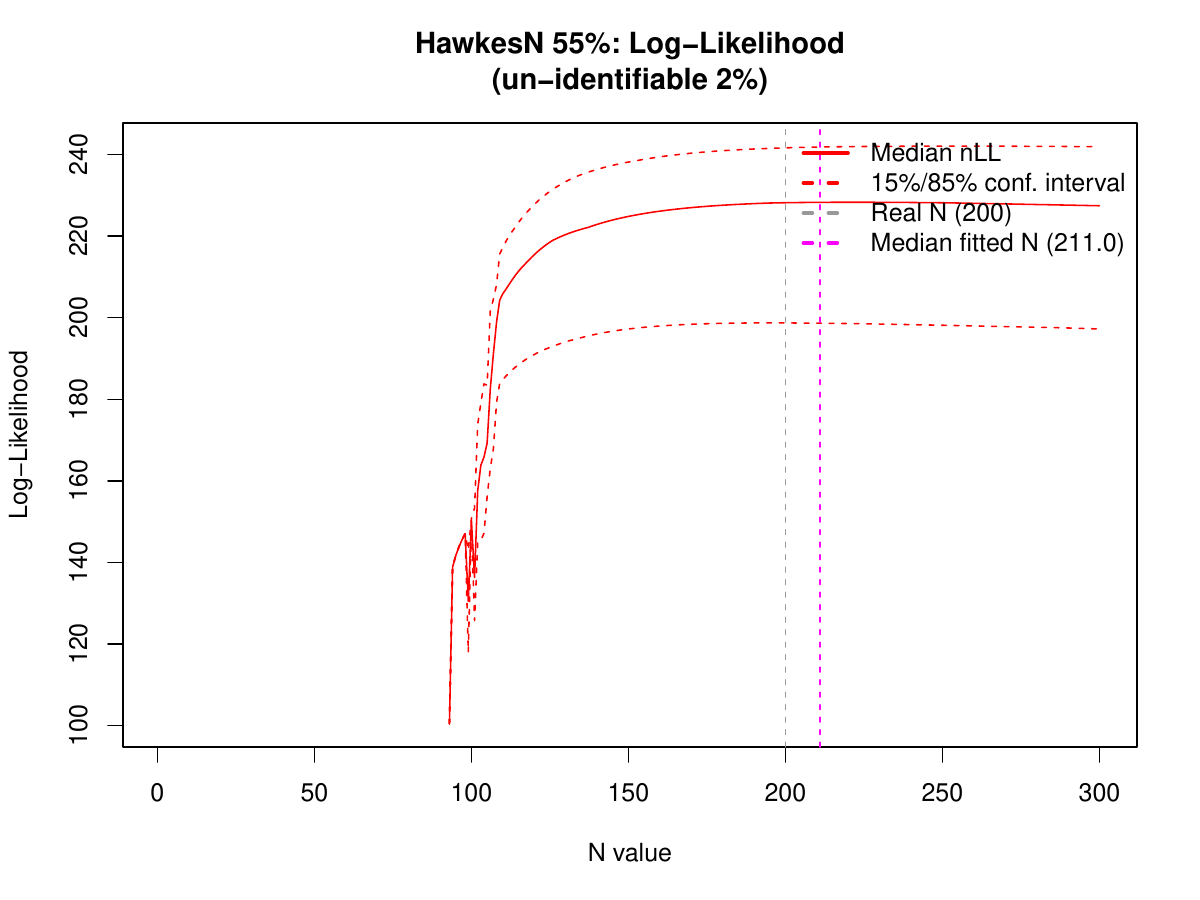}
	}\\
	\subfloat[] {
		\includegraphics[page=8,height=\myheight\textheight]{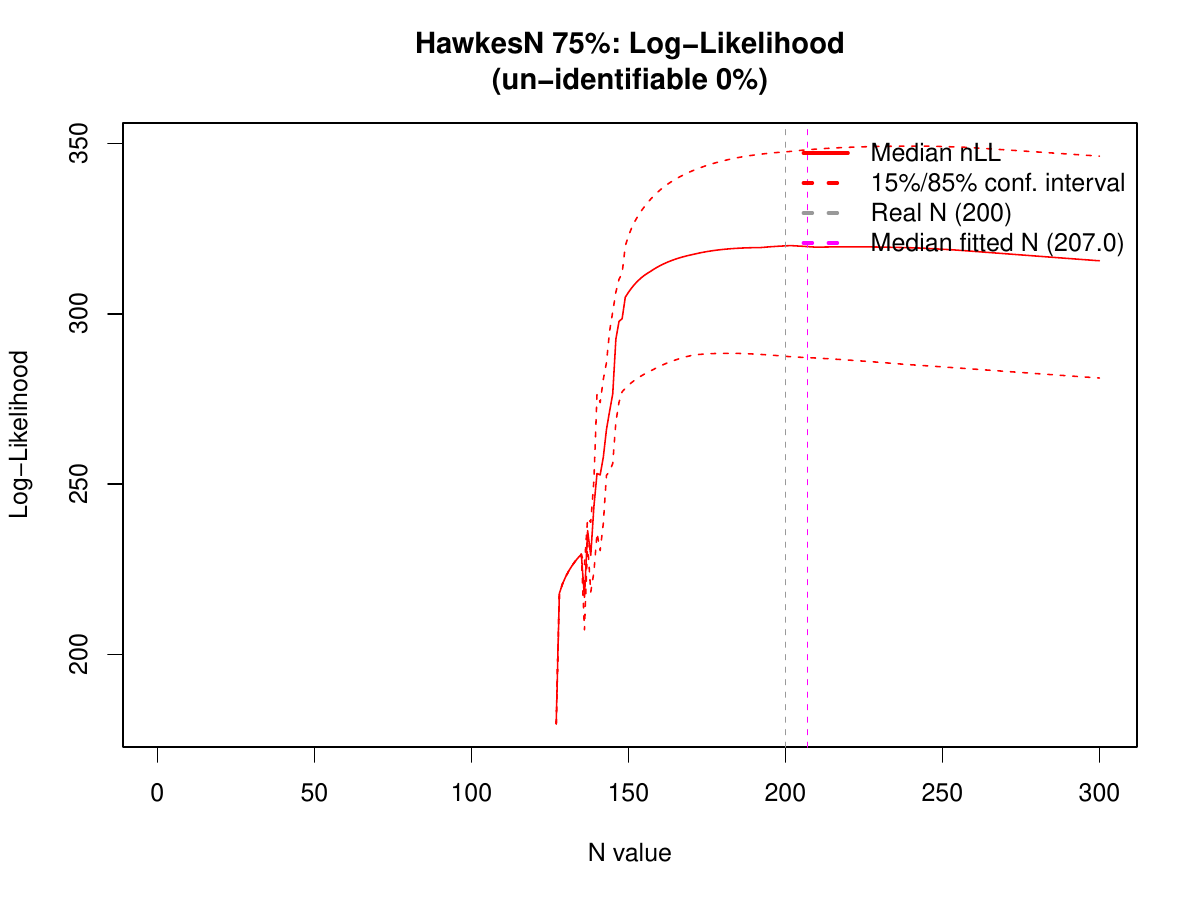}
	}
	\subfloat[] {
		\includegraphics[page=10,height=\myheight\textheight]{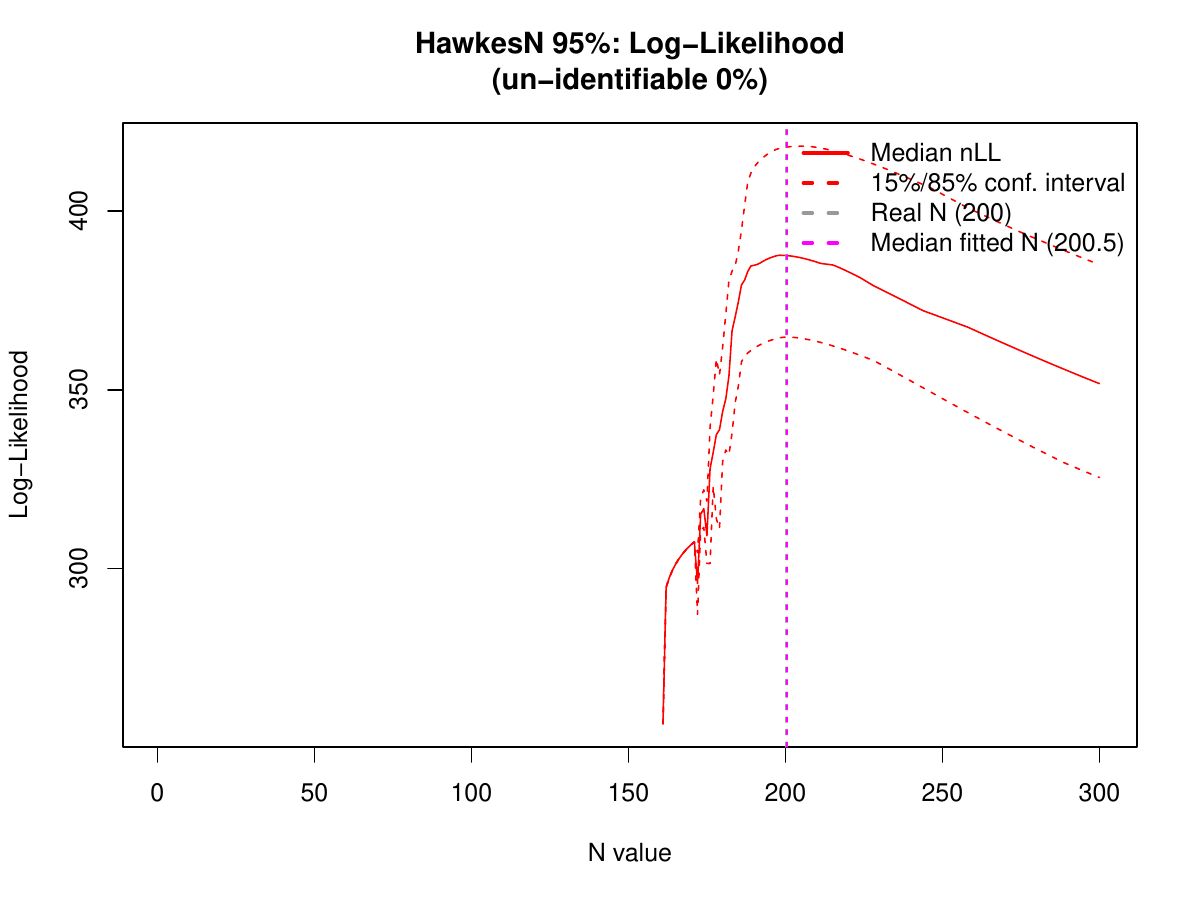}
	}
	\caption{ 
		 Changing of log-likelihood values as values of estimated $N$ change. \emph{(a)-(d)} shows plots for different percentages of cascades are observed. We set $N = 200$ for simulations in this experiment.
	}
	\label{fig:estimated-N-to-LL-hawkesN}
	\captionmoveup
\end{figure}

We show the correlation between likelihood values over different $N$ and the valid root we found by the derivative of likelihood. From Fig.~(\ref{fig:estimated-N-to-LL-hawkesN}) we can verify the trend of likelihood values and the correctness of valid roots showing the maximum likelihood values.

\subsubsection{Difficulty of estimating $N$}
\begin{figure}[tbp]
	\centering
	\newcommand\myheight{0.25}
	\subfloat[] {
		\includegraphics[page=1,height=\myheight\textheight]{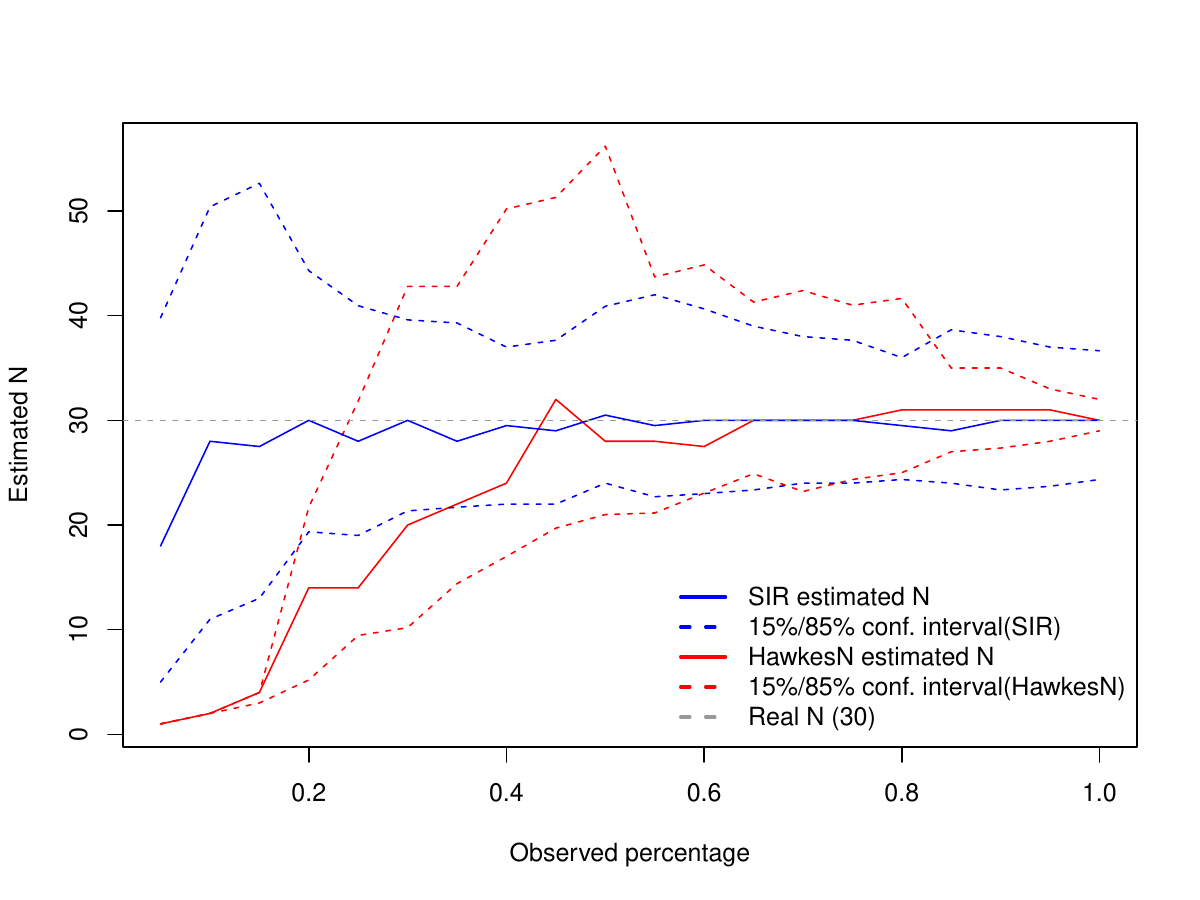}
	}\\
	\subfloat[] {
		\includegraphics[page=4,height=\myheight\textheight]{estimate-N-hardness-hawkesN}
	}
	\caption{ 
		 X axis is the percentages of a cascade observed and Y axis is the estimated $N$ value. \emph{(a)-(b)} shows plots for different real $N$ values used for simulation where $N = 30$ for \emph{(a)} and $N=200$ for \emph{(b)}.
	}
	\label{fig:hardness-of-retrieved-N-hawkesN}
	\captionmoveup
\end{figure}

Fig.~(\ref{fig:hardness-of-retrieved-N-hawkesN}) shows the trend of estimated $N$ values as different percentages of cascades are observed. Both HawkesN and SIR models are tested as comparison. The figure shows that it is hard to retrieve $N$ for the HawkesN model, but it is easy for SIR to estimate $N$ values by only observing a small number of historical events.

\subsection{Estimating $N$ in the Deterministic SIR}
Computing the final size in the deterministic model is straightforward, as it results directly from the differential equations in Eq.~\eqref{eq:sir-ds}-\eqref{eq:sir-dr}.
\citet{Allen2008} shows that dividing Eq.~\eqref{eq:sir-di} by Eq.~\eqref{eq:sir-ds} and integrating, we obtain:
\begin{align}
	&\; \frac{dI}{dS} = -1 + \frac{N \gamma}{\beta S} \nonumber \\
	\Rightarrow &\; I(t) + S(t) = I(0) + S(0) + \frac{N \gamma}{\beta} log\frac{S(t)}{S(0)} \nonumber \\
	^{t \rightarrow \infty, I(\infty) = 0} \Rightarrow &\; S(\infty) = N + \frac{N \gamma}{\beta} log\frac{S(\infty)}{S(0)}. \label{eq:det-sir-size}
	\eqmoveup
\end{align}
Eq~\eqref{eq:det-sir-size} has a root in $[0, N]$, which we find numerically.
The prediction of final size for the deterministic SIR is $R(\infty) = N - S(\infty)$.

\input{appendix_branching}

%% file: fig3-stochastic-vs-deterministic-sir.tex

\begin{figure}[tbp]
	\centering
	\newcommand\myheight{0.18}
	\subfloat[] {
		\includegraphics[page=1,height=\myheight\textheight]{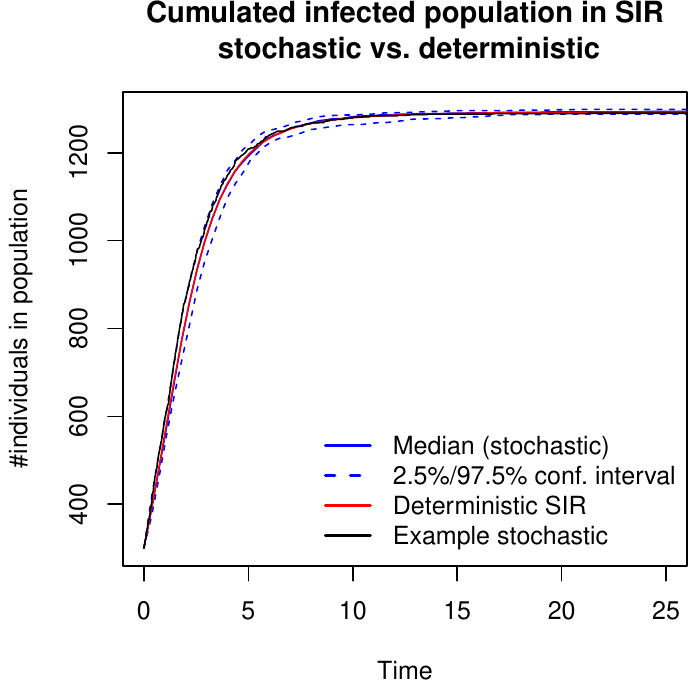}
	}
	\subfloat[] {
		\includegraphics[page=2,height=\myheight\textheight]{stochastic-vs-deterministic-sir}
	}  \\
	\subfloat[] {
		\includegraphics[page=3,height=\myheight\textheight]{stochastic-vs-deterministic-sir}
	}
	\subfloat[] {
		\includegraphics[page=4,height=\myheight\textheight]{stochastic-vs-deterministic-sir}
	}
	\caption{ 
		We simulate 100 stochastic SIR realizations using the parameters $N = 1300, I(0) = 300, \beta = 1, \gamma = 0.2$.
		We show the median and the $2.5\%$ and $97.5\%$ percentile and the deterministic evolution simulated with the same parameters.
		We also show an example of stochastic realization.
	}
	\label{fig:stochastic-vs-deterministic-SIR}
\end{figure}

%% file: 5-simulation-fitting.tex

\section{SIR-HawkesN equivalence through simulation and fitting}

In this section, we show through simulation the equivalence of HawkesN and SIR on synthetic data.
In Sec.~\ref{subsec:likelihood-formulas}, we sketch the fitting procedures for SIR using maximum likelihood.
In Sec.~\ref{subsec:synthetic-equivalence}, we perform a set of experiments of synthetic data: we demonstrate empirically through simulation and subsequent parameter fitting the equivalence between HawkesN and SIR, and we study some of their key quantities.

\subsection{Maximum likelihood estimates for SIR}
\label{subsec:likelihood-formulas}
\verify{Note that this section should be updated with Lexing's observations from \texttt{extra/comments-versions/2017\_10\_31\_Comments\_pages\_7-9--Lexing.pdf}}

The parameters of both of the flavors of SIR described in Sec.~\ref{subsec:sir-model} -- deterministic and stochastic -- can be fitted from observed data using a maximum likelihood procedure.
In the rest of this section, we describe the observed data and we derive the likelihood functions for each model.

\textbf{Likelihood function for stochastic SIR.}
The SIR process is defined by the three parameters $\{ \beta, \gamma, N\}$ and it can be seen as a marked point process observed a set of events $\{ c_j, t_j, j=1,\ldots,n\}$, where $c_j$ is the class of event $t_j$ (infection or recovery).
The likelihood of observing a particular event $\{ c_j, t_j\}$ has two components: 
the likelihood of observing the inter-arrival time $\Delta t_j = t_j - t_{j-1}$ (note that $\Delta t_j$ is different from  $\tau_j$, the time to recovery defined in Sec.~\ref{subsec:sir-model});
the likelihood of observing an event of that particular class.
The event rate of the point process is $\lambda(t) = \lambda^I(t) + \lambda^R(t)$ (defined in Lemma~\ref{lemma:lambda-SIR-definition}), which is piece-wise constant between events (shown in Fig.~\ref{fig:sir-two-processes}).
Consequently, the likelihood of observing an inter-arrival time $\Delta t_j$ is $\lambda(t_{j-1}) e^{-\lambda(t_{j-1}) \Delta t_j}$.
Finally, the probability of observing the given class of event is given by Eq.~\ref{eq:transition-probs}.
Formally, the likelihood function for the stochastic SIR is:
\begin{align}
	\mathcal{L}( \beta, \gamma, N) = &\prod_{j=2}^{n} \left[ \lambda(t_{j-1}) e^{-\lambda(t_{j-1}) \, (t_j - t_{j-1})}  \right] \nonumber \\
	\times &\prod_{j=2}^{n} \left[\frac{\beta S(t_{j-1})}{\beta S(t_{j-1}) + N \gamma } \mathds{1}(c_j = \text{``infection''}) \right]\nonumber \\
	\times &\prod_{j=2}^{n} \left[\frac{N \gamma}{\beta S(t_{j-1}) + N \gamma} \mathds{1}(c_j = \text{``recovery''}) \right] \label{eq:sir-ll}
\end{align}
We minimize the negative logarithm of the function in Eq.~\eqref{eq:sir-ll} using \texttt{L-BFGS-B}~\cite{Liu1989} with the parameter bounds $\beta > 0, \gamma > 0, N > n$.

\textbf{Likelihood function for deterministic SIR.}
Unlike HawkesN and the stochastic SIR, the deterministic SIR observes volumes of population at discrete time intervals -- $\{S[t], I[t], R[t] \}, t = t_1, t_2, \dots$.
$S[t]$, $I[t]$ and $R[t]$ are time-series.
The key to fitting the parameters of the deterministic SIR ($\{ \beta, \gamma, N\}$) is constructing the predicted time-series $\overbar{S}[t], \overbar{I}[t], \overbar{R}[t]$ by simulating forward the system of differential equations \eqref{eq:sir-ds}-\eqref{eq:sir-dr} starting from $\overbar{S}[0] = N - I(0), \overbar{I}[0] = I(0), \overbar{R}[0] = 0$.
Finally, we either minimize a square error loss metric, or we construct a likelihood metric starting from the observation that the random variable counting the number of events in a Poisson process is Poisson distributed.

\begin{table}[tbp]
	\centering
	\caption{Equivalence of Hawkes and SIR through simulation. All parameters are shown as SIR parameters.} 
	\begin{tabular}{r|rrrr}
	  \toprule
	  parameters & N & $\gamma$ & $\beta$ \\ 
 	 \midrule
		simulation (ideal) & 1300.00 &  0.20 & 1.00 \\ 
		SIR $\rightarrow$ HawkesN & 1300.2 $\pm 8.7$ &  0.19 $\pm 0.04$ & 0.95 $\pm 0.05$ \\ 
		HawkesN $\rightarrow$ SIR & 1311.23 $\pm 28.16$ &  0.23 $\pm 0.09$ & 1.01 $\pm 0.08$ \\ 
 	  \bottomrule
	\end{tabular}
	\label{tab:hawkesN-to-sir}
	\captionmoveup
\end{table}

\input{si-fig-observed-expected}

\subsection{Equivalence on Synthetic Data}
\label{subsec:synthetic-equivalence} 


We study the equivalence of HawkesN and SIR on synthetic data, through simulation and fitting.
We simulate 20 realizations of the stochastic SIR model using a fixed set of parameters.
For each realization, we fit the new infection process $t^I_j$ using HawkesN, following the procedure shown in Sec.~\ref{subsec:likelihood-formulas}.
The HawkesN parameters are mapped into SIR parameters using Theorem~\ref{theorem:expected-equivalence}.
We present in Table~\ref{tab:hawkesN-to-sir} the mean and standard deviation for each fitted parameter.
We also perform the inverse operation: we simulate 20 realization of a HawkesN process using the same (equivalent) previous parameters.
Because the recovery times are not observed, the likelihood corresponding to the inter-arrival times $\Delta t_j$ (first term on the r.h.s. of Eq.~\ref{eq:sir-ll}) is not defined and we cannot fit a stochastic SIR process.
However, we can fit a deterministic SIR by computing population sizes ($S(t)$ and $C(t) = I(t) + R(t)$) at fixed intervals of time.
Table~\ref{tab:hawkesN-to-sir} shows the mean and standard deviation of the fitted parameter.
Visibly, the fitted parameters are very close to the simulation parameters, in accordance with the theoretical results in Sec.~\ref{subsec:linking-two-models}.
Fig.~\ref{subfig:obs-expected-populations} shows the relation between observed and expected SIR population sizes ($I(t)$ and $\mathds{E}_\tau[I(t)]$; $\lambda^I(t)$ and $\mathds{E}_\tau[\lambda^I(t)]$), for one SIR stochastic realization.
We can see that the expectation traces closely the observed values.
Similar conclusions can be drawn from Fig.~\ref{subfig:obs-expected-rates}, for the observed infection and recovery rate ($\lambda^I(t)$ and $\lambda^R(t)$) and their expectation when only the new infection events are observed ($\mathds{E}_\tau[\lambda^I(t)]$ and $\mathds{E}_\tau[\lambda^R(t)]$).

%% file: si-fig-observed-expected.tex

\begin{figure}[tbp]
	\centering
	\newcommand\myheight{0.18}
	\subfloat[] {
		\includegraphics[page=1,height=\myheight\textheight]{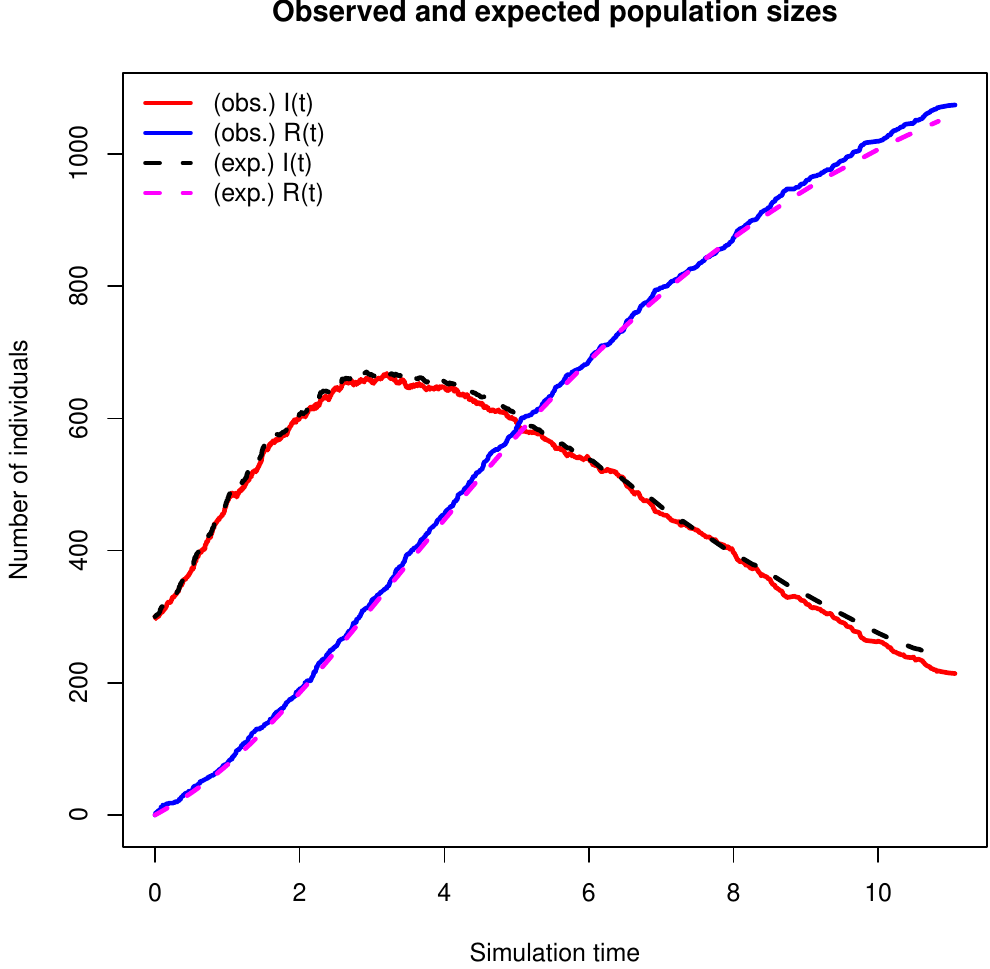}
		\label{subfig:obs-expected-populations}
	}
	\subfloat[] {
		\includegraphics[page=1,height=\myheight\textheight]{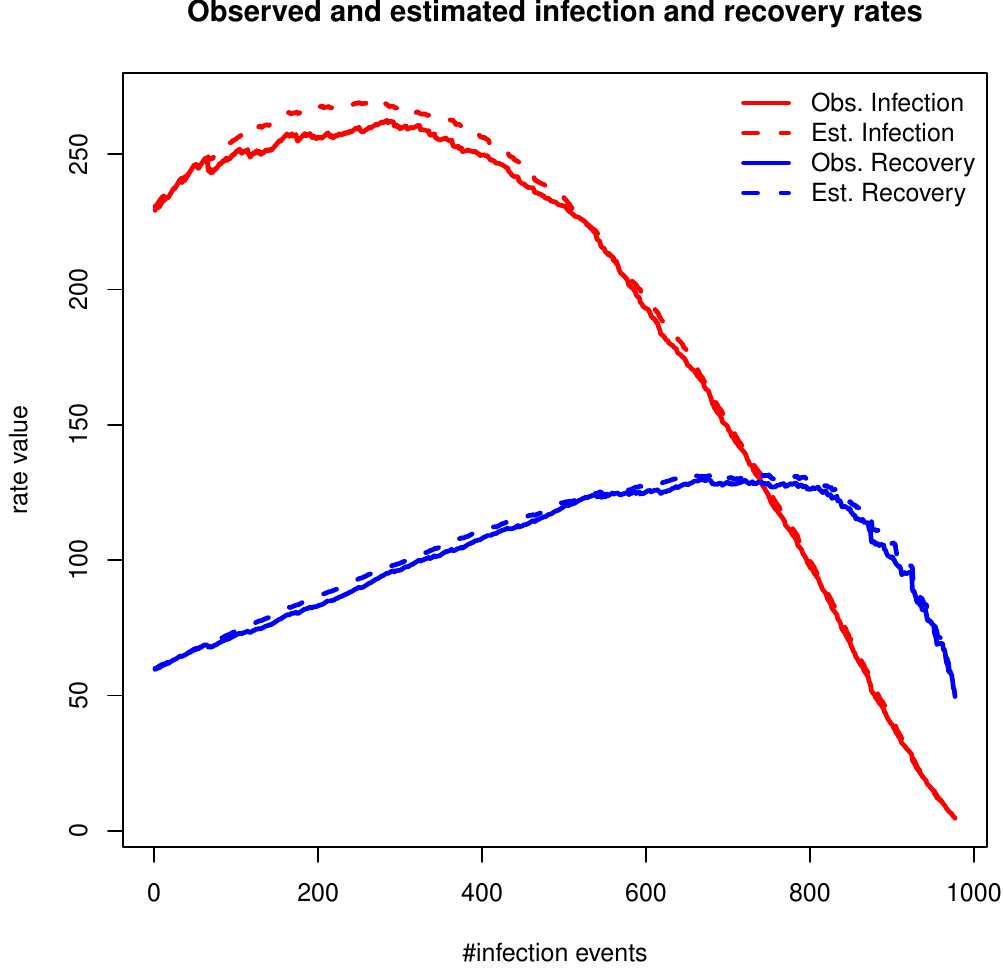}
		\label{subfig:obs-expected-rates}
	}
	\caption{ 
		\textbf{(a)} Observed (continuous lines) and expected (dashed lines) sizes of \textcolor{red}{infected population} ($I(t)$ and $\mathds{E}_{\tau_j}[I(t)]$) and \textcolor{blue}{recovered population} ($R(t)$ and $\mathds{E}_{\tau_j}[R(t)]$);
		\textbf{(b)} Observed (continuous lines) and expected (dashed lines) \textcolor{red}{rate of new infections} ($\lambda^i(t)$ and $\mathds{E}_{\tau_j}[\lambda^i(t)]$) and \textcolor{blue}{rate of new recoveries} ($\lambda^r(t)$ and $\mathds{E}_{\tau_j}[\lambda^r(t)]$). 
		SIR simulated with parameters: $N = 1300, I(0) = 300, \beta = 1, \gamma = 0.2, R_0 = 5$.
	}
	\label{fig:sir-expected-observed}
	\captionmoveup
\end{figure}

%% file: fig4-observed-estimated-quantities-synthetic-data.tex

\begin{figure}[tbp]
	\centering
	\newcommand\myheight{0.16}
	\subfloat[] {
		\includegraphics[height=\myheight\textheight]{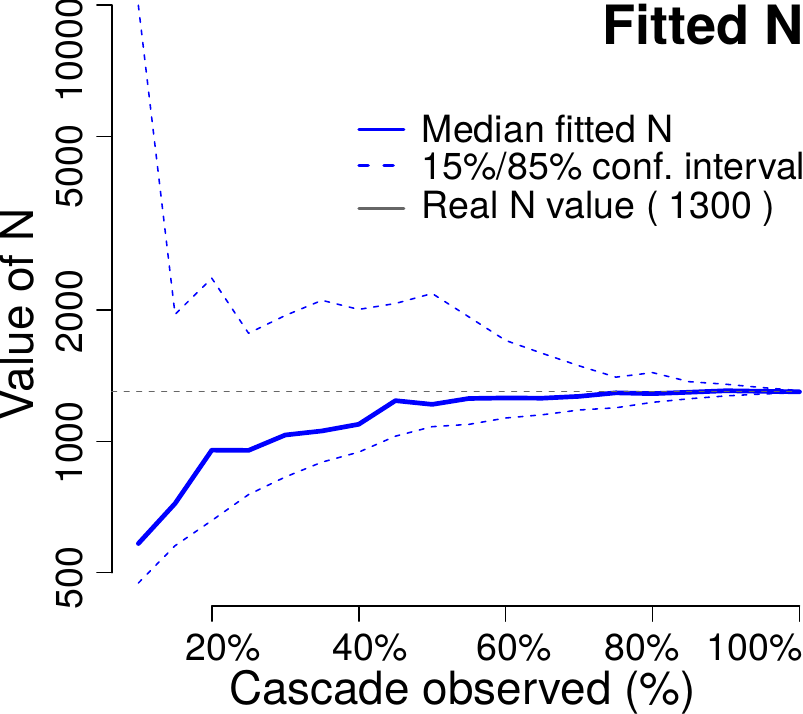}
	} 
	\subfloat[] {
		\includegraphics[height=\myheight\textheight]{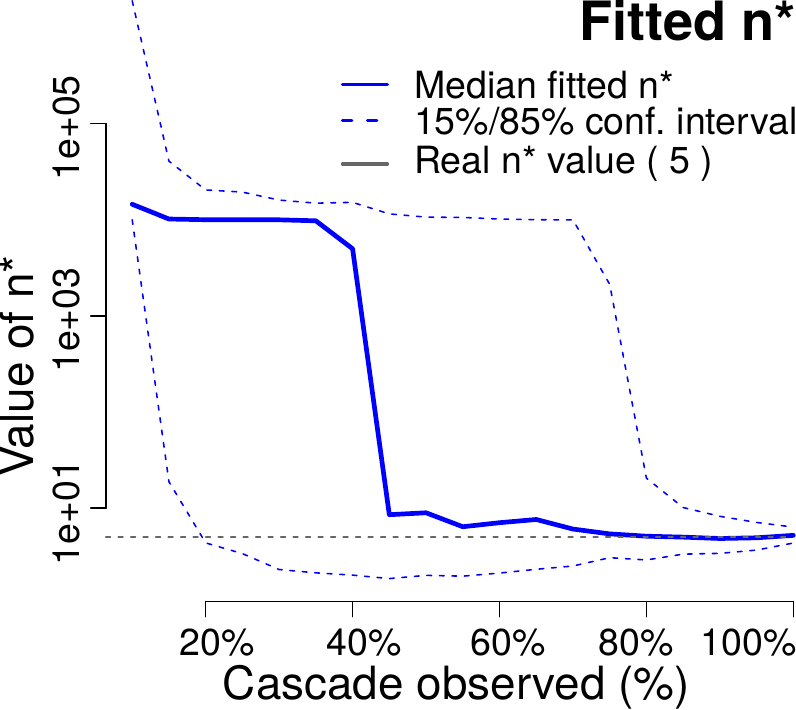}
	}\\
	\subfloat[] {
		\includegraphics[height=\myheight\textheight]{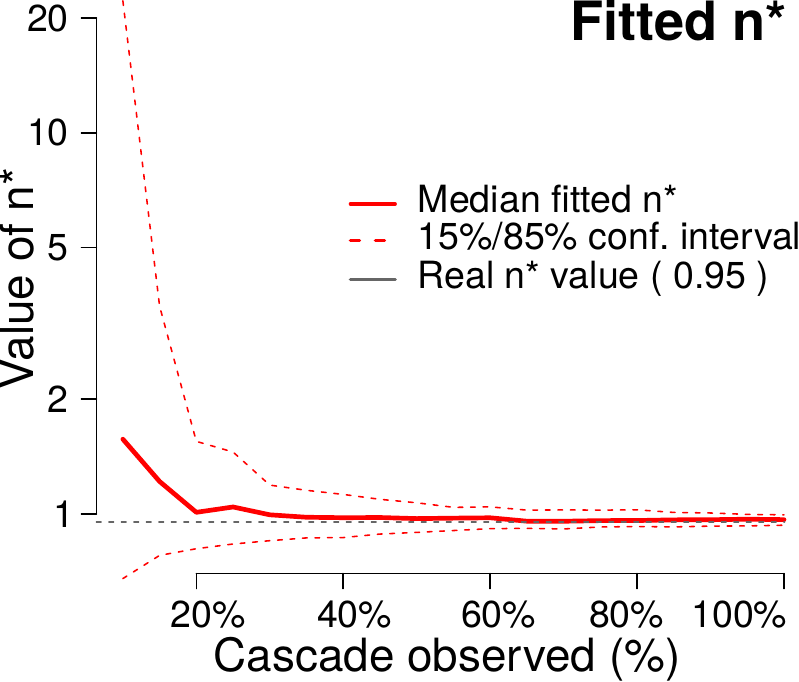}
		\label{subfig:reliability-fit-hawkes}
	}
	\caption{ 
        Robustness of estimating the population size $N$ and the branching factor $n^*$ for HawkesN.
		One set of parameters for each model was simulated 100 times and fitted on increasingly longer prefixes of each simulation.
		One value for $N$ and $n*$ is obtained for each fit and the median and the 15\%/85\% percentile values are shown.
		\verify{\emph{(c)} Reliability of fit for the basic Hawkes model.}
	}
	\label{fig:equivalence-through-fitting}
	\captionmoveup
	\captionmoveup
\end{figure}

%% file: appendix_branching.tex
\section{Narrative on branching factors}

\TODO{MAR to make it consistent after camera-ready submission}

\textbf{Branching factor of Hawkes processes.}
One key quantity that describes the Hawkes process is the branching factor $n^\ast$, defined as the expected number of child events directly spawned by an event.
In a Hawkes process with no immigration ($\mu(t) = 0$), $n^\ast$ is indicative of the expected number of events.
When $n^\ast < 1$, the process in a \emph{subcritical regime}: the number of events is bounded and the event rate $\lambda(t)$ decays to zero over time.
For $n^\ast > 1$, the process is in a \emph{supercritical regime} and the number of events is infinite.

\textbf{Initial branching factor of HawkesN.}
We define the branching factor of HawkesN as the expected number of children events directly spawned \emph{by the first event of the process}.
For large values of $N$ and fast decaying kernel functions $\phi(t)$, we can approximate $\frac{N_t}{N} \approx 0$ and therefore the branching factor for HawkesN is:

\begin{equation} \label{eq:branching-factor-hawkesn}
	n^\ast = \int_0^\infty \kappa \theta e^{-\theta \tau} d\tau = \kappa .
	\eqmoveup
\end{equation}
Note that the branching factor of HawkesN is equivalent to the branching factor of the basic Hawkes process, as in the early stages of the process the population depletion does not play a significant role.
The branching factor
 
is indicative of the speed at which the cascade unfolds and its final size distribution (as shown in Sec.~\ref{subsec:cascade-size}).
%


\textbf{The basic reproduction number} (denoted $\mathcal{R}_0$)
is the expected number of infections caused by a single infected individual at the start of the outbreak. 
Initially, almost all individuals in the population are susceptible $\mathsf{S(0)} \approx N$ and an infectious individual infects others at the constant rate of $\beta\frac{\mathsf{S(t)}}{N} \approx \beta$ for the duration of her infection (which lasts on average $\frac{1}{\gamma}$). 
Consequently, $\mathcal{R}_0=\beta/\gamma$.
$\mathcal{R}_0>1$ is the necessary and sufficient condition to have a growing epidemic:
\begin{align}
	& \frac{d\mathsf{I(0)}}{dt} > 0  \overset{Eq.~\eqref{eq:sir-di}}{\Leftrightarrow} \; \frac{\beta}{\gamma} \frac{\mathsf{S(0)}}{N} > 1 
	\overset{\mathsf{S(0)} \approx N}{\Leftrightarrow} \; \mathcal{R}_0 = \frac{\beta}{\gamma} > 1 \enspace. 
	\eqmoveup
\end{align}

\begin{corollary}  \label{corollary:n-star-r0}
	The Basic Reproduction Number of an SIR process and the branching factor of its equivalent HawkesN process (according to Theorem~\ref{theorem:expected-equivalence}) are equal.
	\vspace{-0.3cm}
\end{corollary}

\begin{equation*} 
	\textit{Proof:}\qquad n^\ast \overset{Eq.~\eqref{eq:branching-factor-hawkesn}}{=} \kappa \overset{Th.~\ref{theorem:expected-equivalence}}{=} \frac{\beta}{\gamma} = \mathcal{R}_0 . 
\end{equation*}

Corollary~\ref{corollary:n-star-r0} is significant because it links two of the most important quantities in the HawkesN and the SIR models, which have been used to address apparently unrelated problems.
For example, the branching factor $n^\ast$ has been used as a threshold in seismology to differentiate between aftershock behavior~\cite{Kagan1991,Helmstetter2002}, in social media analysis to predict information cascade sizes~\cite{Mishra2016,Zhao2015} and to predict the virality and promotion potential of online content~\cite{Rizoiu2017,Rizoiu2017b}.
The basic reproduction number $\mathcal{R}_0$ has be used in epidimiology to quantify the probability of disease extinction, the final size distribution, and expected duration of an epidemic~\cite{Allen2008,Yan2008} and in social media to measure the ``quality'' of retweet cascades~\cite{Martin2016}.
The link shown in this section allows to bring mature techniques employed with SIR into the world of online diffusion modeling with Hawkes processes.

\secmoveup
\subsection{Observations on branching factor}
\label{subsec:branching-factor}

\input{fig6-branching-factor-density}

Here we study the branching factor $n^\ast$.
We fit HawkesN by observing $80\%$ of each cascade, and we compute $n^\ast$ using Eq.~\ref{eq:branching-factor-hawkesn}.
Fig.~\ref{fig:branching-factor-density} (left) shows the density distribution of $n^\ast$  in the three datasets.
For \Seismic and \News, there is a peak around $0.2$, followed by a long tail.
This is consistent with the findings of \citet{Martin2016}.
For \Active however, the density shows a secondary peak around $1.5$, which is probably related to the fact that this dataset contains diffusion about Youtube videos.
We further investigate $n^\ast$ on \Active, by tabulating cascades against the category of the video that the cascade relates to.
Notably, cascades related to \texttt{Sports}, \texttt{People \& Blogs} and \texttt{Film \& Animation} tend to have higher values of $n^\ast$ than the dataset median.
Similarly, cascade in \texttt{Gaming}, \texttt{Howto \& Style} and \texttt{Nonprofit \& Activism} have lower $n^\ast$.
When studying the population size $N$, \texttt{Gaming} stands out as particular category as it has relative high values of $N$.
This is indicative of a large user reach for information relating to \texttt{Gaming} diffusions.

%% file: fig6-branching-factor-density.tex

\begin{figure*}[htbp]
	\centering
	\includegraphics[width=0.95\textwidth]{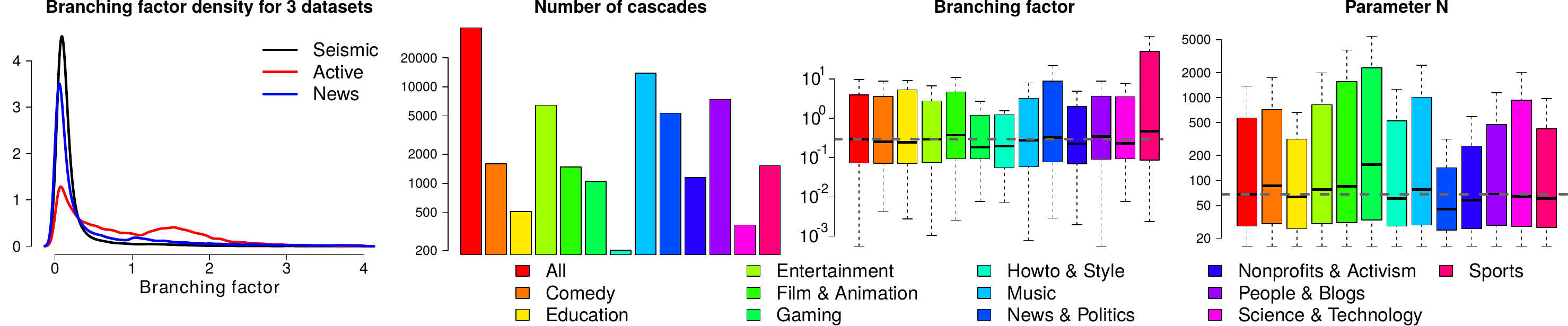}
	\caption{ 
		\textbf{(first panel)} Density distribution for the branching factor of HawkesN, for the three studied datasets (only $n^\ast \leq 4$ is showed here).
		\textbf{(last three panels)} The number of cascades associated with Youtube videos in the \Active dataset, the branching factor $n^\ast$ and population size $N$ (fitted by HawkesN), tabulated against video category.
	}
	\label{fig:branching-factor-density}
	\captionmoveup
\end{figure*}